\PassOptionsToPackage{table,dvipsnames}{xcolor}
\PassOptionsToPackage{noend}{algorithm2e}
\PassOptionsToPackage{skins}{tcolorbox}

\documentclass[format=acmsmall,authorversion,nonacm,screen]{acmart}
\PassOptionsToPackage{inline}{enumitem}

\newif\ifshowcomments
\showcommentsfalse
\showcommentstrue

\newcommand{\commentiff}[1]{%
  \ifshowcomments
    {#1}
  \fi%
}

\makeatletter
\newcommand{\miniscule}{\@setfontsize\miniscule{3.7}{4}}%
\makeatother

\usepackage[ruled,lined,linesnumbered,vlined]{algorithm2e}

\SetCommentSty{mycommfont}
\DontPrintSemicolon %

\usepackage{amsmath}
\usepackage{amsfonts}
\usepackage{amsthm}
\usepackage{balance}
\usepackage{enumitem}
\usepackage{graphicx}
\usepackage{listings}
\usepackage{multicol}
\usepackage{multirow}
\usepackage{natbib}
\usepackage{subcaption}
\usepackage{url}
\usepackage{xspace}
\usepackage{xcolor}
\usepackage{tcolorbox}
\usepackage{colortbl}

\usepackage{tikz}
\usepackage{pgfplots}

\usetikzlibrary{shapes}
\usetikzlibrary{shapes.geometric}
\usetikzlibrary{arrows.meta, positioning}

\newcommand{\eg}{\hbox{\emph{e.g.}}\xspace}

\newcommand{\ie}{\hbox{\emph{i.e.}}\xspace}

\definecolor{BlindColorTolOne}{HTML}{332288}
\definecolor{BlindColorTolTwo}{HTML}{117733} %
\definecolor{BlindColorTolThree}{HTML}{44AA99}
\definecolor{BlindColorTolFour}{HTML}{88CCEE}
\definecolor{BlindColorTolFive}{HTML}{DDCC77}
\definecolor{BlindColorTolSix}{HTML}{CC6677} %
\definecolor{BlindColorTolSeven}{HTML}{AA4499}
\definecolor{BlindColorTolEight}{HTML}{882255}

\definecolor{BlindColorWongOne}{HTML}{000000} %
\definecolor{BlindColorWongTwo}{HTML}{E69F00}
\definecolor{BlindColorWongThree}{HTML}{56B4E9}
\definecolor{BlindColorWongFour}{HTML}{009E73}
\definecolor{BlindColorWongFive}{HTML}{F0E442}
\definecolor{BlindColorWongSix}{HTML}{0072B2} %
\definecolor{BlindColorWongSeven}{HTML}{D55E00}
\definecolor{BlindColorWongEight}{HTML}{CC79A7}

\definecolor{mygreen}{HTML}{02818a}

\mathchardef\mhyphen="2D

\newcommand{\cn}[1]{\NamedCommentTemplate{BlindColorTolSeven}{CN}{#1}}

\newcommand{\yiwen}[1]{\NamedCommentTemplate{BlindColorTolOne}{Yiwen}{#1}}
\newcommand{\victor}[1]{\NamedCommentTemplate{BlindColorTolTwo}{Victor}{#1}}

\newcommand{\zy}[1]{\NamedCommentTemplate{BlindColorTolThree}{Zhenyang}{#1}}

\newcounter{FindingCounter}

\newcommand{\myparagraph}[1]{
  \vspace*{0.04cm}
  \noindent \textit{\textbf{#1.}}\quad
}

\newcommand{\mycode}[1]{\texttt{#1}\xspace}

\newcommand{\projectname}[1]{\textsf{\hbox{#1}}\xspace}
\SetKwData{AlgTrue}{true}
\SetKwData{AlgFalse}{false}

\usepackage{makecell}
\usepackage{booktabs}
\usepackage{comment}
\usepackage{pbalance}
\usepackage{bbding}
\usepackage{tabularx}
\usepackage{arydshln}
\usepackage[normalem]{ulem}
\usepackage[sort,nospace]{cite}
\usepackage{layouts}

\lstdefinelanguage{Snippet}[]{}{
    morekeywords={View, Intent, this, Notification\_morning, class, AlarmManager,
      Activity, PendingIntent, Calendar, timepicker, void}
}

\usetikzlibrary{positioning,fit,calc}

\newcounter{numchallenge}

\newcounter{numrq}

\newcommand{\hide}[1]{}



\pgfplotsset{compat=1.5}

\pgfplotscreateplotcyclelist{color marks}{
black,solid,mark=none\\
red,dashed,mark=none\\
blue,dotted,mark=none\\
violet,dashdotted,mark=none\\
brown,dashdotdotted,mark=none\\
orange,densely dashed,mark=none\\
gray,densely dotted,mark=none\\
}

\NewCommandCopy{\oldcn}{\cn}
\renewcommand{\cn}[1]{\commentiff{\oldcn{#1}}}

\NewCommandCopy{\oldyiwen}{\yiwen}
\renewcommand{\yiwen}[1]{\commentiff{\oldyiwen{#1}}}

\NewCommandCopy{\oldvictor}{\victor}
\renewcommand{\victor}[1]{\commentiff{\oldvictor{#1}}}

\NewCommandCopy{\oldzy}{\zy}
\renewcommand{\zy}[1]{\commentiff{\oldzy{#1}}}

\newcommand*{\rom}[1]{\expandafter{\romannumeral #1\relax}}

\definecolor{specialgreen}{HTML}{70AD47}
\definecolor{lightlightgray}{rgb}{0.9, 0.9, 0.9}

\newcounter{findingbox}
\newtcolorbox{findingbox}{
    colback=gray!10,       %
    colframe=black,        %
    left=2mm,              %
    right=2mm,             %
    top=2mm,               %
    bottom=2mm,            %
    before upper={\stepcounter{findingbox}\emph{Finding~\thefindingbox:}~},                     %
}

\newtcolorbox{examplebox}[1]{
    left=5pt,top=2pt,
    colback=blue!5!gray!10,
    colframe=blue!30!gray,
    coltitle=white,
    sidebyside,
    sidebyside align=top,
    fonttitle=\bfseries,
    title={#1},
}

\newtcolorbox{promptbox}[1]{
    left=5pt,top=2pt,
    colback=teal!10!white,
    colframe=teal!80!black,
    righthand width=100pt,
    segmentation style={solid},
    overlay={\draw[tcbcolframe, line width=1pt] (segmentation.north)--(segmentation.south);},
    skin=bicolor,
    colbacklower=teal!20!white,
    coltitle=white,
    sidebyside,
    sidebyside align=top,
    fonttitle=\bfseries,
    title={#1},
}

\usepackage{xparse}

\NewDocumentCommand{\myspacecode}{v}{%
    \texttt{\obeyspaces\let\ \textvisiblespace #1}\xspace
}

\newif\ifInCopypaper
\InCopypaperfalse

\newcommand{\papertitle}{Unmasking the Genuine Type Inference Capabilities of LLMs for Java Code Snippets}

\newcommand{\snr}{\projectname{SnR}}
\newcommand{\thalia}{\projectname{Thalia}}

\newcommand{\llamas}{Llama3.1:8b\xspace}
\newcommand{\llamam}{Llama3.1:70b\xspace}
\newcommand{\gptfo}{GPT-4o\xspace}
\newcommand{\gptfomini}{GPT-4o-mini\xspace}

\newcommand{\starcoder}{StarCoder2\xspace}
\newcommand{\thestack}{The Stack v2\xspace}
\newcommand{\stacki}{StackV2\xspace}
\newcommand{\starcoderi}{\starcoder{}:15b\xspace}

\newcommand{\stattype}{\projectname{StatType}}

\newcommand{\stattypeso}{\nobreak{StatType-SO}\xspace}
\newcommand{\thaliacs}{\nobreak{ThaliaType}\xspace}
\newcommand{\thaliatype}{\thaliacs}

\newcommand{\sonumcodesnippet}{267\xspace}
\newcommand{\sonumlib}{six\xspace}

\newcommand{\csnumcodesnippet}{300\xspace}

\newcommand{\MyDefine}{\quad\Coloneqq\quad}
\newcommand{\MyAlt}{\;\;\vert\;\;}

\newcommand{\llm}{LLM\xspace}
\newcommand{\llms}{LLMs\xspace}
\newcommand{\fqn}{FQN\xspace}
\newcommand{\fqns}{FQNs\xspace}
\newcommand{\stackoverlow}{StackOverflow\xspace}
\newcommand{\stackoverflow}{StackOverflow\xspace}
\newcommand{\github}{GitHub\xspace}
\newcommand{\llama}{Llama\xspace}
\newcommand{\gpt}{GPT\xspace}

\newcommand{\pdecreasellamasp}{59\%\xspace} %
\newcommand{\pdecreasellamasr}{72\%\xspace} %

\newcommand{\pdecreasedatasetfstar}{66.8\%\xspace} %
\newcommand{\pdecreasedatasetffo}{48.5\%\xspace} %
\newcommand{\pdecreasedatasetfls}{67.2\%\xspace} %
\newcommand{\pdecreasedatasetfsnr}{9.8\%\xspace} %

\newcommand{\priorllmincrease}{7\%\xspace} %

\newcommand{\transformsofolower}{1.7\%\xspace} %
\newcommand{\transformsolslower}{30.0\%\xspace} %
\newcommand{\transformsosclower}{16.8\%\xspace} %

\newcommand{\transformsomaxlslower}{7.9\%\xspace} %
\newcommand{\transformsomaxsclower}{3.0\%\xspace} %

\newcommand{\urldataset}{\url{https://github.com/uw-pluverse/thalia-type}\xspace}

\newcommand{\resultpr}[1]{%
  \ifcase#1\relax%
{}\xspace%
\or
{\multicolumn\{3\}\{c\}\{\snr\}}\xspace%
\or
{\multicolumn\{3\}\{c\}\{\llamas\}}\xspace%
\or
{\multicolumn\{3\}\{c\}\{\llamam\}}\xspace%
\or
{\multicolumn\{3\}\{c\}\{\gptfomini\}}\xspace%
\or
{\multicolumn\{3\}\{c\}\{\gptfo\}}\xspace%
\or
{}\xspace%
\or
{P}\xspace%
\or
{R}\xspace%
\or
{F1}\xspace%
\or
{P}\xspace%
\or
{R}\xspace%
\or
{F1}\xspace%
\or
{P}\xspace%
\or
{R}\xspace%
\or
{F1}\xspace%
\or
{P}\xspace%
\or
{R}\xspace%
\or
{F1}\xspace%
\or
{P}\xspace%
\or
{R}\xspace%
\or
{F1}\xspace%
\or
{\stattypeso}\xspace%
\or
{95.50\%}\xspace%
\or
{91.46\%}\xspace%
\or
{93.44\%}\xspace%
\or
{76.92\%}\xspace%
\or
{69.46\%}\xspace%
\or
{73.00\%}\xspace%
\or
{86.08\%}\xspace%
\or
{83.69\%}\xspace%
\or
{84.87\%}\xspace%
\or
{86.34\%}\xspace%
\or
{89.92\%}\xspace%
\or
{88.09\%}\xspace%
\or
{95.66\%}\xspace%
\or
{95.00\%}\xspace%
\or
{95.33\%}\xspace%
\or
{82.28\%}\xspace%
\or
{81.08\%}\xspace%
\or
{81.67\%}\xspace%
\or
{\thaliacs}\xspace%
\or
{84.15\%}\xspace%
\or
{84.43\%}\xspace%
\or
{84.29\%}\xspace%
\or
{31.27\%}\xspace%
\or
{19.40\%}\xspace%
\or
{23.95\%}\xspace%
\or
{61.58\%}\xspace%
\or
{25.85\%}\xspace%
\or
{36.41\%}\xspace%
\or
{66.64\%}\xspace%
\or
{37.73\%}\xspace%
\or
{48.18\%}\xspace%
\or
{54.74\%}\xspace%
\or
{44.54\%}\xspace%
\or
{49.12\%}\xspace%
\or
{43.46\%}\xspace%
\or
{19.66\%}\xspace%
\or
{27.08\%}\xspace%
  \fi}
\newcommand{\transso}[1]{%
  \ifcase#1\relax%
{}\xspace%
\or
{\multicolumn\{3\}\{c\}\{\snr\}}\xspace%
\or
{\multicolumn\{3\}\{c\}\{\llamas\}}\xspace%
\or
{\multicolumn\{3\}\{c\}\{\llamam\}}\xspace%
\or
{\multicolumn\{3\}\{c\}\{\gptfomini\}}\xspace%
\or
{\multicolumn\{3\}\{c\}\{\gptfo\}}\xspace%
\or
{}\xspace%
\or
{P}\xspace%
\or
{R}\xspace%
\or
{F1}\xspace%
\or
{P}\xspace%
\or
{R}\xspace%
\or
{F1}\xspace%
\or
{P}\xspace%
\or
{R}\xspace%
\or
{F1}\xspace%
\or
{P}\xspace%
\or
{R}\xspace%
\or
{F1}\xspace%
\or
{P}\xspace%
\or
{R}\xspace%
\or
{F1}\xspace%
\or
{\stattypeso}\xspace%
\or
{95.50\%\phantom\{***\}}\xspace%
\or
{91.46\%\phantom\{***\}}\xspace%
\or
{93.44\%\phantom\{***\}}\xspace%
\or
{76.92\%\phantom\{***\}}\xspace%
\or
{69.46\%\phantom\{***\}}\xspace%
\or
{73.00\%\phantom\{***\}}\xspace%
\or
{86.08\%\phantom\{***\}}\xspace%
\or
{83.69\%\phantom\{***\}}\xspace%
\or
{84.87\%\phantom\{***\}}\xspace%
\or
{86.34\%\phantom\{***\}}\xspace%
\or
{89.92\%\phantom\{***\}}\xspace%
\or
{88.09\%\phantom\{***\}}\xspace%
\or
{95.66\%\phantom\{***\}}\xspace%
\or
{95.00\%\phantom\{***\}}\xspace%
\or
{95.33\%\phantom\{***\}}\xspace%
\or
{82.28\%\phantom\{***\}}\xspace%
\or
{81.08\%\phantom\{***\}}\xspace%
\or
{81.67\%\phantom\{***\}}\xspace%
\or
{Rename Variable}\xspace%
\or
{95.50\%\phantom\{***\}}\xspace%
\or
{91.46\%\phantom\{***\}}\xspace%
\or
{93.44\%\phantom\{***\}}\xspace%
\or
{65.28\%***}\xspace%
\or
{56.69\%***}\xspace%
\or
{60.68\%*\phantom\{**\}}\xspace%
\or
{85.34\%\phantom\{***\}}\xspace%
\or
{76.15\%***}\xspace%
\or
{80.49\%***}\xspace%
\or
{89.63\%\phantom\{***\}}\xspace%
\or
{88.46\%**\phantom\{*\}}\xspace%
\or
{89.04\%\phantom\{***\}}\xspace%
\or
{97.03\%\phantom\{***\}}\xspace%
\or
{95.62\%\phantom\{***\}}\xspace%
\or
{96.32\%\phantom\{***\}}\xspace%
\or
{82.09\%\phantom\{***\}}\xspace%
\or
{71.23\%***}\xspace%
\or
{76.28\%\phantom\{***\}}\xspace%
\or
{Lower Code}\xspace%
\or
{95.43\%\phantom\{***\}}\xspace%
\or
{91.62\%\phantom\{***\}}\xspace%
\or
{93.49\%\phantom\{***\}}\xspace%
\or
{69.43\%**\phantom\{*\}}\xspace%
\or
{65.15\%**\phantom\{*\}}\xspace%
\or
{67.22\%***}\xspace%
\or
{83.81\%\phantom\{***\}}\xspace%
\or
{80.46\%*\phantom\{**\}}\xspace%
\or
{82.10\%**\phantom\{*\}}\xspace%
\or
{86.14\%*\phantom\{**\}}\xspace%
\or
{88.92\%\phantom\{***\}}\xspace%
\or
{87.51\%\phantom\{***\}}\xspace%
\or
{94.82\%*\phantom\{**\}}\xspace%
\or
{95.69\%\phantom\{***\}}\xspace%
\or
{95.25\%\phantom\{***\}}\xspace%
\or
{83.72\%\phantom\{***\}}\xspace%
\or
{79.92\%\phantom\{***\}}\xspace%
\or
{81.78\%\phantom\{***\}}\xspace%
\or
{Add Keyword}\xspace%
\or
{95.50\%\phantom\{***\}}\xspace%
\or
{91.46\%\phantom\{***\}}\xspace%
\or
{93.44\%\phantom\{***\}}\xspace%
\or
{78.65\%\phantom\{***\}}\xspace%
\or
{67.46\%\phantom\{***\}}\xspace%
\or
{72.63\%\phantom\{***\}}\xspace%
\or
{85.39\%\phantom\{***\}}\xspace%
\or
{80.92\%**\phantom\{*\}}\xspace%
\or
{83.10\%**\phantom\{*\}}\xspace%
\or
{87.90\%\phantom\{***\}}\xspace%
\or
{89.38\%*\phantom\{**\}}\xspace%
\or
{88.63\%\phantom\{***\}}\xspace%
\or
{96.06\%\phantom\{***\}}\xspace%
\or
{95.77\%\phantom\{***\}}\xspace%
\or
{95.92\%\phantom\{***\}}\xspace%
\or
{83.98\%\phantom\{***\}}\xspace%
\or
{75.00\%***}\xspace%
\or
{79.24\%*\phantom\{**\}}\xspace%
\or
{All}\xspace%
\or
{95.43\%\phantom\{***\}}\xspace%
\or
{91.62\%\phantom\{***\}}\xspace%
\or
{93.49\%\phantom\{***\}}\xspace%
\or
{55.83\%***}\xspace%
\or
{47.15\%***}\xspace%
\or
{51.13\%***}\xspace%
\or
{77.93\%***}\xspace%
\or
{70.08\%***}\xspace%
\or
{73.80\%***}\xspace%
\or
{84.60\%**\phantom\{*\}}\xspace%
\or
{82.85\%***}\xspace%
\or
{83.72\%***}\xspace%
\or
{93.29\%***}\xspace%
\or
{94.15\%*\phantom\{**\}}\xspace%
\or
{93.72\%***}\xspace%
\or
{79.34\%**\phantom\{*\}}\xspace%
\or
{59.38\%***}\xspace%
\or
{67.93\%***}\xspace%
  \fi}
\newcommand{\transcs}[1]{%
  \ifcase#1\relax%
{}\xspace%
\or
{\multicolumn\{3\}\{c\}\{\snr\}}\xspace%
\or
{\multicolumn\{3\}\{c\}\{\llamas\}}\xspace%
\or
{\multicolumn\{3\}\{c\}\{\llamam\}}\xspace%
\or
{\multicolumn\{3\}\{c\}\{\gptfomini\}}\xspace%
\or
{\multicolumn\{3\}\{c\}\{\gptfo\}}\xspace%
\or
{}\xspace%
\or
{P}\xspace%
\or
{R}\xspace%
\or
{F1}\xspace%
\or
{P}\xspace%
\or
{R}\xspace%
\or
{F1}\xspace%
\or
{P}\xspace%
\or
{R}\xspace%
\or
{F1}\xspace%
\or
{P}\xspace%
\or
{R}\xspace%
\or
{F1}\xspace%
\or
{P}\xspace%
\or
{R}\xspace%
\or
{F1}\xspace%
\or
{\thaliacs}\xspace%
\or
{84.15\%\phantom\{***\}}\xspace%
\or
{84.43\%\phantom\{***\}}\xspace%
\or
{84.29\%\phantom\{***\}}\xspace%
\or
{31.27\%\phantom\{***\}}\xspace%
\or
{19.40\%\phantom\{***\}}\xspace%
\or
{23.95\%\phantom\{***\}}\xspace%
\or
{61.58\%\phantom\{***\}}\xspace%
\or
{25.85\%\phantom\{***\}}\xspace%
\or
{36.41\%\phantom\{***\}}\xspace%
\or
{66.64\%\phantom\{***\}}\xspace%
\or
{37.73\%\phantom\{***\}}\xspace%
\or
{48.18\%\phantom\{***\}}\xspace%
\or
{54.74\%\phantom\{***\}}\xspace%
\or
{44.54\%\phantom\{***\}}\xspace%
\or
{49.12\%\phantom\{***\}}\xspace%
\or
{43.46\%\phantom\{***\}}\xspace%
\or
{19.66\%\phantom\{***\}}\xspace%
\or
{27.08\%\phantom\{***\}}\xspace%
\or
{Rename Variable}\xspace%
\or
{84.15\%\phantom\{***\}}\xspace%
\or
{84.43\%\phantom\{***\}}\xspace%
\or
{84.29\%\phantom\{***\}}\xspace%
\or
{31.36\%*\phantom\{**\}}\xspace%
\or
{19.11\%\phantom\{***\}}\xspace%
\or
{23.74\%*\phantom\{**\}}\xspace%
\or
{57.08\%*\phantom\{**\}}\xspace%
\or
{24.47\%*\phantom\{**\}}\xspace%
\or
{34.25\%**\phantom\{*\}}\xspace%
\or
{52.47\%***}\xspace%
\or
{38.44\%\phantom\{***\}}\xspace%
\or
{44.37\%***}\xspace%
\or
{43.88\%***}\xspace%
\or
{45.77\%\phantom\{***\}}\xspace%
\or
{44.80\%*\phantom\{**\}}\xspace%
\or
{46.24\%\phantom\{***\}}\xspace%
\or
{22.42\%***}\xspace%
\or
{30.20\%*\phantom\{**\}}\xspace%
\or
{Lower Code}\xspace%
\or
{84.14\%\phantom\{***\}}\xspace%
\or
{84.39\%\phantom\{***\}}\xspace%
\or
{84.27\%\phantom\{***\}}\xspace%
\or
{27.06\%*\phantom\{**\}}\xspace%
\or
{18.21\%\phantom\{***\}}\xspace%
\or
{21.77\%\phantom\{***\}}\xspace%
\or
{60.16\%\phantom\{***\}}\xspace%
\or
{25.25\%\phantom\{***\}}\xspace%
\or
{35.57\%\phantom\{***\}}\xspace%
\or
{58.04\%***}\xspace%
\or
{36.01\%**\phantom\{*\}}\xspace%
\or
{44.45\%*\phantom\{**\}}\xspace%
\or
{43.45\%\phantom\{***\}}\xspace%
\or
{44.43\%\phantom\{***\}}\xspace%
\or
{43.93\%\phantom\{***\}}\xspace%
\or
{48.09\%*\phantom\{**\}}\xspace%
\or
{23.87\%***}\xspace%
\or
{31.91\%***}\xspace%
\or
{Add Keyword}\xspace%
\or
{84.15\%\phantom\{***\}}\xspace%
\or
{84.43\%\phantom\{***\}}\xspace%
\or
{84.29\%\phantom\{***\}}\xspace%
\or
{25.00\%*\phantom\{**\}}\xspace%
\or
{19.22\%\phantom\{***\}}\xspace%
\or
{21.73\%\phantom\{***\}}\xspace%
\or
{59.56\%\phantom\{***\}}\xspace%
\or
{25.07\%\phantom\{***\}}\xspace%
\or
{35.28\%\phantom\{***\}}\xspace%
\or
{64.27\%\phantom\{***\}}\xspace%
\or
{36.57\%*\phantom\{**\}}\xspace%
\or
{46.62\%\phantom\{***\}}\xspace%
\or
{32.26\%\phantom\{***\}}\xspace%
\or
{44.80\%\phantom\{***\}}\xspace%
\or
{37.51\%\phantom\{***\}}\xspace%
\or
{42.96\%***}\xspace%
\or
{19.66\%\phantom\{***\}}\xspace%
\or
{26.98\%***}\xspace%
\or
{All}\xspace%
\or
{84.14\%\phantom\{***\}}\xspace%
\or
{84.39\%\phantom\{***\}}\xspace%
\or
{84.27\%\phantom\{***\}}\xspace%
\or
{27.57\%*\phantom\{**\}}\xspace%
\or
{17.77\%**\phantom\{*\}}\xspace%
\or
{21.61\%\phantom\{***\}}\xspace%
\or
{45.92\%***}\xspace%
\or
{23.87\%***}\xspace%
\or
{31.41\%*\phantom\{**\}}\xspace%
\or
{47.50\%***}\xspace%
\or
{36.87\%\phantom\{***\}}\xspace%
\or
{41.52\%***}\xspace%
\or
{56.30\%***}\xspace%
\or
{45.29\%\phantom\{***\}}\xspace%
\or
{50.20\%\phantom\{***\}}\xspace%
\or
{40.81\%***}\xspace%
\or
{20.74\%\phantom\{***\}}\xspace%
\or
{27.51\%***}\xspace%
  \fi}
\newcommand{\resultboa}[1]{%
  \ifcase#1\relax%
{\multicolumn\{2\}\{l\}\{Popularity\}}\xspace%
\or
{\multicolumn\{2\}\{l\}\{[0-1e2)\}}\xspace%
\or
{\multicolumn\{2\}\{l\}\{[1e2-1e3)\}}\xspace%
\or
{\multicolumn\{2\}\{l\}\{[1e3-1e4)\}}\xspace%
\or
{\multicolumn\{2\}\{l\}\{[1e4-1e5)\}}\xspace%
\or
{\multicolumn\{2\}\{l\}\{>=1e5\}}\xspace%
\or
{\multirow\{7\}\{*\}\{\rotatebox[origin=c]\{90\}\{\stattypeso\}\}}\xspace%
\or
{Total FQNs}\xspace%
\or
{27}\xspace%
\or
{}\xspace%
\or
{69}\xspace%
\or
{}\xspace%
\or
{312}\xspace%
\or
{}\xspace%
\or
{429}\xspace%
\or
{}\xspace%
\or
{463}\xspace%
\or
{}\xspace%
\or
{}\xspace%
\or
{}\xspace%
\or
{TP}\xspace%
\or
{Recall}\xspace%
\or
{TP}\xspace%
\or
{Recall}\xspace%
\or
{TP}\xspace%
\or
{Recall}\xspace%
\or
{TP}\xspace%
\or
{Recall}\xspace%
\or
{TP}\xspace%
\or
{Recall}\xspace%
\or
{}\xspace%
\or
{\snr}\xspace%
\or
{27}\xspace%
\or
{100.00\%}\xspace%
\or
{60}\xspace%
\or
{86.96\%}\xspace%
\or
{297}\xspace%
\or
{95.19\%}\xspace%
\or
{372}\xspace%
\or
{86.71\%}\xspace%
\or
{433}\xspace%
\or
{93.52\%}\xspace%
\or
{}\xspace%
\or
{\llamas}\xspace%
\or
{2}\xspace%
\or
{7.41\%}\xspace%
\or
{28}\xspace%
\or
{40.58\%}\xspace%
\or
{173}\xspace%
\or
{55.45\%}\xspace%
\or
{322}\xspace%
\or
{75.06\%}\xspace%
\or
{378}\xspace%
\or
{81.64\%}\xspace%
\or
{}\xspace%
\or
{\llamam}\xspace%
\or
{7}\xspace%
\or
{25.93\%}\xspace%
\or
{31}\xspace%
\or
{44.93\%}\xspace%
\or
{228}\xspace%
\or
{73.08\%}\xspace%
\or
{383}\xspace%
\or
{89.28\%}\xspace%
\or
{439}\xspace%
\or
{94.82\%}\xspace%
\or
{}\xspace%
\or
{\gptfomini}\xspace%
\or
{4}\xspace%
\or
{14.81\%}\xspace%
\or
{43}\xspace%
\or
{62.32\%}\xspace%
\or
{272}\xspace%
\or
{87.18\%}\xspace%
\or
{399}\xspace%
\or
{93.01\%}\xspace%
\or
{451}\xspace%
\or
{97.41\%}\xspace%
\or
{}\xspace%
\or
{\gptfo}\xspace%
\or
{6}\xspace%
\or
{22.22\%}\xspace%
\or
{53}\xspace%
\or
{76.81\%}\xspace%
\or
{296}\xspace%
\or
{94.87\%}\xspace%
\or
{421}\xspace%
\or
{98.14\%}\xspace%
\or
{459}\xspace%
\or
{99.14\%}\xspace%
\or
{}\xspace%
\or
{\starcoderi}\xspace%
\or
{7}\xspace%
\or
{25.93\%}\xspace%
\or
{36}\xspace%
\or
{52.17\%}\xspace%
\or
{235}\xspace%
\or
{75.32\%}\xspace%
\or
{372}\xspace%
\or
{86.71\%}\xspace%
\or
{404}\xspace%
\or
{87.26\%}\xspace%
\or
{\multirow\{7\}\{*\}\{\rotatebox[origin=c]\{90\}\{\thaliacs\}\}}\xspace%
\or
{Total FQNs}\xspace%
\or
{994}\xspace%
\or
{}\xspace%
\or
{597}\xspace%
\or
{}\xspace%
\or
{439}\xspace%
\or
{}\xspace%
\or
{324}\xspace%
\or
{}\xspace%
\or
{331}\xspace%
\or
{}\xspace%
\or
{}\xspace%
\or
{}\xspace%
\or
{TP}\xspace%
\or
{Recall}\xspace%
\or
{TP}\xspace%
\or
{Recall}\xspace%
\or
{TP}\xspace%
\or
{Recall}\xspace%
\or
{TP}\xspace%
\or
{Recall}\xspace%
\or
{TP}\xspace%
\or
{Recall}\xspace%
\or
{}\xspace%
\or
{\snr}\xspace%
\or
{735}\xspace%
\or
{73.94\%}\xspace%
\or
{533}\xspace%
\or
{89.28\%}\xspace%
\or
{407}\xspace%
\or
{92.71\%}\xspace%
\or
{300}\xspace%
\or
{92.59\%}\xspace%
\or
{292}\xspace%
\or
{88.22\%}\xspace%
\or
{}\xspace%
\or
{\llamas}\xspace%
\or
{19}\xspace%
\or
{1.91\%}\xspace%
\or
{42}\xspace%
\or
{7.04\%}\xspace%
\or
{64}\xspace%
\or
{14.58\%}\xspace%
\or
{138}\xspace%
\or
{42.59\%}\xspace%
\or
{258}\xspace%
\or
{77.95\%}\xspace%
\or
{}\xspace%
\or
{\llamam}\xspace%
\or
{18}\xspace%
\or
{1.81\%}\xspace%
\or
{80}\xspace%
\or
{13.40\%}\xspace%
\or
{144}\xspace%
\or
{32.80\%}\xspace%
\or
{182}\xspace%
\or
{56.17\%}\xspace%
\or
{270}\xspace%
\or
{81.57\%}\xspace%
\or
{}\xspace%
\or
{\gptfomini}\xspace%
\or
{56}\xspace%
\or
{5.63\%}\xspace%
\or
{161}\xspace%
\or
{26.97\%}\xspace%
\or
{227}\xspace%
\or
{51.71\%}\xspace%
\or
{269}\xspace%
\or
{83.02\%}\xspace%
\or
{300}\xspace%
\or
{90.63\%}\xspace%
\or
{}\xspace%
\or
{\gptfo}\xspace%
\or
{103}\xspace%
\or
{10.36\%}\xspace%
\or
{233}\xspace%
\or
{39.03\%}\xspace%
\or
{256}\xspace%
\or
{58.31\%}\xspace%
\or
{288}\xspace%
\or
{88.89\%}\xspace%
\or
{316}\xspace%
\or
{95.47\%}\xspace%
\or
{}\xspace%
\or
{\starcoderi}\xspace%
\or
{22}\xspace%
\or
{2.21\%}\xspace%
\or
{58}\xspace%
\or
{9.72\%}\xspace%
\or
{94}\xspace%
\or
{21.41\%}\xspace%
\or
{120}\xspace%
\or
{37.04\%}\xspace%
\or
{234}\xspace%
\or
{70.69\%}\xspace%
  \fi}
\newcommand{\reducebothboxmean}[1]{%
  \ifcase#1\relax%
{41.49\%}\xspace%
\or
{9.12\%}\xspace%
  \fi}
\newcommand{\reducebothboxmedian}[1]{%
  \ifcase#1\relax%
{33.33\%}\xspace%
\or
{-0.00\%}\xspace%
  \fi}
\newcommand{\reduceboxmean}[1]{%
  \ifcase#1\relax%
{6.37\%}\xspace%
\or
{14.73\%}\xspace%
  \fi}
\newcommand{\reduceboxmedian}[1]{%
  \ifcase#1\relax%
{4.58\%}\xspace%
\or
{8.74\%}\xspace%
  \fi}
\newcommand{\ratiotoken}[1]{%
  \ifcase#1\relax%
{Average}\xspace%
\or
{4.82\%}\xspace%
\or
{95803}\xspace%
\or
{4619}\xspace%
\or
{Average}\xspace%
\or
{17.90\%}\xspace%
\or
{91166}\xspace%
\or
{16317}\xspace%
\or
{Control and Structural}\xspace%
\or
{3.17\%}\xspace%
\or
{36791}\xspace%
\or
{1165}\xspace%
\or
{Control and Structural}\xspace%
\or
{14.39\%}\xspace%
\or
{37864}\xspace%
\or
{5450}\xspace%
\or
{Inferred Simple Name}\xspace%
\or
{16.88\%}\xspace%
\or
{4954}\xspace%
\or
{836}\xspace%
\or
{Inferred Simple Name}\xspace%
\or
{53.33\%}\xspace%
\or
{4478}\xspace%
\or
{2388}\xspace%
\or
{Whitespace}\xspace%
\or
{6.18\%}\xspace%
\or
{29388}\xspace%
\or
{1816}\xspace%
\or
{Whitespace}\xspace%
\or
{20.95\%}\xspace%
\or
{29144}\xspace%
\or
{6105}\xspace%
\or
{Other}\xspace%
\or
{3.25\%}\xspace%
\or
{24670}\xspace%
\or
{802}\xspace%
\or
{Other}\xspace%
\or
{12.06\%}\xspace%
\or
{19680}\xspace%
\or
{2374}\xspace%
  \fi}

\usepackage{cleveref} %

\Crefname{algocf}{Algorithm}{Algorithms}
\crefname{algocf}{Algorithm}{Algorithms}

\Crefname{algorithm}{Algorithm}{Algorithms}
\crefname{algorithm}{Algorithm}{Algorithms}

\crefname{appendix}{Appendix}{Appendices}
\Crefname{appendix}{Appendix}{Appendices}

\Crefname{figure}{Figure}{Figures}
\crefname{figure}{Figure}{Figures}

\crefname{listing}{Listing}{Listings}
\Crefname{listing}{Listing}{Listings}

\Crefname{table}{Table}{Tables}
\crefname{table}{Table}{Tables}

\crefname{thm}{Theorem}{Theorems}
\Crefname{thm}{Theorem}{Theorems}

\crefname{equation}{Equation}{Equations}
\Crefname{equation}{Equation}{Equations}

\crefformat{chapter}{\S~#2#1#3}
\crefmultiformat{chapter}{\S\S~#2#1#3}{ and~#2#1#3}{, #2#1#3}{, and~#2#1#3}

\crefformat{section}{\S~#2#1#3}
\crefmultiformat{section}{\S\S~#2#1#3}{ and~#2#1#3}{, #2#1#3}{, and~#2#1#3}

\AtBeginDocument{%
  \providecommand\BibTeX{{%
    \normalfont B\kern-0.5em{\scshape i\kern-0.25em b}\kern-0.8em\TeX}}}

\setcopyright{acmlicensed}
\copyrightyear{2018}
\acmYear{2018}
\acmDOI{XXXXXXX.XXXXXXX}

\acmISBN{978-1-4503-XXXX-X/18/06}

\begin{document}

\title{\papertitle}

\author{Yiwen Dong}
\orcid{0000-0002-3205-9010}
\email{yiwen.dong@uwaterloo.ca}
\affiliation{%
    \institution{University of Waterloo}
    \city{Waterloo}
    \state{ON}
    \country{Canada}
}

\author{Zhenyang Xu}
\orcid{0000-0002-9451-4031}
\email{zhenyang.xu@uwaterloo.ca}
\affiliation{%
    \institution{University of Waterloo}
    \city{Waterloo}
    \state{ON}
    \country{Canada}
}

\author{Yongqiang Tian}
\orcid{0000-0003-1644-2965}
\affiliation{%
    \institution{Monash University}
    \country{Australia}}
\email{yongqiang.tian@monash.edu}

\author{Chengnian Sun}
\orcid{0000-0002-0862-2491}
\email{cnsun@uwaterloo.ca}
\affiliation{%
    \institution{University of Waterloo}
    \city{Waterloo}
    \state{ON}
    \country{Canada}
}

\begin{abstract}

Type inference is a crucial task for reusing online code snippets.
Although code snippets are prevalently shared on platforms like \stackoverflow,
they frequently lack essential type information, such as fully qualified names~(\fqns) and required libraries.
Recent studies have leveraged Large Language Models~(\llms)
to perform type inference for such code snippets,
demonstrating promising performance.
However, these evaluations may suffer from
data leakage,
as the benchmark suite, \stattypeso, used for evaluation has been publicly
available on \github since 2017.
Consequently, it remains uncertain
whether the strong performance of \llms
reflects genuine semantic understanding of code or
is due to the ground truth being included in the training set.

This paper strives to comprehensively evaluate the genuine type inference capabilities of \llms
on Java code snippets and identify potential limitations of \llms.
First,
we created \thaliacs{}, a new, previously unreleased benchmark suite
designed for type inference evaluation. %
Second, using the \starcoder \llm as a baseline,
we uncovered data leakage from \stattypeso in
\starcoder's open-source training set
and observed that other
state-of-the-art \llms exhibit similar performance drops
when evaluated on \thaliacs{}, with precision decreasing by up to \pdecreasellamasp
and recall by up to \pdecreasellamasr.
Finally, we developed semantic-preserving code transformations
to further investigate the capabilities of \llms in understanding the execution semantics of code snippets.
Our results showed that the performance of \llms on \stattypeso is far less robust
to these transformations than on \thaliatype,
suggesting that the performance on \stattypeso may be biased by data leakage and have limited generalizability.

These findings highlight the importance of carefully designed, leakage-free benchmarks
for evaluating \llms on type inference tasks.
We recommend future studies adopt \thaliatype to ensure rigorous and reliable assessments of \llms' genuine
type inference capabilities.

\end{abstract}



\maketitle

\section{Introduction}
\label{sec:intro}

Java code snippets found online, such as those from \stackoverlow, are often not compilable or directly
reusable because they lack crucial type information, including fully qualified names
(\fqns, \ie, import statements) and required libraries.
Developers must manually infer and recover this missing information
in order to reuse these code snippets~\citep{Terragni:ISSTA:16,Terragni:ASE:21}.
This process requires understanding the libraries
and identifying the correct \fqns based on the names and functions used in the code snippets.
For instance, if a code snippet references a type with the simple name \mycode{Logger},
the developer must determine whether the appropriate import is \mycode{java.util.logging.Logger}
from Java's utilities library or \mycode{org.slf4j.Logger} from another logging library,
depending on the methods invoked on the objects of \mycode{Logger} in the code snippet.

Type inference for code snippets can assist developers
in recovering missing type information
needed to reuse these code
snippets~\citep{SnR,Subramanian:ICSE:14,Phan:ICSE:18,Saifullah:ASE:19,Huang:ASE:2022,Kabir:TOSEM:24}.
For example,
current approaches to type inference for Java code snippets
generally fall into two  categories:
\textit{constraint-based} and
\textit{machine learning (ML)-based} techniques.
Constraint-based methods build a knowledge base of libraries and the types contained
within~\citep{SnR,Subramanian:ICSE:14}.
The state-of-the-art technique \snr~\citep{SnR}
leverages this knowledge base and uses constraints extracted from the code snippet
to identify the precise types being used.

ML-based approaches~\citep{Phan:ICSE:18,Saifullah:ASE:19,Huang:TOSEM:2023,Kabir:TOSEM:24}
learn from prior examples of type usage in existing code to perform type inference.
They often produce incorrect inferences
due to their limited understanding of code structure and the rules governing Java’s type system.
Recent advancements in large language models (\llms),
such as OpenAI’s proprietary \gpt family and
Meta’s open-weight \llama models,
offer new possibilities for overcoming these limitations.
By leveraging training on diverse and extensive datasets,
\llms have shown potential in performing various software engineering tasks~\citep{Hou:TOSEM:24}.
Prior studies~\citep{Huang:TOSEM:2023,Kabir:TOSEM:24,Chen:Arxiv:24} suggest that
\llms achieved performance comparable to that of the
state-of-the-art \snr.

\begin{table}[t]
	\caption{Overview of the benchmark suites and language models used to evaluate \llm performance.
The early publication of \stattypeso, combined with the late knowledge cutoff dates of the models,
increases the likelihood of data leakage.
}
	\label{tab:model-dataset-stats}
	\scriptsize
	\centering
	\begin{subtable}[t]{\textwidth}
		\caption{The benchmark suites used for evaluation.}
		\label{subtab:dataset-stats}
		\centering
		\begin{tabular}{lcc}
			\toprule
			Benchmark Suite & First Available & \#Code Snippets \\
			\midrule
			\stattypeso & 2017~(see \cref{tab:stack-repos}) & \sonumcodesnippet \\
			\thaliacs & \emph{Always New} & \csnumcodesnippet \\
			\bottomrule
		\end{tabular}
	\end{subtable}

\vspace{0.2cm}

	\begin{subtable}[t]{\textwidth}
		\caption{Evaluated models with their respective version identifiers, knowledge cutoff dates, and whether \stattypeso was present in their training data.
		The question mark denotes that \stattypeso's presence cannot be confirmed.}
		\label{subtab:model-stats}
		\centering
		\begin{tabular}{lllc}
			\toprule
			Model & Version & Knowledge Cutoff & Data Leakage \\
			\midrule
			\starcoderi & instruct-v0.1-q4\_0 & September 2023~\citep{starcoder2} & \Checkmark \\
			\gptfo & 2024-08-06 & October 2023~\citep{OPENAIMODELS} & \textbf{?} \\
			\gptfomini & 2024-07-18 & October 2023~\citep{OPENAIMODELS} & \textbf{?} \\
			\llamas & instruct\_q4\_0 & December 2023~\citep{llama3} & \textbf{?} \\
			\llamam & instruct\_q4\_0 & December 2023~\citep{llama3} & \textbf{?} \\
			\bottomrule
		\end{tabular}
	\end{subtable}
\end{table}

However, the evaluation of these prior studies~\citep{Huang:TOSEM:2023,Kabir:TOSEM:24,Chen:Arxiv:24}
is potentially affected by \emph{data leakage}~\citep{Inan:MS:21} where the training dataset
includes the benchmark dataset together with the ground truth,
and data leakage allows \llms to regurgitate the training
data rather than conducting type inference.
The benchmark suite \stattypeso, used for evaluating type inference techniques has been fully, publicly
available on \github since 2017~\citep{StatTypeSO:GitHub:17:full}.
As shown in
\cref{tab:model-dataset-stats}, recent state-of-the-art \llms have knowledge cutoff dates in late
2023, creating the potential for leakage.
Thus, it remains uncertain whether the \llms' strong performance stems from their ability to understand
the semantics of the code snippets or merely from retrieving the ground truth from their training
data.
While recent work has called attention to this data leakage problem for
software engineering research~\citep{Ozkaya:Software:23,Sallou:ICSENIER:24},
\emph{no study has thoroughly, empirically evaluated \llms' performance on type inference on code snippets}.
Because prior work relied on \gpt models where the exact training data is kept confidential,
detecting such leakage is difficult~\citep{Carlini:USENIX:21,Nasr:Arxiv:23}.

\ifInCopypaper
\else
\begin{figure}
\includegraphics[width=0.85\linewidth, trim={0 290pt 365pt 0},clip]{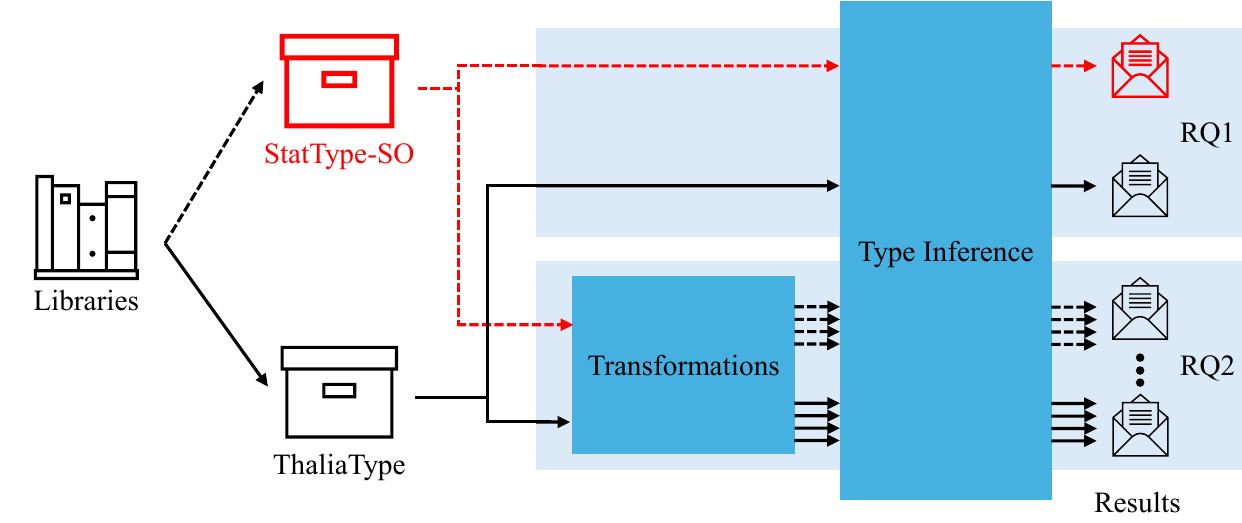}
\caption{The general workflow.
The libraries used to collect code snippets in \stattypeso were also used to generate code snippets in \thaliatype.
These code snippets are used in RQ1, and their transformed versions are used in RQ2 to evaluate various
type inference techniques.
Their results are compared to evaluate the genuine type inference capabilities of \llms for Java code snippets.
Red indicates potential data leakage.
}
\label{fig:rq1-2-workflow}
\end{figure}

\fi

To address this limitation,
we first investigated \thestack~\citep{the-stack-v2}~(simply referred to as \stacki), an open-source dataset used to train the \starcoder model.
Our exploration revealed that all code snippets from \stattypeso with \textit{ground truth} were present in \stacki.
This means that \starcoder was trained using
code snippets along with their expected import statements from \stattypeso.
Thus, \starcoder's performance is affected by data leakage,
and can potentially recall the code snippets and their expected import statements
from training rather than performing type inference
by analyzing the semantics of the code snippets.
To better understand the behavior of \llms, we compare \starcoder's performance against \llama and \gpt models.
We organized our inquiry into the following two research questions as illustrated in \cref{fig:rq1-2-workflow}.

In this paper, we thoroughly evaluate \llms' type inference performance on Java code snippets,
focusing on their ability to leverage type information,
the impact of data leakage,
and their generalizability to unseen code.

\noindent\textbf{\emph{RQ1: How well do \llms perform type inference on unseen code snippets?}}
\quad
To address data leakage, and investigate \llms' genuine ability to perform type inference using the semantic
information present in code snippets,
we generated a new benchmark suite named \thaliacs using \thalia,
a program synthesis technique originally designed for testing static type systems~\citep{Sotiropoulos:POPL:24}.
\thaliacs comprises \csnumcodesnippet code snippets that are guaranteed to be unseen by the evaluated \llms,
thereby avoiding data leakage.
Unlike \stattypeso, which was published in 2017~(\cref{subtab:dataset-stats}),
new code snippets can always be generated using our replication package,
enabling future evaluations without data leakage.
To generate \thaliacs, we used the same set of libraries used in \stattypeso to
\begin{enumerate*}
\item
enable a direct comparison of type inference performance between the two benchmark suites and
\item
ensure the evaluated \llms were consistently and sufficiently exposed to the types used in both benchmark suites during training.
\end{enumerate*}
By evaluating \llms on \thaliacs and contrasting the results with those from \stattypeso,
we can gain a better understanding of the impact of data leakage on performance.

\noindent\textbf{\emph{RQ2: To what extent do \llms understand the execution semantics of code snippets?}}
\quad
To investigate whether \llms truly understand
the execution semantics of code snippets during type inference,
we designed \emph{semantic-preserving} code transformations
to create syntactically different versions of the input code (\cref{sec:transformations}).
These transformations were intended to hinder \llms' reliance
on memorized training data by making direct recall more challenging.
By observing how \llm' performance changes on these semantically equivalent
but syntactically distinct versions, we can gain
insights into their ability to grasp the underlying semantic meaning of the code.

Following recommendations from prior software engineering research~\citep{Sallou:ICSENIER:24},
we evaluate \llm's performance using three open-weight models \starcoder, \llamas, and \llamam,
along with state-of-the-art closed-source models \gptfo and \gptfomini~(see \cref{subtab:model-stats}).
Among these, \starcoder serves as a baseline model with confirmed data leakage~(see \cref{subsec:training-data-starcoder}).
For additional comparison, we include \snr~\citep{SnR}, a state-of-the-art
constraint-based type inference technique, as a non-LLM baseline.
This evaluation aims to provide deeper insights into the potential limitations of using \llms and
to guide future research evaluating \llms' performance on other software engineering tasks.

\myparagraph{Findings of RQ1}
In RQ1, we found that \llms are prone to making incorrect
inferences on code snippets from \thaliatype,
likely because these snippets are truly unseen and not affected by data leakage.
All tested \llms exhibited a similar decline in performance as \starcoderi which experienced a \pdecreasedatasetfstar decrease in F1-score.
For example, \llamas and \gptfo showed comparable declines of \pdecreasedatasetfls and \pdecreasedatasetffo respectively.
On both \stattypeso and \thaliatype, \starcoderi outperformed \llamas but was outperformed by \llamam,
\gptfomini, and \gptfo.
In contrast, \snr, which is not affected by data leakage, obtained a precision of \resultpr{42}
and a recall of \resultpr{43} on \thaliacs.
The consistent drop in \llms' performance between the baseline \starcoderi and other models
suggests that all tested \llms are potentially affected by data leakage.
In addition, although \starcoderi was trained on \stattypeso snippets,
it only achieved an F1-score of \resultpr{40},
indicating that even in the presence of data leakage, models do not perfectly recall the training data.

\myparagraph{Findings of RQ2}
In RQ2, \llms showed robustness to individual transformations on both benchmark suites,
and also to the combined transformation on \thaliatype code snippets.
\starcoderi showed only a \transformsomaxsclower decrease in F1-score in the worst case,
while \llamas experienced a \transformsomaxlslower drop.
\emph{However}, combined transformations on \stattypeso consistently led to performance degradation across all models,
including a \transformsosclower reduction for \starcoder and up to \transformsolslower for \llamas.
Notably, these same combined transformations did not consistently degrade performance on \thaliatype.
In fact, \starcoderi experienced a slight improvement in F1-score.
One possible explanation is the presence of data leakage, which may allow models to recall answers from training.
When transformations introduce enough variation to prevent direct recall,
the models are forced to rely on semantic reasoning.
Under these conditions, their performance on \stattypeso aligned more closely with that on truly unseen data.

Our results suggest that future evaluations of \llm-based techniques should also assess generalizability,
defined as the ability for \llm-based techniques to maintain consistent performance across unseen code snippets.
This can be achieved by incorporating both transformed code snippets and \thaliatype,
rather than relying solely on \stattypeso, which may suffer from data leakage.

\myparagraph{Contribution}
This paper makes the following contributions.
\begin{itemize}[leftmargin=*]
\item
We introduce \thaliatype,
a new benchmark suite constructed through a novel adaptation of \thalia for rigorously evaluating
the type inference performance of \llms while mitigating potential bias from data leakage.
Using \thaliatype, we demonstrate that prior evaluations based on \stattypeso are likely affected by
data leakage, leading to inflated and potentially misleading performance results.

\item
We evaluate whether \llms understand the execution semantics of code snippets by applying transformations that
preserve the semantics and comparing their effects.
Our results on \stattypeso and \thaliatype show that \llms are generally able to maintain performance
across individual transformations.
However, complex transformations on \stattypeso result in disproportionately
large performance drops compared to both their effects on \thaliatype and the individual transformations.
This lack of generalizability can be caused by data leakage where models learned the correct answer from training.

\item
To enable generation of new code snippets for future research on \llm type inference for code snippets,
we have made our benchmark suite and code publicly available at \urldataset.

\end{itemize}

\section{Type Inference on Java Code Snippets}
\label{sec:background}

\textit{Type inference on code snippets}
aims to automatically determine the precise types within the code,
specifically their \fqns~\citep{Subramanian:ICSE:14,SnR,Phan:ICSE:18, Saifullah:ASE:19,Huang:TOSEM:2023,Kabir:TOSEM:24}.
This process is essential for effectively reusing code snippets from sources like \stackoverlow,
where type information is often incomplete or missing entirely.

\begin{figure}[t]
\begin{subfigure}[b]{0.6\textwidth}
\centering
\begin{lstlisting}[xleftmargin=14pt]
    String s = "Example";
    Logger.getLogger(s);
\end{lstlisting}
\caption{An example code snippet using a type with the simple name \mycode{Logger} and a method called \mycode{getLogger} that takes a \mycode{String}.}
\label{subfig:example-code}
\end{subfigure}%
\hfill
\begin{subfigure}[b]{0.350\textwidth}
\begin{center}
\scriptsize
\setlength{\tabcolsep}{2pt}
\begin{tabular}{ll}
\toprule
Simple Name & Inferred FQN \\
\midrule
String & java.lang.String \\
Logger & java.util.logger.Logger \\
\bottomrule
\end{tabular}
\end{center}
\caption{The inferred FQNs for the simple names used in the code snippet.}
\label{subfig:example-out}
\end{subfigure}%
\vspace*{1em}
\begin{subfigure}[t]{\textwidth}
\begin{center}
\scriptsize
\begin{tabular}{lllll}
\toprule
Simple Name & FQN & Method Name & Return Type & Argument Types \\
\midrule
Logger & java.util.logging.Logger & getLogger & java.util.logging.Logger & java.lang.String \\
Logger & java.util.logging.Logger & info & void & java.lang.String \\
Logger & org.slf4j.Logger & info & void  & java.lang.String \\
Logger & org.slf4j.Logger & debug & void  & java.lang.String \\
String & java.lang.String & length & int & $\emptyset$ \\
\ldots & \ldots & \ldots & \ldots & \ldots \\
\bottomrule
\end{tabular}
\end{center}
\caption{A simplified knowledge base containing mappings of simple names, methods, return types, and argument types.
The symbol $\emptyset$ represents that the method takes in no arguments.
Additional types and methods are omitted with \ldots for brevity.
}
\label{subfig:example-kb}
\end{subfigure}

\caption{An example code snippet with example output and example knowledge base used by
constraint-based type inference techniques for resolving ambiguities in types.}
\label{fig:example}
\end{figure}

A key challenge in type inference lies in resolving ambiguities with simple names.
For instance, as shown in \cref{subfig:example-code}, the simple name \mycode{Logger} could refer to types defined in various libraries,
such as the
\mycode{JDK}~\citep{JDKLogger}, \mycode{SLF4J}~\citep{SLFLogger}, \mycode{Log4j}~\citep{LogLogger}, or \mycode{Logback}~\citep{LogBackLogger}.
Disambiguating these references necessitates a deep semantic analysis of the snippet, including method signatures~(\eg, \mycode{getLogger}),
and their argument types.
The complexity of this task is significantly increased by the vast search space of possible type combinations across numerous libraries,
making the inference process computationally demanding.

This section provides an overview of state-of-the-art machine learning-based approaches to type inference,
along with our prompting strategy~(\cref{subsec:background-ml}).
Additionally, we briefly introduce \snr,
a constraint-based type inference technique, as a baseline for comparison against \llms~(\cref{subsec:snr}).

\subsection{Machine Learning-Based Type Inference}
\label{subsec:background-ml}
Machine learning-based methods use statistical models and learned representations of code to
conduct type inference and resolve ambiguous simple names in code snippets~\citep{Phan:ICSE:18, Saifullah:ASE:19,Huang:TOSEM:2023,Kabir:TOSEM:24}.
Of these, \llms have recently achieved state-of-the-art performance on type
inference on the \stattypeso~\citep{Huang:TOSEM:2023,Kabir:TOSEM:24,Chen:Arxiv:24} benchmark suite.
Existing approaches typically employ a simple but effective prompt to guide \llms in
generating \fqns for types in the code snippet~\citep{Huang:TOSEM:2023,Kabir:TOSEM:24,Chen:Arxiv:24}.
Notably, \citet{Kabir:TOSEM:24} extended the single-prompt approach by incorporating a secondary prompt
to address compilation errors in the code snippet generated by the first prompt.
This iterative process achieved 99.5\% precision on \stattypeso.
In contrast, our work focuses on the issue of \emph{data leakage} in type inference
rather than repairing code snippets.
Data leakage is a fundamental problem that compromises the validity of results,
regardless of whether a single-prompt or iterative approach is used.
To ensure consistency and comparability with prior research,
we adopted a single-prompt approach for type inference.
Following established practices~\citep{Huang:TOSEM:2023,Kabir:TOSEM:24},
we developed a straightforward but highly effective prompt for \llm-based type inference.

\myparagraph{Prompt Template}
\label{subsec:prompt}
Our prompt, shown in \cref{fig:prompt},
is designed to guide the \llm in inferring the type information in Java code snippets.
It includes two key components:
\begin{enumerate*}
\item
a system message that sets the \llm's role as a helpful assistant, and
\item
a user instruction explicitly requesting the addition of import statements for Java code.
\end{enumerate*}
The placeholder \texttt{\{input\_code\}} is replaced with the actual code snippet during inference.
\cref{subfig:prompt-text} presents the exact wording of the prompt,
while \cref{subfig:prompt-example} demonstrates the corresponding output generated by the \llm for an example code snippet.
Despite its simplicity, our prompt achieves high accuracy
and aligns with best practices for \llm prompt engineering~\citep{LLMPratice}.
Notably, our results using \gptfo on the \stattypeso benchmark suite represent a \priorllmincrease improvement
over the prior results~\citep{Kabir:TOSEM:24},
demonstrating the robustness of our prompt while maintaining a similar level of complexity.

\begin{figure}
\begin{subfigure}{0.67\columnwidth}
\footnotesize

\begin{tabular}{l|p{219pt}}
\hline
  \textbf{System} &  You are a helpful programming assistant.\\
  \textbf{User}  & Add import statements to the following Java code. Do not use wildcard imports. Include only the necessary import statements. Do not import nonexistent types. Please note that you need to pay close attention and your response should be specific and accurate. Avoid repetition. Reply with the import statements only.\\
                 & \vspace*{1px}\texttt{\{input\_code\}}\\
\hline
\end{tabular}

\caption{Example prompt used to infer types on code snippets with \llms. The \texttt{\{input\_code\}} is replaced by the code snippet.
}
\label{subfig:prompt-text}
\end{subfigure}
\hfill
\begin{subfigure}{0.3\columnwidth}
\small
\begin{lstlisting}[basicstyle=\footnotesize,numbers=none]
```java
import java.util.List;
```
\end{lstlisting}
\caption{Example response from \llms using the example prompt.
}
\label{subfig:prompt-example}
\end{subfigure}
\caption{Example prompt used to infer types on code snippets with \llms, along with an example response.
The placeholder \texttt{\{input\_code\}} is substituted with the given code snippet.
}
\label{fig:prompt}
\end{figure}

\subsection{Constraint-Based Type Inference with \snr}
\label{subsec:snr}
\snr~\citep{SnR} is the state-of-the-art technique that uses constraint-based type inference~\citep{SnR,Subramanian:ICSE:14}.
It has proven highly effective for inferring types in code snippets from \stattypeso.
To illustrate the information \snr leverages,
consider the example code in \cref{subfig:example-code}.
\snr processes the snippet by extracting constraints, such as identifying a type with the simple
name \mycode{Logger}, which has a \mycode{getLogger} method that accepts a \mycode{java.lang.String} argument.
Using the knowledge
base shown in \cref{subfig:example-kb},
\snr assigns \mycode{java.util.logging.Logger} to \mycode{Logger}, satisfying the constraints.
This approach precisely eliminates incorrect candidates, such as \mycode{org.slf4j.Logger},
which lacks the required \mycode{getLogger} method.

\snr demonstrates how semantic relationships in the code snippet alone can solve type inference problems.
By evaluating \llms alongside \snr,
we aim to analyze the extent to which \llms leverage such semantic information for type inference.
These findings will provide deeper insights into \llms' type inference capabilities
and highlight opportunities for future research.

\section{Benchmark Suites for Evaluating Type Inference}

We utilize two benchmark suites \stattypeso and \thaliacs for evaluating \llms' type inference performance on code snippets.

\subsection{\stattypeso}
The \stattypeso benchmark suite is prevalent for testing type inference tools~\citep{Phan:ICSE:18,Saifullah:ASE:19,SnR,Huang:ASE:2022}.
It is constructed from real-world, manually repaired Java code snippets from Stack Overflow.
\stattypeso contains \sonumcodesnippet code snippets from \sonumlib popular Java libraries~(Android,
GWT, Hibernate, JDK, JodaTime, XStream).
The snippets were extracted from 50 randomly selected Stack Overflow posts for each library.
To prepare the benchmark suite, all parsing errors in the code snippets were first corrected.
Next, the missing import statements were manually inferred and added to the code snippet,
serving as the ground truth for evaluation.
\cref{fig:ex-so} shows an example from \stattypeso.
The code snippets from \stattypeso are generally short and make a small number of calls to the libraries
and assign the intermediate results to variables.

While \stattypeso is a widely adopted benchmark suite for evaluating type inference techniques,
it is important to acknowledge the potential for data leakage.
Given its long-standing availability since 2017 at GitHub,
\stattypeso may have been inadvertently incorporated
into the training data of some \llms,
potentially leading to artificially inflated performance scores.

\ifInCopypaper
\else
\begin{figure}[t]
\begin{lstlisting}[xleftmargin=20pt,breaklines=true,breakatwhitespace=true]
package jodatime;

import org.joda.time.Chronology;
import org.joda.time.DateTime;
import org.joda.time.DateTimeZone;
import org.joda.time.chrono.GJChronology;

//ID = 2182921
public class JodaTime05 {
	public static void main(String[] args) {
				DateTimeZone zone = DateTimeZone.forID("Europe/London");
				Chronology coptic = GJChronology.getInstance(zone);

				DateTime dt = new DateTime(coptic);
				DateTime minusOneDay = dt.minusDays(1);

				System.out.println(minusOneDay );
	}
}
\end{lstlisting}
\caption{Example code snippet from \stattypeso using the JodaTime and JDK libraries.
Excessive newlines have been removed for clarity of presentation.
These code snippets are generally short.
The import statements serve as the ground truth and are removed before the code snippet is used for evaluating type inference.
}
\label{fig:ex-so}
\end{figure}

\fi

\ifInCopypaper
\else
\begin{algorithm}[t]
\SetAlgoLined
\footnotesize
\SetInd{0em}{1.5em}
\caption{Generating \thaliatype from the given set of libraries.
Similar to \stattypeso, each \thaliatype code snippet primarily uses a single API.
}
\label{algo:generate-thaliatype}
\SetAlFnt{\footnotesize}
\SetAlCapFnt{\footnotesize}
\SetAlCapNameFnt{\footnotesize}
\SetKwInOut{Input}{Input}
\SetKwInOut{Output}{Output}
\SetKw{Continue}{continue}
\SetKw{In}{in}
\SetKw{Or}{or}
\SetKw{Var}{var}
\SetKwComment{Comment}{\# }{}
\SetKwProg{Fn}{def}{:}{end}
\SetKwFor{For}{for}{:}{end}
\SetKwIF{If}{ElseIf}{Else}{if}{:}{else if}{else}{end}
\SetStartEndCondition{ }{}{}
\SetKwFunction{FGenerate}{generate}
\SetKwFunction{FExtractAPIs}{ExtractAPIs}
\SetKwFunction{FThalia}{Thalia}
\SetKwData{L}{libraries}
\SetKwData{D}{dependencies}
\SetKwData{N}{N}
\SetKwData{CodeSnippets}{code\_snippets}
\SetKwData{Lib}{library}
\SetKwData{APIs}{apis}
\SetKwData{Libs}{libs}
\SetKwData{Snippets}{snippets}

\Fn(){\FGenerate{\L, \D}}{
\KwIn{\L.
    A set of jar libraries, \ie, Android, GWT, Hibernate, JDK, JodaTime, and XStream.}
\KwIn{\D: Additional libraries needed to compile the generated code snippets.}
\KwOut{\CodeSnippets: The list of generated code snippets.}

\BlankLine
\CodeSnippets $\leftarrow$ [\,]

\For{\Lib \In \L}{
    \APIs $\leftarrow$ \FExtractAPIs{\Lib} \Comment*{Extracts public classes, fields, and methods from the given \Lib.}
    \Libs $\leftarrow$ \{\texttt{JDK}, \Lib, \texttt{...dependencies}\} \Comment*{Libraries required for compilation.}
    \Snippets $\leftarrow$ \FThalia{\APIs, \Libs, 50} \Comment*{Generates 50 code snippets for the given \APIs using \thalia.}
    \CodeSnippets.append(\Snippets)%
}

\Return \CodeSnippets
}
\end{algorithm}

\fi

\subsection{\thaliacs: A New Benchmark Suite for Type Inference}
\label{subsec:datasets-thalia}
We introduce \thaliacs, a new benchmark suite specifically designed to \emph{rigorously evaluate}
the type inference capabilities of \llms while \emph{mitigating potential biases} from data leakage.
Code snippets in \thaliacs require reasoning over type-related semantic information to accurately infer the types used.
Through this design, \thaliatype evaluates how effectively \llms can leverage type information in code snippets,
ensuring that the measured performance is generalizable to previously unseen code.
If an \llm understands the type information provided,
then it can correctly infer the types for any given code snippet.

\thaliacs is generated using \thalia~\citep{Chaliasos:PLDI:22,Sotiropoulos:POPL:24},
a tool originally developed for testing the type system of compilers.
\thalia has uncovered
117 confirmed or fixed type-related bugs across four major compilers.
This large number of bugs demonstrates that
compiler developers value those found bugs
though triggered by automatically generated programs,
which indirectly demonstrates
the typing relations in \thalia-generated code snippets
might resemble the typing relations in real-world code snippets.
Recognizing its value for \llm evaluation,
we repurposed \thalia to create a diverse set of syntactically valid, well-typed Java programs.
Given a set of types, fields, and methods,
\thalia systematically generates programs that utilize a subset of these components.
By reasoning with the type information within the code snippets,
\llms can infer the types used in \thaliatype.

The general process for generating \thaliacs code snippets is summarized in \cref{algo:generate-thaliatype}.
Six libraries that were used in \stattypeso were given to \thalia as input.
By selecting the same libraries, we ensured that the types in the generated code snippets were familiar to the \llms,
eliminating performance differences caused by varying familiarity with the libraries.
For each library, 50 code snippets are generated.
Each snippet is based on a single library, following the convention in \stattypeso.
In total, \csnumcodesnippet code snippets were generated.

\cref{fig:ex-thalia} shows an example code snippet generated by \thaliatype.
Each code snippet consists of a \mycode{Main} class and a \mycode{test} method.
In this code snippet, the code creates a new \mycode{LocalTime} from a given time represented by
a \mycode{long} value of -58.
The chronology from this \mycode{LocalTime}, which in JodaTime is \mycode{ISOChronology},
is then used to construct the current \mycode{DateMidnight}.
Finally, the \mycode{getDayOfYear()} method is called to retrieve the ordinal day number of the year.

\myparagraph{Similarity between \thaliacs and \stattypeso}
The code snippets in \thaliatype are similar to the code snippets from \stattypeso.
\cref{fig:code-snippet-stats} compares the statistics of the two benchmark suites.
The code snippets in both suites are similar in terms of lines of code and the number of assignments.
While \thalia-generated programs are type-intensive, \thaliacs includes fewer method calls
but utilizes more types (via import statements) in each snippet.
More importantly, our experiments show that \snr achieved comparable performance on both \thaliacs and \stattypeso~(\ifInCopypaper\S5.2\else\cref{subsec:llm-type-inference-results}\fi),
demonstrating that \thaliacs provides
as much semantic information as \stattypeso for the type inference task.
If \llms genuinely understand the semantics of code snippets,
\llms should be able to perform similarly on both suites.

To the best of our knowledge, \thaliacs is the first benchmark suite designed for evaluating \llms
on code snippet type inference while addressing data leakage.
Future research can leverage our replication package to generate additional
code snippets using \thalia using their desired libraries to evaluate newer state-of-the-art \llms.

\ifInCopypaper
\else
\begin{figure}[t]
\begin{lstlisting}[xleftmargin=20pt]
package src.toady;

import org.joda.time.chrono.ZonedChronology;
import org.joda.time.Chronology;
import org.joda.time.LocalTime;
import org.joda.time.DateMidnight;

class Main {
  static public final <K, I extends ZonedChronology, X> void test()
      throws Exception {
    long elf = (long)-58;
    Chronology pantsuits = new LocalTime(elf).getChronology();
    DateMidnight fitness = DateMidnight.now(pantsuits);
    int liner = fitness.getDayOfYear();
  }
}

interface Function0<R> { public R apply(); }
interface Function1<A1, R> { public R apply(A1 a1); }
interface Function2<A1, A2, R> { public R apply(A1 a1, A2 a2); }
interface Function3<A1, A2, A3, R> { public R apply(A1 a1, A2 a2, A3 a3); }
\end{lstlisting}
\caption{Formatted code snippet generated by \thalia using the JodaTime and JDK libraries.
The variable names are randomly generated.
The function interfaces were generated by \thalia but are unused in \thaliatype code snippets.
}
\label{fig:ex-thalia}
\end{figure}

\fi

\ifInCopypaper
\else
\begin{figure}[t]
\input{data-raw/code-snippet-stats-boxplots.pgf}
\caption{Box plots comparing the features found in \stattypeso and \thaliacs.
}
\label{fig:code-snippet-stats}
\end{figure}

\fi

\section{Data Leakage in \llm-Based Type Inference}

In this section, we aim to investigate whether
the evaluation of \llm-based type inference techniques
may be affected by the data leakage issue~\citep{Inan:MS:21}.
Data leakage occurs when
the training datasets overlap with the benchmark datasets used for evaluation,
leading models to achieve artificially high performance by
memorizing examples rather than demonstrating genuine generalization.
The evaluation of the prior studies that either compared against~\citep{Huang:TOSEM:2023,Chen:Arxiv:24} or incorporated \llms~\citep{Kabir:TOSEM:24}
largely relied on the \stattypeso dataset as the benchmark suite,
which has been publicly available on \github since 2017~\citep{StatTypeSO:GitHub:17:full}.
Because the \gpt-based models used in these evaluations were likely trained on datasets containing \stattypeso,
their reported performance may be affected by data leakage.

However, directly detecting data leakage from \llms' training is challenging,
as most datasets are proprietary and undisclosed,
with only a few released to the broader community.
Two widely used \llm families, \gpt~\citep{gpt4} and \llama~\citep{llama3},
are trained on proprietary datasets with only partial disclosure of their training data.
\gpt's creator, OpenAI, has stated that the models are trained on a mixture of publicly
available data~(such as internet text) and data licensed from third-party providers.
However, the exact composition of the dataset remains undisclosed,
and access to the model is limited to APIs.
Similarly, while \llama models are open-weight and thus freely available for download and use,
the data used to train them is not publicly released.
Meta discloses that \llama is trained on publicly available data,
primarily sourced from the internet, but the exact sources are not specified.
This lack of transparency makes it difficult to assess the risk of data leakage in either model.

\label{subsec:training-data-starcoder}
In contrast, \starcoder~\citep{starcoder2} is trained entirely on the openly available dataset \stacki~\citep{the-stack-v2},
which comprises 67.5TB of source code across 658 programming languages.
Because this dataset is openly accessible,
it is possible to directly examine the training corpus and assess the extent of any data leakage.

\ifInCopypaper
\else
\begin{algorithm}[t]
\SetAlgoLined
\footnotesize
\SetInd{0em}{1.5em}
\SetAlFnt{\footnotesize}
\SetAlCapFnt{\footnotesize}
\SetAlCapNameFnt{\footnotesize}
\SetKwInOut{Input}{Input}
\SetKwInOut{Output}{Output}
\SetKw{Continue}{continue}
\SetKw{In}{in}
\SetKw{Or}{or}
\SetKw{Var}{var}
\SetKw{Not}{not}
\SetKwComment{Comment}{\# }{}
\SetKwProg{Fn}{def}{:}{end}
\SetKwFor{For}{for}{:}{end}
\SetKwIF{If}{ElseIf}{Else}{if}{:}{else if}{else}{end}
\SetStartEndCondition{ }{}{}
\SetKwFunction{FgetStackVTwo}{StackV2Dataset}
\SetKwFunction{FgetRepos}{getReposInStackV2}
\SetKwFunction{Fdict}{dict}
\SetKwFunction{Fbasename}{basename}
\SetKwFunction{Fpath}{path}
\SetKwFunction{Frepo}{repo}
\SetKwFunction{Fauthor}{author}
\SetKwData{Dataset}{dataset}
\SetKwData{Filenames}{filenames}
\SetKwData{Data}{data}
\SetKwData{RepoCounts}{repo\_counts}
\SetKwData{AuthorCounts}{author\_counts}
\Fn(){\FgetRepos{\Filenames}}{
\Input{\Filenames. The list of file names for code snippets in the \stattypeso benchmark suite.}
\Output{\RepoCounts and \AuthorCounts. Repositories and authors that are potentially sources of \stattypeso code snippets in the \Dataset are returned.}
\BlankLine

\RepoCounts $\gets$ \Fdict{} \Comment*{Initialize repository frequency dictionary.}
\AuthorCounts $\gets$ \Fdict{} \Comment*{Initialize author frequency dictionary.}
\For(\Comment*[f]{Iterate over the data in the \stacki dataset.}){\Data \In \FgetStackVTwo{}}{
  \If(\Comment*[f]{Match the file in the dataset using the file name.}){\Fbasename{\Data.\Fpath{}} \In \Filenames}{
    \If{\Data.\Frepo{} \Not \In \RepoCounts}{
      \RepoCounts[\Data.\Frepo{}] = 0 \;
    }
    \RepoCounts[\Data.\Frepo{}] += 1 \Comment*{Increment count for the repo.}
    \If{\Data.\Fauthor{} \Not \In \AuthorCounts}{
      \AuthorCounts[\Data.\Fauthor{}] = 0 \;
    }
    \AuthorCounts[\Data.\Fauthor{}] += 1 \Comment*{Increment count for author.}
  }
}
\Return \RepoCounts, \AuthorCounts \;
}
\caption{Finding \stattypeso code snippets in the \stacki dataset based on file names.
}
\label{algo:finding-training}
\end{algorithm}

\fi

\ifInCopypaper
\else
\begin{table}
\caption{Repositories found in the \stacki, along with the number of code snippets and the repository's last commit date.}
\label{tab:stack-repos}
\begin{tabular}{llll}
\toprule
Repository Owner & Repository Name & \# of code snippets & Last Commit Date \\
\midrule
miketran238 & AndroidOracle & 50 & 2017-02-14 \\
miketran238 & JodatimeOracle & 50 & 2017-02-12 \\
mrthlinh & Oracle-GWT & 50 & 2017-02-13 \\
mrthlinh & Oracle-Hibernate & 50 & 2017-02-13 \\
mrthlinh & TypeResolution\_Oracle & 219 & 2017-02-12\\
pdhung3012 & TypeResolution\_Oracle & 267 & 2017-06-30 \\
\bottomrule
\end{tabular}
\end{table}

\fi

To investigate the content of the \stacki, the metadata of the files are examined.
Using the matching procedure described in \cref{algo:finding-training},
file names in \stacki were compared against those in the \stattypeso benchmark suite,
which identified repositories containing overlapping code snippets.
This process produced 26 repositories and three
repository owners with five or more matching file names,
forming the basis for manual inspection.

Manual analysis identified six repositories~(shown in \cref{tab:stack-repos})
containing code snippets alongside their ground truth from the \stattypeso benchmark suite.
Notably, the \mycode{TypeResolution\_\allowbreak{}Oracle} repository by \mycode{mrthlinh}
includes 219 code snippets with ground truth,
excluding only those dependent on the \mycode{Android} library.
In addition, \mycode{pdhung3012}'s \mycode{TypeResolution\_\allowbreak{}Oracle} repository contains the complete \stattypeso benchmark suite.
All identified repositories were committed in 2017,
well before the knowledge cutoffs of \gpt and \llama,
making it highly likely that these models used for evaluation
were trained on the full \stattypeso benchmark suite.

\begin{findingbox}
Multiple public \github repositories contain code snippets with the ground truth from the \stattypeso benchmark suite.
Critically, these repositories were discovered within the \stacki training set.
This inclusion strongly suggests that \llms trained on internet data have likely been exposed to \stattypeso,
raising concerns regarding data leakage.
\end{findingbox}

\section{RQ1: How well do \llms perform type inference on unseen code snippets?}
\label{sec:llm-type-inference}

Given the strong possibility of data leakage, we sought to rigorously assess the
genuine type inference capabilities of \llms.
To do this, RQ1 investigates their performance on two benchmark suites:
the original \stattypeso~(which may have been seen during training),
and \thaliatype~(which contains only unseen code snippets).
This comparison allows us to isolate the impact of potential training overlap and better understand
how well these models generalize to genuinely novel code.

\subsection{Method}
Code snippets from \stattypeso and \thaliacs are given to \llms using the prompt pattern in \cref{fig:prompt} and also our baseline \snr, one at a time,
\emph{without} import statements.
To check the output, import statements in \llms' response are extracted and compared against the original import
statements in the code snippets.
To ensure reproducibility, all the inferences are performed with a fixed seed~(set to one) and a temperature of zero.
\gptfomini and \gptfo are accessed via the OpenAI API~\citep{OPENAIPLATFORM}, while
\starcoderi, \llamas, and \llamam are accessed through a self-hosted Ollama API~\citep{OLLAMAPLATFORM} running on an A100 GPU.
The total cost of using the \gpt models for RQ1 was \$4.47,
based on OpenAI's pricing as of early 2025.

Precision, recall, and F1-scores are computed to evaluate the performance of type inference.
Precision measures the proportion of correctly inferred \fqns among all inferred \fqns,
while recall measures the proportion of expected \fqns that are correctly inferred.

\begin{center}
$\text{Precision} = \frac{\text{Correctly Inferred \fqns}}{\text{All Inferred \fqns}}$\hfil
$\text{Recall} = \frac{\text{Correctly Inferred \fqns}}{\text{All Expected \fqns}}$\hfil
$\text{F1} = \frac{2 \times \text{Precision} \times \text{Recall}}{\text{Precision}+\text{Recall}}$
\end{center}

\usetikzlibrary{patterns}
\begin{figure}[t]

\begin{subfigure}[t]{\textwidth}
\begin{tikzpicture}
\scriptsize
    \begin{axis}[
        ybar,
        bar width=0.5cm,
        width=14.6cm,
        height=6cm,
        ymin=0,
        ymax=120,
        enlarge x limits=0.12,
        legend style={at={(0.5,-0.15)},anchor=north,legend columns=-1},
        xtick={1,2,3,4,5,6},
        xticklabels={\snr, \starcoderi, \llamas, \llamam, \gptfomini, \gptfo},
        xtick style={draw=none},
        ylabel={Score (\%)},
        ylabel style={
            yshift=-5pt,
        },
        ytick pos=left,
        ytick align=outside,
        nodes near coords={\pgfmathprintnumber[fixed,precision=2]{\pgfplotspointmeta}\%},
        nodes near coords align={vertical},
        every node near coord/.append style={rotate=55, anchor=west, xshift=-3pt, yshift=6pt},
    ]

    \addplot[fill=gray] coordinates {(1,95.50) (2,82.28) (3,76.92) (4,86.08) (5,86.34) (6,95.66)};
    \addplot[pattern=north east lines] coordinates {(1,91.46) (2,81.08) (3,69.46) (4,83.69) (5,89.92) (6,95.00)};
    \addplot[pattern=dots] coordinates {(1,93.44) (2,81.67) (3,73.00) (4,84.87) (5,88.09) (6,95.33)};

    \legend{Precision, Recall, F1-Score}
    \end{axis}
\end{tikzpicture}
\caption{Type inference performance on \stattypeso.
}
\label{fig:generated-compare-so}
\end{subfigure}

\begin{subfigure}[t]{\textwidth}
\begin{tikzpicture}
\scriptsize
    \begin{axis}[
        ybar,
        bar width=0.5cm,
        width=14.6cm,
        height=6cm,
        ymin=0,
        ymax=120,
        enlarge x limits=0.12,
        legend style={at={(0.5,-0.15)},anchor=north,legend columns=-1},
        xtick={1,2,3,4,5,6},
        xticklabels={\snr, \starcoderi, \llamas, \llamam, \gptfomini, \gptfo},
        xtick style={draw=none},
        ylabel={Score (\%)},
        ylabel style={
            yshift=-5pt,
        },
        ytick pos=left,
        ytick align=outside,
        nodes near coords={\pgfmathprintnumber[fixed,precision=2]{\pgfplotspointmeta}\%},
        nodes near coords align={vertical},
        every node near coord/.append style={rotate=55, anchor=west, xshift=-3pt, yshift=6pt},
    ]

    \addplot[fill=gray] coordinates {(1,84.15) (2,43.46) (3,31.27) (4,61.58) (5,66.64) (6,54.74)};
    \addplot[pattern=north east lines] coordinates {(1,84.43) (2,19.66) (3,19.40) (4,25.85) (5,37.73) (6,44.54)};
    \addplot[pattern=dots] coordinates {(1,84.29) (2,27.08) (3,23.95) (4,36.41) (5,48.18) (6,49.12)};

    \legend{Precision, Recall, F1-Score}
    \end{axis}
\end{tikzpicture}
\caption{Type inference performance on \thaliacs.}
\label{fig:generated-compare-cs}
\end{subfigure}
\caption{Type inference performance~(precision, recall, and F1-score) of \snr, \starcoderi, \llamas, \llamam, \gptfomini, and \gptfo, on the \stattypeso and \thaliacs benchmark suites.
\starcoderi performed between \llamas and \llamam for both suites.
All \llms experienced a large drop in type inference performance on \thaliatype compared to \stattypeso.}
\label{fig:generated-compare-both}
\end{figure}

\subsection{Results}
\label{subsec:llm-type-inference-results}

Despite achieving strong results in \stattypeso~(\cref{fig:generated-compare-both}), all evaluated
\llms demonstrated lower performance on unseen code snippets in \thaliatype.
\starcoderi consistently ranked between \llamas and \llamam across both benchmark suites.
Specifically, as shown in \cref{fig:generated-compare-so}, \starcoderi achieved an F1 score of \resultpr{40} on \stattypeso,
higher than \llamas, but lower than \llamam, \gptfo, and \gptfomini.
However, its performance dropped sharply to \resultpr{59} on \thaliatype,
representing a \pdecreasedatasetfstar decrease.
Despite this decline, it still outperformed \llamas but remained below the other larger models.
Notably, even the best performing model, \gptfo, experienced a
\pdecreasedatasetffo decrease in F1 score when evaluated on \thaliatype.
The consistent performance drop across all models,
including \starcoderi with confirmed data leakage,
suggests that \llms are likely affected by data leakage, which potentially inflated \llms' type inference
performance on \stattypeso.

Unlike \llms, \snr relies solely on the names of types, method names, and
relationships between types in code snippets.
\snr only experienced a \pdecreasedatasetfsnr decrease in F1 score when evaluated on \thaliatype,
due to the difference in available information in the code snippets.
The fact that \llms outperformed \snr on \stattypeso while performing poorly on \thaliatype
suggests that \llms leverage additional information in \stattypeso
to perform type inference.
These performance discrepancies motivated further investigations into the specific factors contributing to
\llms' type inference capabilities.
In particular, in RQ2~(\cref{sec:transformations}),
we investigate what types of information \llms might leverage for type inference,
and whether their performance depends on matching exact syntax of the code snippets seen during training.

\begin{findingbox}
\llm type inference performance declined dramatically when applied to generated, unseen code snippets,
potentially as a consequence of data leakage. %
In these cases, \llms performed substantially worse than the constraint-based method,
highlighting a promising area for future improvement in \llm-based type inference.
\end{findingbox}

\definecolor{bestnum}{HTML}{a6bddb}
\begin{table}[t]
\centering
\caption{Recall grouped by document frequency of \fqns on GitHub for types in \stattypeso and \thaliacs.
The best result for each frequency group is colored using \fcolorbox{black}{bestnum}{\rule{0pt}{3pt}\rule{3pt}{0pt}}.
The TP column lists the number of correctly inferred \fqns by each tool for each frequency group.
}
\label{tab:rare-boa-recall}
\scriptsize
\begin{tabular}{llrrrrrrrrrr}
\toprule
\multicolumn{2}{l}{Document Frequency} &  \multicolumn{2}{l}{[0,1e2)} & \multicolumn{2}{l}{[1e2,1e3)} & \multicolumn{2}{l}{[1e3,1e4)} & \multicolumn{2}{l}{[1e4,1e5)} & \multicolumn{2}{l}{>=1e5} \\
\midrule
\multirow{8}{*}{\rotatebox[origin=c]{90}{\stattypeso}} & Total FQNs & 27 &  & 69 &  & 312 &  & 429 &  & 463 &  \\
\cmidrule(lr){2-12}\morecmidrules\cmidrule(lr){2-12}
 &  & TP & Recall & TP & Recall & TP & Recall & TP & Recall & TP & Recall \\
\cmidrule(lr){3-4}\cmidrule(lr){5-6}\cmidrule(lr){7-8}\cmidrule(lr){9-10}\cmidrule(lr){11-12}
 & \snr & \cellcolor{bestnum}27 & \cellcolor{bestnum}100.00\% & \cellcolor{bestnum}60 & \cellcolor{bestnum}86.96\% & \cellcolor{bestnum}297 & \cellcolor{bestnum}95.19\% & 372 & 86.71\% & 433 & 93.52\% \\
 & \starcoderi & 7 & 25.93\% & 36 & 52.17\% & 235 & 75.32\% & 372 & 86.71\% & 404 & 87.26\% \\
 & \llamas & 2 & 7.41\% & 28 & 40.58\% & 173 & 55.45\% & 322 & 75.06\% & 378 & 81.64\% \\
 & \llamam & 7 & 25.93\% & 31 & 44.93\% & 228 & 73.08\% & 383 & 89.28\% & 439 & 94.82\% \\
 & \gptfomini & 4 & 14.81\% & 43 & 62.32\% & 272 & 87.18\% & 399 & 93.01\% & 451 & 97.41\% \\
 & \gptfo & 6 & 22.22\% & 53 & 76.81\% & 296 & 94.87\% & \cellcolor{bestnum}421 & \cellcolor{bestnum}98.14\% & \cellcolor{bestnum}459 & \cellcolor{bestnum}99.14\% \\
\midrule
\multirow{8}{*}{\rotatebox[origin=c]{90}{\thaliacs}} & Total FQNs & 994 &  & 597 &  & 439 &  & 324 &  & 331 &  \\
\cmidrule(lr){2-12}\morecmidrules\cmidrule(lr){2-12}
 &  & TP & Recall & TP & Recall & TP & Recall & TP & Recall & TP & Recall \\
\cmidrule(lr){3-4}\cmidrule(lr){5-6}\cmidrule(lr){7-8}\cmidrule(lr){9-10}\cmidrule(lr){11-12}
 & \snr & \cellcolor{bestnum}735 & \cellcolor{bestnum}73.94\% & \cellcolor{bestnum}533 & \cellcolor{bestnum}89.28\% & \cellcolor{bestnum}407 & \cellcolor{bestnum}92.71\% & \cellcolor{bestnum}300 & \cellcolor{bestnum}92.59\% & 292 & 88.22\% \\
 & \starcoderi & 22 & 2.21\% & 58 & 9.72\% & 94 & 21.41\% & 120 & 37.04\% & 234 & 70.69\% \\
 & \llamas & 19 & 1.91\% & 42 & 7.04\% & 64 & 14.58\% & 138 & 42.59\% & 258 & 77.95\% \\
 & \llamam & 18 & 1.81\% & 80 & 13.40\% & 144 & 32.80\% & 182 & 56.17\% & 270 & 81.57\% \\
 & \gptfomini & 56 & 5.63\% & 161 & 26.97\% & 227 & 51.71\% & 269 & 83.02\% & 300 & 90.63\% \\
 & \gptfo & 103 & 10.36\% & 233 & 39.03\% & 256 & 58.31\% & 288 & 88.89\% & \cellcolor{bestnum}316 & \cellcolor{bestnum}95.47\% \\
\bottomrule
\end{tabular}

\end{table}

\subsubsection{Impact of Document Frequency on \llm Type Inference}
To better understand the large performance disparity between \stattypeso and \thaliacs,
we investigated how the document frequency of a type impacts the performance of \llms.
Document frequency measures the number of source files that reference a specific type.
Since \llm outputs are biased by training data~\citep{Huang:arxiv:24,Liu:NIPS:23},
their performance may vary depending on the prevalence of a type in the training corpus.
Understanding this bias helps ensure that \llms deliver similar performance across all potential types
used in the real world.
To quantify a type's document frequency, we analyzed source files from the
Boa~\citep{Dyer:ICSE:13,Dyer:GPCE:13} \github dataset.
Specifically, the document frequency $F(T)$ is calculated as follows for a type $T$.
\[
F(T) = \sum_{f \in \mathcal{F}}
\begin{cases}
1 & \text{if } T \in f, \\
0 & \text{otherwise.}
\end{cases}
\quad (\mathcal{F} \text{ is the set of all source files in the \github dataset.})
\]
Based on the document frequency, we categorized the \fqns in the \stattypeso
and \thaliacs benchmark suites into distinct frequency ranges.
For each group,
we calculated the recall to investigate how a type's real-world frequency impacts \llm performance.
This categorization and analysis provide insights into the extent to which \llms are biased
toward more frequently encountered types.

\myparagraph{Results}
\llms performed poorly on less frequently used \fqns and worse on the unseen benchmark \thaliacs than on
\stattypeso, even for frequently used \fqns.
To analyze how performance varies across \fqns with differing usage frequencies,
we grouped \fqns into frequency ranges by orders of magnitude~(0-100, 100-1,000,
1,000-10,000, 10,000-100,000, and over 100,000 \github files).
The results, shown in \cref{tab:rare-boa-recall}, indicate that \llm performance degrades
as the document frequency of \fqns decreases.
Specifically, for the least frequently used \fqns,
\gptfo only achieved a recall of \resultboa{81} on \stattypeso and \resultboa{177} on \thaliacs.
In contrast, \snr demonstrated consistent performance regardless of a type's popularity,
achieving recalls of \resultboa{33} and \resultboa{129} on \stattypeso and \thaliacs respectively.
On \thaliacs, \llms only outperformed \snr on the most frequently used \fqns, specifically those appearing in over 100,000
\github source files.
This disparity may arise from \llms preferring more frequent \fqns for a given simple name
regardless of the context in the code snippet.
For example, for a type with the simple name \mycode{View}, if \mycode{android.view.View} is the
most frequent type in the >=1e5 category and \mycode{javax.swing.text.View} belongs to a less frequent
category, then always recommending \mycode{android.view.View} will result
in a high recall in the >=1e5 category but a lower recall in other categories.

\begin{findingbox}
\llm type inference performance diminishes
for less frequently used \fqns,
highlighting a key
limitation in applying \llms to real-world applications.
On unseen code snippets in \thaliacs, \snr outperformed all \llms except for \gptfo and \gptfomini,
but only for the most frequently used \fqns.
This shortcoming may hinder the effectiveness of \llms
for developers seeking assistance with these less common \fqns.

\end{findingbox}

Interestingly, despite \starcoderi being trained on code snippets from \stattypeso,
it achieved a maximum recall of only \resultboa{101}.
As in the previous analysis,
its performance generally fell between \llamas and \llamam on both \stattypeso and \thaliatype.
Although training data may inflate performance, \starcoderi does not reproduce code snippets verbatim,
as models are designed to generalize beyond specific examples.
This generalization, while desirable, makes detecting instances of data leakage difficult.

\section{RQ2: To what extent do \llms understand the execution semantics of code snippets?}
\label{sec:transformations}

The performance decline on unseen data observed in RQ1
highlights the need to investigate how \llms perform type inference.
To explore whether this decline stems from a lack of semantic understanding in \llms,
we conducted an experiment using semantic-preserving transformations that alter syntax without changing the semantics.
Our hypothesis is that if \llms possess a deep understanding of code snippet semantics,
their performance should remain stable despite such transformations.
Conversely, significant performance degradation would suggest that \llms rely heavily on superficial
syntactic patterns rather than a genuine semantic comprehension.

\ifInCopypaper
\else

\begin{algorithm}[t]
\caption{Procedure for the three code transformations applied in RQ2.
}
\label{algo:transform}
\SetAlgoLined
\footnotesize
\SetInd{0em}{1.5em}
\SetAlFnt{\footnotesize}
\SetAlCapFnt{\footnotesize}
\SetAlCapNameFnt{\footnotesize}
\SetKwInOut{Input}{Input}
\SetKwInOut{Output}{Output}
\SetKw{Continue}{continue}
\SetKw{In}{in}
\SetKw{Or}{or}
\SetKw{Var}{var}
\SetKwComment{Comment}{\# }{}
\SetKwProg{Fn}{def}{:}{end}
\SetKwFor{For}{for}{:}{end}
\SetKwIF{If}{ElseIf}{Else}{if}{:}{else if}{else}{end}
\SetStartEndCondition{ }{}{}
\SetKwFunction{FRename}{RenameIdentifier}
\SetKwFunction{FId}{FindVariableDeclaration}
\SetKwFunction{FMd}{FindMethodDeclaration}
\SetKwFunction{FEm}{FindExpressionMethodCalls}
\SetKwFunction{FGetName}{getName}
\SetKwFunction{FGetAnnotations}{getAnnotations}
\SetKwFunction{FHasExpression}{hasExpression}
\SetKwFunction{FHasInit}{hasInitialization}
\SetKwFunction{FGetInit}{getInitializer}
\SetKwFunction{FRemoveInit}{removeInitializer}
\SetKwFunction{FReplaceAllSimpleName}{ReplaceAllVariableName}
\SetKwFunction{FReplaceAllMethodName}{ReplaceAllMethodName}
\SetKwFunction{FRenamingClass}{RenameClassNames}
\SetKwFunction{FRenamingPackage}{RenamePackageNames}
\SetKwFunction{FAppend}{add}
\SetKwFunction{FLower}{LowerCode}
\SetKwFunction{FExpr}{findExpressions}
\SetKwFunction{FField}{findFields}
\SetKwFunction{FIsLoopCondition}{IsLoopCondition}
\SetKwFunction{FIsExprStatement}{IsExprStatement}
\SetKwFunction{FIsLambdaExpression}{IsLambdaExpression}
\SetKwFunction{FGetParentBlock}{getParentBlock}
\SetKwFunction{FInsertBefore}{InsertBefore}
\SetKwFunction{FReplace}{Replace}
\SetKwFunction{FParent}{parent}
\SetKwFunction{FGetRandomName}{GetRandomName}
\SetKwData{AST}{p}
\SetKwData{Id}{variable\_declaration}
\SetKwData{Md}{method\_declaration}
\SetKwData{Em}{expression\_method}
\SetKwData{vi}{random\_name}
\SetKwData{mi}{random\_name}
\SetKwData{BlackList}{skip\_list}
\SetKwData{Expression}{expression}
\SetKwData{Field}{field}
\SetKwData{FieldName}{field\_name}
\SetKwData{Init}{initializer}
\SetKwData{ParentBlock}{parentBlock}
\SetKwData{LB}{\symbol{123}}
\SetKwData{RB}{\symbol{125}}
\Fn(){\FRename{\AST}}{
\Input{\AST, representing the parsed code snippet.}
\Output{Transformed code snippet with renamed identifiers.}
\BlankLine
\For(\Comment*[f]{Rename all variables.}){\Id \In\ \FId{\AST}}{ \label{line:traverse:variable}
  \vi $\gets$ \FGetRandomName{} \; \label{line:random:name:variable}
  \FReplaceAllSimpleName{\AST, \vi, \Id.\FGetName{}} \; \label{line:variable:replace}
}
\BlackList $\gets$ \{\,\} \;
\For(\Comment*[f]{Skip method calls with expression.}){\Em \In\ \FEm{\AST}}{
  \If(){\Em.\FHasExpression{}}{ \label{line:filter:begin}
    \BlackList.\FAppend{\Em.\FGetName{}} \; \label{line:filter:end}
  }
}
\For(\Comment*[f]{Rename all method names.}){\Md \In\ \FMd{\AST}}{ \label{line:traverse:method}
  \If(){\Md.\FGetName{} \In \BlackList \Or "@Override" \In \Md.\FGetAnnotations{}}{ \label{line:method:skip:begin}
    \Continue\Comment*{Skip methods that potentially override parent methods.}
  } \label{line:method:skip:end}
  \mi $\gets$ \FGetRandomName{} \; \label{line:method:replace:begin}
  \FReplaceAllMethodName{\AST, \mi, \Md} \; \label{line:method:replace:end}
}
\FRenamingClass{\AST} \Comment*{Rename class names.} \label{line:renaming:class}
\FRenamingPackage{\AST} \Comment*{Rename package names.} \label{line:renaming:package}
\Return{\AST} \;
}

\BlankLine

\Fn(){\FLower{\AST}}{
\Input{\AST, representing the parsed code snippet.}
\Output{Transformed code snippet with lowered expressions.}
\BlankLine
\For(\Comment*[f]{Lower expressions.}){\Expression\ \In\ \FExpr{\AST}}{ \label{line:traverse:expr}
  \If(){\FIsExprStatement{\Expression.\FParent{}} \Or \FIsLoopCondition{\Expression} \Or \FIsLambdaExpression{\Expression}}{ \label{line:not:statement:expression:begin}
  \Continue\Comment*{Skip expression statements, loop conditions, or lambda expressions.}
  } \label{line:not:statement:expression:end}
  \vi $\gets$ \FGetRandomName{} \; \label{line:insert:assignment:begin}
  \FInsertBefore{"var \textnormal\{\ \vi\textnormal\} = \textnormal\{\ \Expression\textnormal\}", \Expression} \; \label{line:insert:assignment:end}
  \FReplace{\AST, \vi, \Expression} \; \label{line:replace:expr}
}
\For(\Comment*[f]{Lower field initializations.}){\Field\ \In\ \FField{\AST}}{ \label{line:traverse:field}
  \If(){\Field.\FHasInit{}}{ \label{line:has:initializer}
    \FieldName $\gets$ \Field.\FGetName{} \; \label{line:split:field:begin}
    \Init $\gets$ \Field.\FGetInit{} \;
    \Field.\FRemoveInit{} \;
    \FInsertBefore{"\{\symbol{92}n \textnormal\{\ \FieldName\textnormal\} = \textnormal\{\ \Init\textnormal\} ; \symbol{92}n \}", \Field} \; \label{line:split:field:end}
  }
}
\Return{\AST}
}

\BlankLine

\SetKwFunction{FKeyword}{AddKeyword}
\SetKwFunction{FSplit}{split}
\SetKwFunction{FRandomKeyword}{GetRandomKeyword}
\SetKwData{Code}{snippet}
\SetKwData{Lines}{new\_lines}
\SetKwData{Line}{line}
\Fn(){\FKeyword{\Code}}{
\Input{\Code, representing the original code snippet.}
\Output{Code snippet with added keyword comments.}
\BlankLine
\Lines $\gets$ [\,]\;
\For(\Comment*[f]{Append a comment to every line.}){\Line\ \In\ \Code.\FSplit{"\textbackslash n"}}{
  \Lines.\FAppend{"\textnormal\{\ \Line\textnormal\}\ //\textnormal\{\ \FRandomKeyword{}\ \textnormal\}"} \;
}
\Return{\Lines}
}
\end{algorithm}

\fi

\subsection{Method}
The transformations, detailed in \cref{algo:transform},
explore the extent to which syntactic changes affect LLMs’ semantic understanding of code.
Three specific transformations were employed to assess the impact of modifications to identifier names
(\cref{subsub:transformations-rename}),
code structures (\cref{subsub:transformations-lower}),
and Java keywords in comments (\cref{subsub:transformations-keyword}).
These were selected to highlight potential limitations in \llms' semantic understanding,
though future research could explore additional transformations.
Their impact was then evaluated using the methodology outlined in \cref{sec:llm-type-inference}.
The \emph{Wilcoxon signed-rank test}~\citep{Wilcoxon:92} was used to determine whether the
transformations caused significant changes in precision, recall, and F1 scores.
The average running cost per transformation for \gpt models was \$6.30 in RQ2,
slightly higher than in RQ1 because of longer code snippets.

\begin{figure}
\footnotesize
\begin{alignat*}{2}
  \mathit{Class} & \MyDefine &&
  \{\mathit{Modifier}\}\ \texttt{class}\ \mathit{SimpleName}\ [\texttt{extends}\ \mathit{Name}]\ [\texttt{implements}\ \{\mathit{Name}\}] \\
  &&&
  \texttt{`\{'} \{\mathit{ClassMember}\} \texttt{`\}'} \\
  \mathit{Modifier} & \MyDefine &&
  \mathit{Annotation} \MyAlt \texttt{public} \MyAlt \texttt{protected} \MyAlt \texttt{private}
  \MyAlt \texttt{static} \MyAlt \texttt{abstract} \MyAlt \texttt{final} \\
  \mathit{ClassMember} & \MyDefine &&
  \mathit{Field} \MyAlt \mathit{Method} \MyAlt \mathit{Block} \\
  \mathit{Field}  & \MyDefine && \{\mathit{Modifier}\}\ \mathit{Type}\ \mathit{SimpleName}\ [\texttt{=}\ \mathit{Expr}]\ \texttt{;} \\
  \mathit{Method} & \MyDefine && \{\mathit{Modifier}\}\ \mathit{Type}\,\mathit{SimpleName}\ \texttt{(}\{\mathit{Type}\,\mathit{SimpleName}\}\texttt{)}\ \mathit{Block} \\
  \mathit{Annotation} & \MyDefine && \texttt{@}\,\mathit{Name} \\
  \mathit{Block} & \MyDefine && \texttt{`\{'} \{\mathit{Statement}\} \texttt{`\}'} \\
  \mathit{Statement} & \MyDefine && \mathit{Expr}\ \texttt{;} \MyAlt \mathit{Expr}\,\texttt{=}\,\mathit{Expr}\ \texttt{;} \MyAlt \mathit{Block}
  \MyAlt\texttt{return}\ [\mathit{Expr}]\,\texttt{;} \\
  &&&
  \MyAlt\mathit{Type}\ \mathit{SimpleName}\ [\texttt{=}\,\mathit{Expr}]\ \texttt{;} \\
  &&&
  \MyAlt\texttt{if (}\,\mathit{Expr}\,\texttt{)}\ \mathit{Statement}\ \texttt{else}\ \mathit{Statement} \\
  &&&
  \MyAlt\texttt{while (}\,\mathit{Expr}\,\texttt{)}\ \mathit{Statement} \\
  &&&
  \MyAlt\texttt{for (}\,\mathit{Expr}\,\texttt{;}\,\mathit{Expr}\,\texttt{;}\,\mathit{Expr}\,\texttt{)}\ \mathit{Statement} \\
  &&&
  \MyAlt\texttt{for (}\,\{\mathit{Modifier}\}\ \mathit{Type}\ \mathit{SimpleName}\ [\texttt{=}\,\mathit{Expr}]\,\texttt{;}\,\mathit{Expr}\,\texttt{;}\,\mathit{Expr}\,\texttt{)}\ \mathit{Statement} \\
  \mathit{Expr} & \MyDefine && \mathit{Name} \MyAlt \mathit{Literal} \MyAlt \texttt{this} \MyAlt \texttt{super} \MyAlt \mathit{Expr}\ \mathit{Op}\ \mathit{Expr} \\
  &&&
  \MyAlt \texttt{(}\,\mathit{Type}\,\texttt{)}\ \mathit{Expr} \MyAlt \mathit{Expr}\ \texttt{instanceof}\ \mathit{Name} \\
  &&&
  \MyAlt [\mathit{Expr}\,\texttt{.}]\,\mathit{SimpleName}
  \MyAlt [\mathit{Expr}\,\texttt{.}]\,\mathit{SimpleName}\,\texttt{(}\{Expr\}\texttt{)} \\
  &&&
  \MyAlt \texttt{new}\ \mathit{Name}\ \texttt{(}\{\mathit{Expr}\,\}\texttt{)} \MyAlt \texttt{!}\,\mathit{Expr} \MyAlt \texttt{-}\,\mathit{Expr}\\
  &&&
  \MyAlt \texttt{(}\,\{\,[\mathit{Type}\,]\ \mathit{SimpleName}\}\texttt{)}\ \texttt{->}\ \mathit{Block} \MyAlt \mathit{SimpleName}\ \texttt{->}\ \mathit{Block} \\
  &&&
  \MyAlt \texttt{(}\,\{\,[\mathit{Type}\,]\ \mathit{SimpleName}\}\texttt{)}\ \texttt{->}\ \mathit{Expr} \MyAlt \mathit{SimpleName}\ \texttt{->}\ \mathit{Expr} \\
  \mathit{Literal} & \MyDefine && \texttt{null} \MyAlt \mathit{NumberLiteral} \MyAlt \mathit{StringLiteral} \MyAlt \mathit{BooleanLiteral} \\
  \mathit{Op} & \MyDefine && \texttt{+} \MyAlt \texttt{-} \MyAlt \texttt{*} \MyAlt \texttt{/} \MyAlt \texttt{\%} \MyAlt \texttt{>} \MyAlt \texttt{==} \MyAlt \texttt{>=} \MyAlt \texttt{!=}\\
  \mathit{Name}  & \MyDefine && \mathit{FQN}\MyAlt\mathit{SimpleName}\\
  \mathit{FQN} & \MyDefine && \mathit{Name}\,\texttt{.}\,\mathit{SimpleName}\\
  \mathit{Type}  & \MyDefine && \mathit{PrimitiveType} \MyAlt\mathit{Name} \MyAlt\texttt{var} \MyAlt\texttt{void} \\
  \mathit{PrimitiveType} & \MyDefine && \texttt{byte}\MyAlt\texttt{short}\MyAlt\texttt{int}\MyAlt\texttt{long}\MyAlt\texttt{float}\MyAlt\texttt{double}\MyAlt\texttt{boolean}\MyAlt\texttt{char} \\
\end{alignat*}
\caption{Simplified Java grammar to illustrate our transformations. \{*\} denotes that the enclosed term
occurs zero or more times. [*] denotes that the enclosed term occurs zero or one time.}
\label{fig:java-grammar}
\end{figure}

\subsubsection{Background: Java Grammar}

The Java grammar depicted in \cref{fig:java-grammar} is used to illustrate the key transformations
on the Java code snippet.
Each code snippet contains one or more classes along with a set of import statements.
Classes may contain fields, methods, and blocks,
with methods comprising blocks of statements and expressions.
Highly repetitive Java statements and expressions were omitted as they follow the same transformation principles.

\subsubsection{Identifier Renaming}
\label{subsub:transformations-rename}

This transformation investigates whether \llms rely on specific identifier names for type inference.
Although identifier names can provide useful signals,
overreliance on unique identifiers or specific identifier sequences seen during training may
limit a model's ability to generalize.
In real-world code snippets, identifiers are often lowercase variants of type names or abbreviated forms,
providing no additional semantic information.

The \FRename function in \cref{algo:transform} details the process,
which systematically renames identifiers such as variables, methods, classes, and packages in the code snippet.
First, all variable declarations are identified using the \FId function (\cref{line:traverse:variable}),
which extracts statements matching the pattern $\mathit{Type}\ \allowbreak\mathit{SimpleName}\ \allowbreak[\texttt{=}\,\mathit{Expr}]\ \texttt{;}$.
The algorithm then traverses each declaration, generating a unique,
random three-word name for each variable using the \FGetRandomName function (\cref{line:random:name:variable}),
from a predefined word list from prior work~\citep{Sotiropoulos:POPL:24} to ensure uniqueness.
The algorithm replaces every occurrence of the variable name in the code~(extracted using the \FGetName function in \cref{line:variable:replace}).

Next, the algorithm processes method declarations.
To avoid renaming references that could alter semantics,
all the method call expressions
($\mathit{Expr}$s with the form $[\mathit{Expr}\,\allowbreak\texttt{.}]\,\allowbreak\mathit{SimpleName}\,\allowbreak\texttt{(}\{Expr\}\texttt{)}$)
that have the first optional part ($[\mathit{Expr}\,\texttt{.})]$) are potentially external,
thus added to a \mycode{skip\_list}
(lines~\ref{line:filter:begin}-\ref{line:filter:end}) to be filtered out (\cref{line:method:skip:begin}).
Additionally, overridden methods annotated with \mycode{@Override} are skipped, as their names must match those in the super class (\cref{line:method:skip:begin}).
Lastly, the algorithm renames classes and packages
(lines~\ref{line:renaming:class}-\ref{line:renaming:package}), following the same procedure used for variables.
The details are omitted here for brevity.

\ifInCopypaper
\else
\begin{figure}[t]
\centering
\scriptsize
\begin{subfigure}[b]{0.24\textwidth}
\begin{lstlisting}[backgroundcolor=\color{gray!10},basicstyle=\scriptsize\ttfamily,frame=single,xleftmargin=17pt]
package p;

class C extends A {
  A a;
  void m() {@\label{line:m-declare}@
    a.m();
  }
  int n() {
    int n = 1;
    return 1;
  }
}
\end{lstlisting}
\caption{Before  renaming.}
\end{subfigure}
\hfill
\begin{subfigure}[b]{0.26\textwidth}
\begin{tabular}{@{}ll@{}}
\toprule
Before & After \\
\midrule
p & GrouseScabsShelley \\
C & TashaMonroviaTimbers \\
a & RowsGarlickyThump \\
n() & ListerineStupefiesFetlock() \\
n & CrackerSherbertsPlod \\
\bottomrule
\end{tabular}
\caption{Renamed identifiers.}
\end{subfigure}
\hfill
\hfill
\begin{subfigure}[b]{0.40\textwidth}
\begin{lstlisting}[backgroundcolor=\color{gray!10},basicstyle=\scriptsize\ttfamily,frame=single]
package GrouseScabsShelley;

class TashaMonroviaTimbers extends A {
  A RowsGarlickyThump;
  void m() {
    RowsGarlickyThump.m();
  }
  int ListerineStupefiesFetlock() {
    int CrackerSherbertsPlod = 1;
    return 1;
  }
}
\end{lstlisting}
\caption{After applying identifier renaming.}
\end{subfigure}
\caption{Example identifier renaming on a simplified code snippet.
}
\label{fig:sample-rename}
\end{figure}

\fi

\cref{fig:sample-rename} provides an example of this transformation,
showing how variables, methods, classes, and packages are renamed.
The method \mycode{m} declared at \cref{line:m-declare} likely overrides a method on the super class, \mycode{A.m()}.
Out of an abundance of caution to preserve semantic consistency, \mycode{m} is not renamed.
Furthermore, it should be noted that different types of names are renamed separately.
For example, the method name \mycode{n} is renamed differently from the variable name \mycode{n}.

\subsubsection{Code Lowering}
\label{subsub:transformations-lower}

The code lowering transformation examines the impact of structural changes on \llms' type inference.
A key aspect of type inference is understanding the types present in a program and how they relate to one another.
Code lowering does not alter these type relationships but instead disrupts the exact sequence of tokens presented to the model.
If a model relies heavily on specific token sequences seen during training,
its generalizability may be impaired,
leading to failures in type inference when the same program is presented in a different structure.

The algorithm first extracts and traverses all the expressions (\cref{line:traverse:expr}) in the code snippet.
For each expression, if its parent is not an expression statement
($\mathit{Statement}$s with the form $\mathit{Expr}\ \texttt{;}$)
and it is not a loop condition
(lines~\ref{line:not:statement:expression:begin}-\ref{line:not:statement:expression:end}),
the algorithm creates a new variable with a random name, assigns the expression to the variable,
and inserts this assignment statement before the expression
(lines~\ref{line:insert:assignment:begin}-\ref{line:insert:assignment:end}).
Next, the original expression is replaced with the newly created variable
assigned with the value of the original expression (\cref{line:replace:expr}).
Such transformations modify the code structure while preserving the semantics of the code snippet.
The expressions in expression statements must be skipped because transforming them may
result in invalid code.
For example, an expression in an expression statement can be a method call
with no return value (\ie, with a \mycode{void} return type).
In such a case, the created assignment statement would be invalid.
Conditions in loops cannot be transformed either since
such transformations could change the semantics of the code.

In addition to expressions, code lowering also applies to fields with an initializer.
For each field in the code snippet (\cref{line:traverse:field}),
if it has an initializer (\cref{line:has:initializer}),
the algorithm splits it into a declaration and a block initializing the field
(lines~\ref{line:split:field:begin}-\ref{line:split:field:end}).

\ifInCopypaper
\else
\begin{figure}[t]
\centering
\scriptsize
\begin{subfigure}{0.45\textwidth}
\begin{lstlisting}[backgroundcolor=\color{gray!10},basicstyle=\scriptsize\ttfamily,frame=single,xleftmargin=15pt]
class C {
  C c = null;
  void m() {
    int a = 0;
    if (a == 0) {
      return;
    }
  }
}
\end{lstlisting}
\caption{Before code lowering.}
\end{subfigure}
\hfill
\begin{subfigure}{0.45\textwidth}
\begin{lstlisting}[backgroundcolor=\color{gray!10},basicstyle=\scriptsize\ttfamily,frame=single]
class C {
  {
    c = null;
  }
  C c;
  void m() {
    int a = 0;
    var GrouseScabsShelley = a == 0;
    if (GrouseScabsShelley) {
      return;
    }
  }
}
\end{lstlisting}
\caption{After applying code lowering.}
\end{subfigure}
\caption{Example code lowering on a simplified code snippet.}
\label{fig:sample-lower}
\end{figure}

\fi

\cref{fig:sample-lower} demonstrates this transformation.
In this example, the field \mycode{C c = null} is split into a declaration and
an initializer block with \mycode{c = null}.
Additionally, the expression in the \mycode{if} statement is replaced with a newly created variable,
and an assignment statement is inserted before its use to assign the original value to that variable.

\ifInCopypaper
\else
\begin{figure}[t]
\centering
\scriptsize
\begin{subfigure}{0.45\textwidth}
\begin{lstlisting}[backgroundcolor=\color{gray!10},basicstyle=\scriptsize\ttfamily,frame=single,xleftmargin=15pt]
class C {
  void m() {
    return;
  }
}
\end{lstlisting}
\caption{Before adding keyword comments.}
\end{subfigure}
\hfill
\begin{subfigure}{0.45\textwidth}
\begin{lstlisting}[backgroundcolor=\color{gray!10},basicstyle=\scriptsize\ttfamily,frame=single]
class C { // continue
  void m() { // void
    return; // int
  } // opens
} // case
\end{lstlisting}
\caption{After adding keyword comments.}
\end{subfigure}
\caption{Example of adding keyword comments on a simplified code snippet.
}
\label{fig:sample-keyword}
\end{figure}

\fi

\subsubsection{Adding Keyword Comments}
\label{subsub:transformations-keyword}

The keyword comment transformation evaluates whether comments containing Java keywords distract \llms.
If a model relies on specific sequences of keywords seen during training,
rather than on the actual content of the code,
then this transformation may lead to reduced performance and an inability to generalize to unseen code snippets.
To perform this transformation,
the \FKeyword function in \cref{algo:transform} appends a single Java keyword~\citep{JavaKeywords}
as a line comment at the end of each line in the code snippet.
An example of the transformed snippet is presented in \cref{fig:sample-keyword}.

\ifInCopypaper
\else
\begin{figure}
\includegraphics[width=0.95\linewidth, trim={0 279pt 190pt 0},clip]{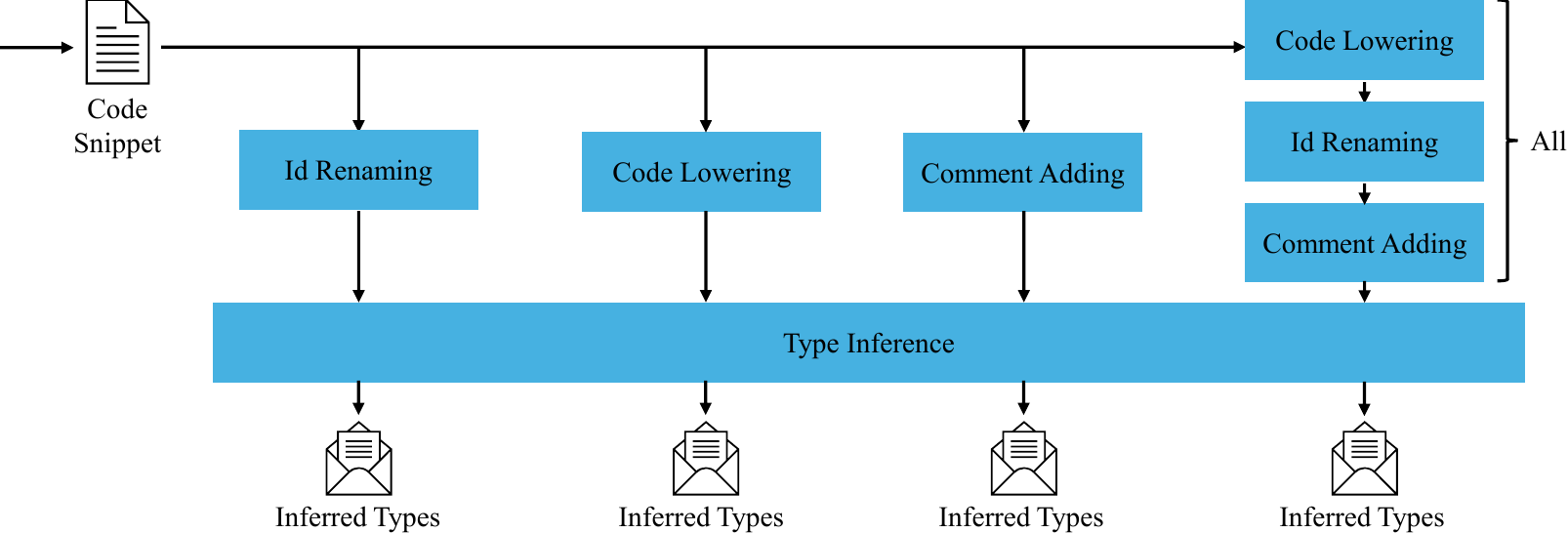}
\caption{RQ2 workflow overview.
Three code transformations were designed and applied individually to assess their isolated impact.
A fourth transformation combines all three sequentially to evaluate potential non-linear interactions.
}
\label{fig:rq2-transformations}
\end{figure}
\fi

\subsubsection{Putting Everything Together}
\cref{fig:rq2-transformations} illustrates the complete set of transformations used in RQ2.
Each of the first three transformations was applied independently to measure its individual impact.
The final transformation combines all three sequentially to more strongly perturb the input,
thereby making it more difficult for models to recall \stattypeso code snippets,
if they were seen during training.
These transformed code snippets are then passed to \llms.
The inferred types are analyzed to see if transformations significantly impacted \llms understanding
of the types used in the code snippet, affecting performance.

\subsection{Results}
\label{subsec:transformations-results}

The effects of applying transformations to the \stattypeso and \thaliacs benchmark suites are shown in
\cref{tab:transformation-so,tab:transformation-thalia},
respectively.
These transformations significantly affected the precision, recall, and F1-scores of all evaluated \llms.
However, the results are complex and merit detailed examination.

\definecolor{siglow}{HTML}{ece7f2}
\definecolor{sigmid}{HTML}{a6bddb}
\definecolor{sighigh}{HTML}{2b8cbe}
\definecolor{sigimplow}{HTML}{e6d2c9}
\definecolor{sigimpmid}{HTML}{dbbda6}
\definecolor{sigimphigh}{HTML}{be8c2b}
\begin{table}[t]
\caption{The precision, recall, and F1-scores of tools after transformations identifier renaming~(Id Renaming),
code lowering, and adding keyword comments~(comment adding) were applied.
Note that \fcolorbox{black}{siglow}{\rule{0pt}{3pt}\rule{3pt}{0pt}} represents $p < 0.05$, \fcolorbox{black}{sigmid}{\rule{0pt}{3pt}\rule{3pt}{0pt}} represents $p < 0.01$, and \fcolorbox{black}{sighigh}{\rule{0pt}{3pt}\rule{3pt}{0pt}} represents $p < 0.001$ when performance decreases, and \fcolorbox{black}{sigimplow}{\rule{0pt}{3pt}\rule{3pt}{0pt}} represents $p < 0.05$, and \fcolorbox{black}{sigimphigh}{\rule{0pt}{3pt}\rule{3pt}{0pt}} represents $p < 0.001$ when performance improves.
}
\centering
\scriptsize
\setlength{\tabcolsep}{1pt}

\begin{subtable}[t]{\textwidth}
\caption{The precision, recall, and F1-scores of tools on \stattypeso and transformed code snippets.}
\centering
\resizebox{\linewidth}{!}{
\begin{tabular}{@{}lrrrrrrrrrrrrrrrrrr@{}}
\toprule
 & \multicolumn{3}{c}{\snr} & \multicolumn{3}{c}{\starcoderi} & \multicolumn{3}{c}{\llamas} & \multicolumn{3}{c}{\llamam} & \multicolumn{3}{c}{\gptfomini} & \multicolumn{3}{c}{\gptfo} \\
\cmidrule(lr){2-4}\cmidrule(lr){5-7}\cmidrule(lr){8-10}\cmidrule(lr){11-13}\cmidrule(lr){14-16}\cmidrule(lr){17-19}
 & Precision & Recall & F1 & Precision & Recall & F1 & Precision & Recall & F1 & Precision & Recall & F1 & Precision & Recall & F1 & Precision & Recall & F1 \\
\midrule
\stattypeso & 95.50\% & 91.46\% & 93.44\% & 82.28\% & 81.08\% & 81.67\% & 76.92\% & 69.46\% & 73.00\% & 86.08\% & 83.69\% & 84.87\% & 86.34\% & 89.92\% & 88.09\% & 95.66\% & 95.00\% & 95.33\% \\
\midrule
Id Renaming & 95.50\% & 91.46\% & 93.44\% & 82.09\% & 71.23\%\cellcolor{sighigh} & 76.28\% & 65.28\%\cellcolor{sighigh} & 56.69\%\cellcolor{sighigh} & 60.68\%\cellcolor{siglow} & 85.34\% & 76.15\%\cellcolor{sighigh} & 80.49\%\cellcolor{sighigh} & 89.63\% & 88.46\%\cellcolor{sigmid} & 89.04\% & 97.03\% & 95.62\% & 96.32\%  \\
Code Lowering & 95.43\% & 91.62\% & 93.49\% & 83.72\% & 79.92\% & 81.78\% & 69.43\%\cellcolor{sigmid} & 65.15\%\cellcolor{sigmid} & 67.22\%\cellcolor{sighigh} & 83.81\% & 80.46\%\cellcolor{siglow} & 82.10\%\cellcolor{sigmid} & 86.14\%\cellcolor{siglow} & 88.92\% & 87.51\% & 94.82\%\cellcolor{siglow} & 95.69\% & 95.25\% \\
Comment Adding & 95.50\% & 91.46\% & 93.44\% & 83.98\% & 75.00\%\cellcolor{sighigh} & 79.24\%\cellcolor{siglow} & 78.65\% & 67.46\% & 72.63\% & 85.39\% & 80.92\%\cellcolor{sigmid} & 83.10\%\cellcolor{sigmid} & 87.90\% & 89.38\%\cellcolor{siglow} & 88.63\% & 96.06\% & 95.77\% & 95.92\% \\
\arrayrulecolor[gray]{0.8}
\midrule
\arrayrulecolor{black}
All & 95.43\% & 91.62\% & 93.49\% & 79.34\%\cellcolor{sigmid} & 59.38\%\cellcolor{sighigh} & 67.93\%\cellcolor{sighigh} & 55.83\%\cellcolor{sighigh} & 47.15\%\cellcolor{sighigh} & 51.13\%\cellcolor{sighigh} & 77.93\%\cellcolor{sighigh} & 70.08\%\cellcolor{sighigh} & 73.80\%\cellcolor{sighigh} & 84.60\%\cellcolor{sigmid} & 82.85\%\cellcolor{sighigh} & 83.72\%\cellcolor{sighigh} & 93.29\%\cellcolor{sighigh} & 94.15\%\cellcolor{siglow} & 93.72\%\cellcolor{sighigh} \\
\bottomrule
\end{tabular}

}
\label{tab:transformation-so}
\end{subtable}

\vspace{0.2cm}

\begin{subtable}[t]{\textwidth}
\caption{The precision, recall, and F1-scores of tools on \thaliacs and transformed code snippets.}
\centering
\resizebox{\linewidth}{!}{
\begin{tabular}{@{}lrrrrrrrrrrrrrrrrrr@{}}
\toprule
 & \multicolumn{3}{c}{\snr} & \multicolumn{3}{c}{\starcoderi} & \multicolumn{3}{c}{\llamas} & \multicolumn{3}{c}{\llamam} & \multicolumn{3}{c}{\gptfomini} & \multicolumn{3}{c}{\gptfo} \\
\cmidrule(lr){2-4}\cmidrule(lr){5-7}\cmidrule(lr){8-10}\cmidrule(lr){11-13}\cmidrule(lr){14-16}\cmidrule(lr){17-19}
 & Precision & Recall & F1 & Precision & Recall & F1 & Precision & Recall & F1 & Precision & Recall & F1 & Precision & Recall & F1 & Precision & Recall & F1 \\
\midrule
\thaliacs & 84.15\% & 84.43\% & 84.29\% & 43.46\% & 19.66\% & 27.08\% & 31.27\% & 19.40\% & 23.95\% & 61.58\% & 25.85\% & 36.41\% & 66.64\% & 37.73\% & 48.18\% & 54.74\% & 44.54\% & 49.12\%  \\
\midrule
Id Renaming & 84.15\% & 84.43\% & 84.29\% & 46.24\% & 22.42\%\cellcolor{sigimphigh} & 30.20\%\cellcolor{sigimplow} & 31.36\%\cellcolor{sigimplow} & 19.11\% & 23.74\%\cellcolor{siglow} & 57.08\%\cellcolor{siglow} & 24.47\%\cellcolor{siglow} & 34.25\%\cellcolor{sigmid} & 52.47\%\cellcolor{sighigh} & 38.44\% & 44.37\%\cellcolor{sighigh} & 43.88\%\cellcolor{sighigh} & 45.77\% & 44.80\%\cellcolor{siglow} \\
Code Lowering & 84.14\% & 84.39\% & 84.27\% & 48.09\%\cellcolor{sigimplow} & 23.87\%\cellcolor{sigimphigh} & 31.91\%\cellcolor{sigimphigh} & 27.06\%\cellcolor{siglow} & 18.21\% & 21.77\% & 60.16\% & 25.25\% & 35.57\% & 58.04\%\cellcolor{sighigh} & 36.01\%\cellcolor{sigmid} & 44.45\%\cellcolor{siglow} & 43.45\% & 44.43\% & 43.93\% \\
Comment Adding & 84.15\% & 84.43\% & 84.29\% & 42.96\%\cellcolor{sighigh} & 19.66\% & 26.98\%\cellcolor{sighigh} & 25.00\%\cellcolor{siglow} & 19.22\% & 21.73\% & 59.56\% & 25.07\% & 35.28\% & 64.27\% & 36.57\%\cellcolor{siglow} & 46.62\% & 32.26\% & 44.80\% & 37.51\% \\
\arrayrulecolor[gray]{0.8}
\midrule
\arrayrulecolor{black}
All & 84.14\% & 84.39\% & 84.27\% & 40.81\%\cellcolor{sighigh} & 20.74\% & 27.51\%\cellcolor{sigimphigh} & 27.57\%\cellcolor{siglow} & 17.77\%\cellcolor{sigmid} & 21.61\% & 45.92\%\cellcolor{sighigh} & 23.87\%\cellcolor{sighigh} & 31.41\%\cellcolor{siglow} & 47.50\%\cellcolor{sighigh} & 36.87\% & 41.52\%\cellcolor{sighigh} & 56.30\%\cellcolor{sigimphigh} & 45.29\% & 50.20\% \\
\bottomrule
\end{tabular}

}
\label{tab:transformation-thalia}
\end{subtable}

\end{table}

Focusing on each transformation in isolation~(identifier renaming, code lowering, comment adding),
the results show that \llms generally exhibit resilience,
meaning that simple transformations often do not significantly change a model's performance.
This was particularly true for more advanced models such as \gptfo but also held for smaller models like \starcoderi.
Notably, \starcoderi, known to have been trained on \stattypeso code snippets,
exhibited significant performance drops only under identifier renaming (recall)
and comment adding (recall and F1-score) when evaluated on \stattypeso.
The isolated transformation results on \stattypeso and \thaliatype indicate a degree of generalizability in \llms' understanding of code.

\vspace{-5pt}
\begin{findingbox}
\llms, especially more advanced models,
demonstrate generalizability to simple transformations that preserve execution semantics,
maintaining overall performance despite syntactic changes such as identifier renaming, code lowering, and comment adding.
This shows that the models tested do not rely on any one syntactic element for type inference.
\end{findingbox}
\vspace{-5pt}

However, when all three transformations were applied simultaneously,
performance changes were amplified, clearly, significantly, and negatively affecting all \llms' precision,
recall, and F1-scores on \stattypeso.
For example, F1-scores dropped by \transformsofolower on \gptfo,
\transformsosclower on \starcoderi, and \transformsolslower on \llamas.
Similar trends were observed for precision and recall.
In contrast, applying the same three transformations on \thaliatype code snippets produced less consistent
and generally smaller effects.
In some cases, performance even improved.
For instance, \starcoderi achieved a higher F1-score despite a lowered precision,
and \gptfo demonstrated higher precision,
with no significant changes in other metrics on \thaliatype code snippets.

Focusing on \gptfo on \stattypeso, despite identifier renaming and comment adding
causing insignificant performance increases, and code lowering causing a small but significant decrease in precision
without impacting F1, when all combined, significantly decreased performance in precision,
recall, and F1-scores.
These effects were not observed for any model using \thaliatype code snippets.

The discrepancies in performance indicate that the results in \stattypeso may be influenced by data leakage
and have limited generalizability compared to the results in \thaliatype.
A potential explanation is that transformations reduced the ability of \llms to recognize
code patterns from \stattypeso snippets from training, leading to degraded performance.
In contrast, \thaliatype consists of newly generated code snippets that were not in the training data.
As a result, models must rely more on reasoning over execution semantics rather than recall from training.
The finding that performance on \thaliatype does not consistently decline,
and sometimes even improves,
suggests that current \llms can withstand simple, execution-semantic preserving transformations.
\llms generalize more effectively on \thaliatype than their \stattypeso performance suggests.

In contrast, \snr, which analyzes the semantic meaning behind the code snippets,
was unaffected by these transformations, experiencing only minor, statistically insignificant variations
under the code lowering transformation.
These variations occurred when multiple types satisfied
all constraints, with the order of type variables serving as a tie-breaker.

\vspace{-5pt}
\begin{findingbox}
In contrast to \snr,
which is designed to extract semantic meaning from code snippets and remains unaffected by semantic-preserving transformations,
\llms' type inference generalizability is negatively impacted by combined transformations on \stattypeso,
a pattern not observed in \thaliatype.
This contrast suggests that while \llms are capable of generalizing in \thaliatype and its transformed variants,
their generalization ability diminishes in \stattypeso, potentially due in part to data leakage.
\end{findingbox}

\section{Threats to Validity}
\label{sec:threats}

There are four main threats to validity.

First, the \emph{representativeness of \thaliatype} may not fully
capture certain qualities of real-world Java code snippets, such as variable naming.
However, this limitation is mitigated by the primary objective of our evaluation,
which is to assess the type inference capabilities of \llms.
Specifically,
\thaliatype is designed to evaluate whether \llms can infer correct type information by reasoning
about the execution semantics of the code snippets,
rather than merely recalling examples from training.
We acknowledge that \llms may leverage additional semantic cues present in \stattypeso
but not found in \thaliatype for type inference.
To account for these potential differences,
RQ2 applies transformations to both benchmark suites.
The identifier renaming transformation on \stattypeso even slightly increased type inference
performance of \gptfo,
which means that factors such as identifier names do not significantly affect \gptfo's performance on
\stattypeso and cannot explain \llms' diminished performance on \thaliatype.
Furthermore, the strong performance of \snr on both \stattypeso and \thaliatype indicates that,
\begin{enumerate*}[label=(\arabic*),]
\item \thaliatype shares key features that are necessary for type inference with \stattypeso,
\item it presents a meaningful challenge for type inference techniques, and
\item it exposes areas where \llms require further improvement.
\end{enumerate*}

Second, variations in the \emph{document frequency of \fqns} between \thaliacs and \stattypeso
could influence \llm type inference performance,
since \llms' outputs are biased by the training data~\citep{Huang:arxiv:24,Liu:NIPS:23}.
\llm's performance may vary depending on the prevalence of a type
within \llm's training data.
To address this, we analyzed the performance of \llms across different document frequency levels.
Despite these efforts, \llms consistently exhibited lower performance on unseen code snippets from \thaliacs compared to \stattypeso.

Third, our findings may \emph{not generalize to other \llms}.
We mitigated this by following best practices~\citep{Sallou:ICSENIER:24} and evaluated a diverse set of models,
including both open-weight~(\starcoderi, \llamas, \llamam) and state-of-the-art closed models~(\gptfo, \gptfomini)
of varying sizes, providing a representative sample of current \llm capabilities.

Fourth, \emph{prompt engineering} may enable \llm to achieve better performance for type inference
on code snippets.
To mitigate this threat, we followed best practices to
design a simple but effective prompt~\citep{LLMPratice}, which achieved high performance on the \stattypeso suite.
Regardless, data leakage remains a fundamental issue that can influence \llm evaluations regardless of prompt quality.
Our evaluation highlights how data leakage may have impacted prior assessments of \llm type inference performance.

\section{Related Work}
\label{sec:related_work}
In this section, we briefly discuss different techniques for type inference on code snippets
and \llm data leakage in other fields including automated program repair,
code generation, refactoring, and more.

\subsection{Type Inference for Code Snippets}
Existing type inference approaches for code snippets mainly fall into two categories,
constraint-based approaches and
machine-learning-based approaches.

\myparagraph{Constraint-Based Approaches}
The general idea of constraint-based approaches is to first analyze the code snippet and
derive a collection of constraints on the types that need to be inferred.
With the derived constraints, a set of APIs that satisfies all the constraints can be obtained
by performing a constraint-solving algorithm.
Baker is the first work that applied constraint-based type inference for code snippets~\citep{Subramanian:ICSE:14}.
A subsequent study, \snr, outperformed Baker by improving the handling of parameterized types and
leveraged Datalog to efficiently solve the derived constraints against a knowledge base of known types~\citep{SnR}.

\myparagraph{Machine-Learning-Based Approaches}
Machine-learning-based approaches typically leverage models trained on a large set of programs
from open-source projects.
\citet{Phan:ICSE:18} proposed \stattype, which learns the \fqns that often co-occur from a large corpus.
With such knowledge, \stattype can derive the \fqn for an API based on the neighboring API names.
A subsequent work by \citet{Saifullah:ASE:19} made improvements by leveraging both local and global
contexts.
\citet{Huang:TOSEM:2023} employed the pre-trained CodeBert~\citep{Feng:EMNLP:20},
a transformer-based masked language model to predict \fqns in the code snippet.
Compared to \llms evaluated in this study, CodeBert is much smaller with 125 million parameters,
whereas even \llamas's contains 8 billion parameters.
\citet{Chen:Arxiv:24} proposed a hybrid approach that iteratively combines constraint-based techniques,
such as \snr, with machine learning methods to refine the results further.
\citet{Kabir:TOSEM:24} introduced ZS4C,
an LLM-based approach that employs multiple prompts and incorporated compiler error messages to infer and correct
compilation errors in the code snippet using LLMs.
However,
since our work focuses on mitigating data leakage,
which fundamentally undermines inference reliability regardless of the prompting strategy used,
we used a single prompt.
More broadly, data leakage can also diminish the contribution of ZS4C's use of compiler error messages,
as it enables LLMs to achieve artificially high scores on \stattypeso,
thereby masking the substantial improvements gained when compiler error messages are incorporated.

\subsection{Type Inference for Dynamically Typed Languages}
Type inference for dynamically typed languages faces extra challenges as there can be
variables with insufficient static type constraints in programs written in these languages,
which static type inference cannot soundly handle.
Many existing studies resort to deep learning-based approaches to overcome this challenge.
For example, \citet{hellendoorn2018deep} propose DeepTyper,
a deep learning model that is trained to provide type suggestions
given certain contexts and relations.
\citet{pradel2020typewriter} propose TypeWriter, which uses a deep learning-based approach
to predict the types and then utilizes a search-based approach to validate the predicted types.
\citet{peng2022static} propose HiTyper, a deep learning-based approach that combines with static type inference.
It conducts static inference and DL-based prediction iteratively to construct a type dependency graph,
which records type dependency information among variables.
This paper focuses on the performance of \llms in type inference for Java code snippets,
leaving the investigation of type inference for dynamically typed languages to future work.

\subsection{\llm Data Leakage}
Data leakage is not limited to type inference but also affects other software engineering tasks.
Currently, \llms have been integrated into a wide variety of software engineering tools such as
automated code repair~\citep{Yuan:ISSTA:22,Zhang:ASE:23,Xia:ICSE:23,Xia:ISSTA:24,Ouyang:ISSTA:24,Bouzenia:ICSE:25,Ye:ISSTA:25,Kong:TOSEM:25},
code generation~\citep{Jiang:TOSEM:24,Mu:FSE:25,Li:TOSEM:25,Fan:arxiv:25},
code refactoring~\citep{Shirafuji:APSEC:23,Pomian:ICSME:24},
and code completion~\citep{Guo:ICML:23,Wu:ICML:24,Semenkin:arxiv:25},
which can all be potentially affected by data leakage either currently or in the future.

\citet{Xia:ICSE:23} discovered that 15\% of the bug-fixing patches for automated program
repair generated by earlier \llms~(\ie, CodeT5, GPT-Neo, GPT-J, and GPT-NeoX) were already present
in the training data.
\citet{Sainz:ChatGPT:cheet,Sainz:EMNLP:23} examined several academic datasets and reported that most of them
were either used as training data for ChatGPT or likely exposed to it.
A subsequent work by~\citet{golchin2023time} proposes a more advanced detection technique and
detects the presence of test data of several datasets in the training data of \llms.
\citet{Marone:NEURIPS:23} propose data portraits to efficiently check whether a given string is present in the training data,
assuming the training data is available.
Recently, \citet{Kong:FSE:25}, detected memorization in \llms using low-probability events such as exact code repair matches.
However, for type inference on code snippets, there is generally one ground truth of types that are expected.
Thus, exact matches do not precisely indicate data leakage.
As \llm training datasets grow larger and increasingly opaque, detecting instances of data leakage
during evaluation becomes progressively more challenging.

Nevertheless, data leakage remains an open challenge for evaluating \llms.
Some studies have been conducted in other fields~\citep{Magar:ACL:22,Balloccu:EACL:24,Mirzadeh:Arxiv:24,cao2024javabench,Uddin:ACL:25,Elangovan:ACL:21}
regarding data leakage and \llm evaluation.
\citet{Mirzadeh:Arxiv:24} showed that \llms have significant limitations when conducting
genuine mathematical reasoning, which suggests that \llms are conducting sophisticated pattern matching
rather than true logical reasoning.
A recent study by \citep{cao2024javabench} demonstrated that the performance of \llms is far behind undergraduate students
on the proposed project-level Java benchmark that exercises object-oriented programming features.

Our evaluation on type inference rather than math questions or code generation
also showed a significant drop in performance with
generated code snippets which suggests that understanding the type system in code is challenging
for \llms.
So far, there is no conclusive evidence showing that \llms are conducting genuine type inference on
code snippets.
Small changes to the inputs can drastically alter model outputs~\citep{Shi:PMLR:23,Mirzadeh:Arxiv:24,Jiang:Arxiv:24}.
While some guidelines have been raised for software engineering research using
\llms~\citep{Ozkaya:Software:23,Sallou:ICSENIER:24}, protecting against data leakage remains an open challenge.
We hope our work will shed further light on the data leakage issue in software engineering.

\section{Conclusion}
\label{sec:conclusion}
This paper conducted a comprehensive assessment of \llms' type inference capabilities on Java code snippets.
First, to address potential data leakage we introduced a new benchmark suite, \thaliacs,
and compared it against the widely used \stattypeso.
Second, using \starcoder as a baseline~(which we confirmed to suffer from data leakage),
our evaluation revealed that all tested \llms exhibited similar performance degradations on unseen code snippets.
Specifically, we observed up to a \pdecreasellamasp decrease in precision
and a \pdecreasellamasr decrease in recall.
Moreover, \llm performance diminished when inferring less commonly used \fqns,
suggesting potential bias and presenting opportunities for future research.
Finally, we designed and applied three semantic-preserving code transformations to both benchmark suites,
to investigate \llms' understanding of the execution semantics.
While \llms maintained consistent performance under simple transformations,
their performance significantly declined when transformations are combined on \stattypeso.
These results on \stattypeso, not observed on \thaliatype,
indicate that \llm performance on \stattypeso may be inflated by data leakage and may not be generalizable.
Overall, our findings underscore the need for future evaluations
to incorporate unseen benchmarks, such as newly generated \thaliatype code snippets and transformed variants,
while discontinuing \stattypeso.
This approach provides a more reliable assessment of \llms' performance in type inference tasks.
All code and datasets used in this paper are publicly available at \urldataset.

\bibliographystyle{ACM-Reference-Format}
\bibliography{acmart}


\begin{thebibliography}{71}


\ifx \showCODEN    \undefined \def \showCODEN     #1{\unskip}     \fi
\ifx \showISBNx    \undefined \def \showISBNx     #1{\unskip}     \fi
\ifx \showISBNxiii \undefined \def \showISBNxiii  #1{\unskip}     \fi
\ifx \showISSN     \undefined \def \showISSN      #1{\unskip}     \fi
\ifx \showLCCN     \undefined \def \showLCCN      #1{\unskip}     \fi
\ifx \shownote     \undefined \def \shownote      #1{#1}          \fi
\ifx \showarticletitle \undefined \def \showarticletitle #1{#1}   \fi
\ifx \showURL      \undefined \def \showURL       {\relax}        \fi
\providecommand\bibfield[2]{#2}
\providecommand\bibinfo[2]{#2}
\providecommand\natexlab[1]{#1}
\providecommand\showeprint[2][]{arXiv:#2}

\bibitem[JDK(2014)]%
        {JDKLogger}
 \bibinfo{year}{2014}\natexlab{}.
\newblock \bibinfo{title}{Logger (Java Platform SE 8 )}.
\newblock
\urldef\tempurl%
\url{https://docs.oracle.com/javase/8/docs/api/java/util/logging/Logger.html}
\showURL{%
\tempurl}


\bibitem[Sta(2017)]%
        {StatTypeSO:GitHub:17:full}
 \bibinfo{year}{2017}\natexlab{}.
\newblock \bibinfo{title}{Github}.
\newblock
\urldef\tempurl%
\url{https://github.com/pdhung3012/TypeResolution_Oracle}
\showURL{%
\tempurl}


\bibitem[SLF(2021)]%
        {SLFLogger}
 \bibinfo{year}{2021}\natexlab{}.
\newblock \bibinfo{title}{Logger (SLF4J javadoc)}.
\newblock
\urldef\tempurl%
\url{https://www.slf4j.org/api/org/slf4j/Logger.html}
\showURL{%
\tempurl}


\bibitem[LLM(2024)]%
        {LLMPratice}
 \bibinfo{year}{2024}\natexlab{}.
\newblock \bibinfo{title}{LLM prompting guide}.
\newblock
\urldef\tempurl%
\url{https://huggingface.co/docs/transformers/en/tasks/prompting}
\showURL{%
\tempurl}


\bibitem[Log(2024a)]%
        {LogLogger}
 \bibinfo{year}{2024}\natexlab{a}.
\newblock \bibinfo{title}{Logger (Apache Log4j API 2.24.3 API)}.
\newblock
\urldef\tempurl%
\url{https://logging.apache.org/log4j/2.x/javadoc/log4j-api/org/apache/logging/log4j/Logger.html}
\showURL{%
\tempurl}


\bibitem[Log(2024b)]%
        {LogBackLogger}
 \bibinfo{year}{2024}\natexlab{b}.
\newblock \bibinfo{title}{Logger (Logback-Parent 1.5.15 API)}.
\newblock
\urldef\tempurl%
\url{https://logback.qos.ch/apidocs/ch.qos.logback.classic/ch/qos/logback/classic/Logger.html}
\showURL{%
\tempurl}


\bibitem[OPE(2024a)]%
        {OPENAIMODELS}
 \bibinfo{year}{2024}\natexlab{a}.
\newblock \bibinfo{title}{Models - OpenAI API}.
\newblock
\urldef\tempurl%
\url{https://platform.openai.com/docs/models}
\showURL{%
\tempurl}


\bibitem[OLL(2024)]%
        {OLLAMAPLATFORM}
 \bibinfo{year}{2024}\natexlab{}.
\newblock \bibinfo{title}{Ollama}.
\newblock
\urldef\tempurl%
\url{https://ollama.com/}
\showURL{%
\tempurl}


\bibitem[OPE(2024b)]%
        {OPENAIPLATFORM}
 \bibinfo{year}{2024}\natexlab{b}.
\newblock \bibinfo{title}{Overview - OpenAI API}.
\newblock
\urldef\tempurl%
\url{https://platform.openai.com/docs/overview}
\showURL{%
\tempurl}


\bibitem[Balloccu et~al\mbox{.}(2024)]%
        {Balloccu:EACL:24}
\bibfield{author}{\bibinfo{person}{Simone Balloccu}, \bibinfo{person}{Patrícia
  Schmidtová}, \bibinfo{person}{Mateusz Lango}, {and} \bibinfo{person}{Ondřej
  Dušek}.} \bibinfo{year}{2024}\natexlab{}.
\newblock \showarticletitle{Leak, Cheat, Repeat: Data Contamination and
  Evaluation Malpractices in Closed-Source LLMs}. In
  \bibinfo{booktitle}{\emph{Proceedings of the 18th Conference of the European
  Chapter of the Association for Computational Linguistics}}.
  \bibinfo{publisher}{Association for Computational Linguistics}.
\newblock


\bibitem[BigCode(2024)]%
        {the-stack-v2}
\bibfield{author}{\bibinfo{person}{BigCode}.} \bibinfo{year}{2024}\natexlab{}.
\newblock \bibinfo{title}{The Stack v2 dedup}.
\newblock
  \bibinfo{howpublished}{\url{https://huggingface.co/datasets/bigcode/the-stack-v2-dedup}}.
\newblock


\bibitem[Bouzenia et~al\mbox{.}(2025)]%
        {Bouzenia:ICSE:25}
\bibfield{author}{\bibinfo{person}{Islem Bouzenia}, \bibinfo{person}{Premkumar
  Devanbu}, {and} \bibinfo{person}{Michael Pradel}.}
  \bibinfo{year}{2025}\natexlab{}.
\newblock \showarticletitle{{ RepairAgent: An Autonomous, LLM-Based Agent for
  Program Repair }}. In \bibinfo{booktitle}{\emph{2025 IEEE/ACM 47th
  International Conference on Software Engineering (ICSE)}}.
  \bibinfo{publisher}{IEEE Computer Society}, \bibinfo{address}{Los Alamitos,
  CA, USA}, \bibinfo{pages}{2188--2200}.
\newblock
\href{https://doi.org/10.1109/ICSE55347.2025.00157}{doi:\nolinkurl{10.1109/ICSE55347.2025.00157}}


\bibitem[Cao et~al\mbox{.}(2024)]%
        {cao2024javabench}
\bibfield{author}{\bibinfo{person}{Jialun Cao}, \bibinfo{person}{Zhiyong Chen},
  \bibinfo{person}{Jiarong Wu}, \bibinfo{person}{Shing-Chi Cheung}, {and}
  \bibinfo{person}{Chang Xu}.} \bibinfo{year}{2024}\natexlab{}.
\newblock \showarticletitle{JavaBench: A Benchmark of Object-Oriented Code
  Generation for Evaluating Large Language Models}. In
  \bibinfo{booktitle}{\emph{Proceedings of the 39th IEEE/ACM International
  Conference on Automated Software Engineering}}. \bibinfo{pages}{870--882}.
\newblock


\bibitem[Carlini et~al\mbox{.}(2021)]%
        {Carlini:USENIX:21}
\bibfield{author}{\bibinfo{person}{Nicholas Carlini}, \bibinfo{person}{Florian
  Tram{\`e}r}, \bibinfo{person}{Eric Wallace}, \bibinfo{person}{Matthew
  Jagielski}, \bibinfo{person}{Ariel Herbert-Voss}, \bibinfo{person}{Katherine
  Lee}, \bibinfo{person}{Adam Roberts}, \bibinfo{person}{Tom Brown},
  \bibinfo{person}{Dawn Song}, \bibinfo{person}{{\'U}lfar Erlingsson},
  \bibinfo{person}{Alina Oprea}, {and} \bibinfo{person}{Colin Raffel}.}
  \bibinfo{year}{2021}\natexlab{}.
\newblock \showarticletitle{Extracting Training Data from Large Language
  Models}. In \bibinfo{booktitle}{\emph{30th USENIX Security Symposium (USENIX
  Security 21)}}. \bibinfo{publisher}{USENIX Association},
  \bibinfo{pages}{2633--2650}.
\newblock
\showISBNx{978-1-939133-24-3}
\urldef\tempurl%
\url{https://www.usenix.org/conference/usenixsecurity21/presentation/carlini-extracting}
\showURL{%
\tempurl}


\bibitem[Chaliasos et~al\mbox{.}(2022)]%
        {Chaliasos:PLDI:22}
\bibfield{author}{\bibinfo{person}{Stefanos Chaliasos},
  \bibinfo{person}{Thodoris Sotiropoulos}, \bibinfo{person}{Diomidis
  Spinellis}, \bibinfo{person}{Arthur Gervais}, \bibinfo{person}{Benjamin
  Livshits}, {and} \bibinfo{person}{Dimitris Mitropoulos}.}
  \bibinfo{year}{2022}\natexlab{}.
\newblock \showarticletitle{Finding typing compiler bugs}. In
  \bibinfo{booktitle}{\emph{Proceedings of the 43rd ACM SIGPLAN International
  Conference on Programming Language Design and Implementation}} (San Diego,
  CA, USA) \emph{(\bibinfo{series}{PLDI 2022})}.
  \bibinfo{publisher}{Association for Computing Machinery},
  \bibinfo{address}{New York, NY, USA}, \bibinfo{pages}{183–198}.
\newblock
\showISBNx{9781450392655}
\href{https://doi.org/10.1145/3519939.3523427}{doi:\nolinkurl{10.1145/3519939.3523427}}


\bibitem[Chen et~al\mbox{.}(2024)]%
        {Chen:Arxiv:24}
\bibfield{author}{\bibinfo{person}{Zhixiang Chen}, \bibinfo{person}{Anji Li},
  \bibinfo{person}{Neng Zhang}, \bibinfo{person}{Jianguo Chen},
  \bibinfo{person}{Yuan Huang}, {and} \bibinfo{person}{Zibin Zheng}.}
  \bibinfo{year}{2024}\natexlab{}.
\newblock \bibinfo{title}{iJTyper: An Iterative Type Inference Framework for
  Java by Integrating Constraint- and Statistically-based Methods}.
\newblock
\showeprint[arxiv]{2402.09995}~[cs.SE]
\urldef\tempurl%
\url{https://arxiv.org/abs/2402.09995}
\showURL{%
\tempurl}


\bibitem[Dong et~al\mbox{.}(2022)]%
        {SnR}
\bibfield{author}{\bibinfo{person}{Yiwen Dong}, \bibinfo{person}{Tianxiao Gu},
  \bibinfo{person}{Yongqiang Tian}, {and} \bibinfo{person}{Chengnian Sun}.}
  \bibinfo{year}{2022}\natexlab{}.
\newblock \showarticletitle{SnR: Constraint-Based Type Inference for Incomplete
  Java Code Snippets}. In \bibinfo{booktitle}{\emph{Proceedings of the 44th
  International Conference on Software Engineering}} (Pittsburgh, Pennsylvania)
  \emph{(\bibinfo{series}{ICSE '22})}. \bibinfo{publisher}{Association for
  Computing Machinery}, \bibinfo{address}{New York, NY, USA},
  \bibinfo{pages}{1982–1993}.
\newblock
\showISBNx{9781450392211}
\href{https://doi.org/10.1145/3510003.3510061}{doi:\nolinkurl{10.1145/3510003.3510061}}


\bibitem[Dyer et~al\mbox{.}(2013a)]%
        {Dyer:ICSE:13}
\bibfield{author}{\bibinfo{person}{Robert Dyer}, \bibinfo{person}{Hoan~Anh
  Nguyen}, \bibinfo{person}{Hridesh Rajan}, {and} \bibinfo{person}{Tien~N.
  Nguyen}.} \bibinfo{year}{2013}\natexlab{a}.
\newblock \showarticletitle{Boa: A Language and Infrastructure for Analyzing
  Ultra-Large-Scale Software Repositories}. In
  \bibinfo{booktitle}{\emph{Proceedings of the 35th International Conference on
  Software Engineering}} (San Francisco, CA)
  \emph{(\bibinfo{series}{{ICSE}'13})}. \bibinfo{pages}{422--431}.
\newblock


\bibitem[Dyer et~al\mbox{.}(2013b)]%
        {Dyer:GPCE:13}
\bibfield{author}{\bibinfo{person}{Robert Dyer}, \bibinfo{person}{Hridesh
  Rajan}, {and} \bibinfo{person}{Tien~N. Nguyen}.}
  \bibinfo{year}{2013}\natexlab{b}.
\newblock \showarticletitle{Declarative Visitors to Ease Fine-grained Source
  Code Mining with Full History on Billions of {AST} Nodes}. In
  \bibinfo{booktitle}{\emph{Proceedings of the 12th International Conference on
  Generative Programming: Concepts \& Experiences}} (Indianapolis, IN)
  \emph{(\bibinfo{series}{GPCE})}. \bibinfo{pages}{23--32}.
\newblock


\bibitem[Elangovan et~al\mbox{.}(2021)]%
        {Elangovan:ACL:21}
\bibfield{author}{\bibinfo{person}{Aparna Elangovan}, \bibinfo{person}{Jiayuan
  He}, {and} \bibinfo{person}{Karin Verspoor}.}
  \bibinfo{year}{2021}\natexlab{}.
\newblock \showarticletitle{Memorization vs. Generalization : Quantifying Data
  Leakage in {NLP} Performance Evaluation}. In
  \bibinfo{booktitle}{\emph{Proceedings of the 16th Conference of the European
  Chapter of the Association for Computational Linguistics: Main Volume}},
  \bibfield{editor}{\bibinfo{person}{Paola Merlo}, \bibinfo{person}{Jorg
  Tiedemann}, {and} \bibinfo{person}{Reut Tsarfaty}} (Eds.).
  \bibinfo{publisher}{Association for Computational Linguistics},
  \bibinfo{address}{Online}, \bibinfo{pages}{1325--1335}.
\newblock
\href{https://doi.org/10.18653/v1/2021.eacl-main.113}{doi:\nolinkurl{10.18653/v1/2021.eacl-main.113}}


\bibitem[Fan et~al\mbox{.}(2025)]%
        {Fan:arxiv:25}
\bibfield{author}{\bibinfo{person}{Lishui Fan}, \bibinfo{person}{Zhongxin Liu},
  \bibinfo{person}{Haoye Wang}, \bibinfo{person}{Lingfeng Bao},
  \bibinfo{person}{Xin Xia}, {and} \bibinfo{person}{Shanping Li}.}
  \bibinfo{year}{2025}\natexlab{}.
\newblock \bibinfo{title}{FAIT: Fault-Aware Fine-Tuning for Better Code
  Generation}.
\newblock
\showeprint[arxiv]{2503.16913}~[cs.SE]
\urldef\tempurl%
\url{https://arxiv.org/abs/2503.16913}
\showURL{%
\tempurl}


\bibitem[Feng et~al\mbox{.}(2020)]%
        {Feng:EMNLP:20}
\bibfield{author}{\bibinfo{person}{Zhangyin Feng}, \bibinfo{person}{Daya Guo},
  \bibinfo{person}{Duyu Tang}, \bibinfo{person}{Nan Duan},
  \bibinfo{person}{Xiaocheng Feng}, \bibinfo{person}{Ming Gong},
  \bibinfo{person}{Linjun Shou}, \bibinfo{person}{Bing Qin},
  \bibinfo{person}{Ting Liu}, \bibinfo{person}{Daxin Jiang}, {and}
  \bibinfo{person}{Ming Zhou}.} \bibinfo{year}{2020}\natexlab{}.
\newblock \showarticletitle{{C}ode{BERT}: A Pre-Trained Model for Programming
  and Natural Languages}. In \bibinfo{booktitle}{\emph{Findings of the
  Association for Computational Linguistics: EMNLP 2020}},
  \bibfield{editor}{\bibinfo{person}{Trevor Cohn}, \bibinfo{person}{Yulan He},
  {and} \bibinfo{person}{Yang Liu}} (Eds.). \bibinfo{publisher}{Association for
  Computational Linguistics}, \bibinfo{address}{Online},
  \bibinfo{pages}{1536--1547}.
\newblock
\href{https://doi.org/10.18653/v1/2020.findings-emnlp.139}{doi:\nolinkurl{10.18653/v1/2020.findings-emnlp.139}}


\bibitem[Golchin and Surdeanu(2023)]%
        {golchin2023time}
\bibfield{author}{\bibinfo{person}{Shahriar Golchin} {and}
  \bibinfo{person}{Mihai Surdeanu}.} \bibinfo{year}{2023}\natexlab{}.
\newblock \showarticletitle{Time travel in llms: Tracing data contamination in
  large language models}.
\newblock \bibinfo{journal}{\emph{arXiv preprint arXiv:2308.08493}}
  (\bibinfo{year}{2023}).
\newblock


\bibitem[Gosling et~al\mbox{.}(2023)]%
        {JavaKeywords}
\bibfield{author}{\bibinfo{person}{James Gosling}, \bibinfo{person}{Bill Joy},
  \bibinfo{person}{Guy Steele}, \bibinfo{person}{Gilad Bracha},
  \bibinfo{person}{Alex Buckley}, \bibinfo{person}{Daniel Smith}, {and}
  \bibinfo{person}{Gavin Bierman}.} \bibinfo{year}{2023}\natexlab{}.
\newblock \bibinfo{booktitle}{\emph{The Java® Language Specification, Java SE
  21 Edition}}.
\newblock Oracle.
\newblock
\urldef\tempurl%
\url{https://docs.oracle.com/javase/specs/jls/se21/html/jls-3.html#jls-3.9}
\showURL{%
\tempurl}


\bibitem[Guo et~al\mbox{.}(2023)]%
        {Guo:ICML:23}
\bibfield{author}{\bibinfo{person}{Daya Guo}, \bibinfo{person}{Canwen Xu},
  \bibinfo{person}{Nan Duan}, \bibinfo{person}{Jian Yin}, {and}
  \bibinfo{person}{Julian McAuley}.} \bibinfo{year}{2023}\natexlab{}.
\newblock \showarticletitle{LongCoder: a long-range pre-trained language model
  for code completion}. In \bibinfo{booktitle}{\emph{Proceedings of the 40th
  International Conference on Machine Learning}} (Honolulu, Hawaii, USA)
  \emph{(\bibinfo{series}{ICML'23})}. \bibinfo{publisher}{JMLR.org}, Article
  \bibinfo{articleno}{486}, \bibinfo{numpages}{10}~pages.
\newblock


\bibitem[Hellendoorn et~al\mbox{.}(2018)]%
        {hellendoorn2018deep}
\bibfield{author}{\bibinfo{person}{Vincent~J Hellendoorn},
  \bibinfo{person}{Christian Bird}, \bibinfo{person}{Earl~T Barr}, {and}
  \bibinfo{person}{Miltiadis Allamanis}.} \bibinfo{year}{2018}\natexlab{}.
\newblock \showarticletitle{Deep learning type inference}. In
  \bibinfo{booktitle}{\emph{Proceedings of the 2018 26th acm joint meeting on
  european software engineering conference and symposium on the foundations of
  software engineering}}. \bibinfo{pages}{152--162}.
\newblock


\bibitem[Hou et~al\mbox{.}(2024)]%
        {Hou:TOSEM:24}
\bibfield{author}{\bibinfo{person}{Xinyi Hou}, \bibinfo{person}{Yanjie Zhao},
  \bibinfo{person}{Yue Liu}, \bibinfo{person}{Zhou Yang},
  \bibinfo{person}{Kailong Wang}, \bibinfo{person}{Li Li},
  \bibinfo{person}{Xiapu Luo}, \bibinfo{person}{David Lo},
  \bibinfo{person}{John Grundy}, {and} \bibinfo{person}{Haoyu Wang}.}
  \bibinfo{year}{2024}\natexlab{}.
\newblock \showarticletitle{Large Language Models for Software Engineering: A
  Systematic Literature Review}.
\newblock \bibinfo{journal}{\emph{ACM Trans. Softw. Eng. Methodol.}}
  (\bibinfo{date}{Sept.} \bibinfo{year}{2024}).
\newblock
\showISSN{1049-331X}
\href{https://doi.org/10.1145/3695988}{doi:\nolinkurl{10.1145/3695988}}
\newblock
\shownote{Just Accepted}.


\bibitem[Huang et~al\mbox{.}(2024)]%
        {Huang:arxiv:24}
\bibfield{author}{\bibinfo{person}{Dong Huang}, \bibinfo{person}{Qingwen Bu},
  \bibinfo{person}{Jie Zhang}, \bibinfo{person}{Xiaofei Xie},
  \bibinfo{person}{Junjie Chen}, {and} \bibinfo{person}{Heming Cui}.}
  \bibinfo{year}{2024}\natexlab{}.
\newblock \bibinfo{title}{Bias Testing and Mitigation in LLM-based Code
  Generation}.
\newblock
\showeprint[arxiv]{2309.14345}~[cs.SE]
\urldef\tempurl%
\url{https://arxiv.org/abs/2309.14345}
\showURL{%
\tempurl}


\bibitem[Huang et~al\mbox{.}(2023a)]%
        {Huang:TOSEM:2023}
\bibfield{author}{\bibinfo{person}{Qing Huang}, \bibinfo{person}{Zhiqiang
  Yuan}, \bibinfo{person}{Zhenchang Xing}, \bibinfo{person}{Xin Peng},
  \bibinfo{person}{Xiwei Xu}, {and} \bibinfo{person}{Qinghua Lu}.}
  \bibinfo{year}{2023}\natexlab{a}.
\newblock \showarticletitle{FQN Inference in Partial Code by Prompt-tuned
  Language Model of Code}.
\newblock \bibinfo{journal}{\emph{ACM Trans. Softw. Eng. Methodol.}}
  \bibinfo{volume}{33}, \bibinfo{number}{2}, Article \bibinfo{articleno}{31}
  (\bibinfo{date}{dec} \bibinfo{year}{2023}), \bibinfo{numpages}{32}~pages.
\newblock
\showISSN{1049-331X}
\href{https://doi.org/10.1145/3617174}{doi:\nolinkurl{10.1145/3617174}}


\bibitem[Huang et~al\mbox{.}(2023b)]%
        {Huang:ASE:2022}
\bibfield{author}{\bibinfo{person}{Qing Huang}, \bibinfo{person}{Zhiqiang
  Yuan}, \bibinfo{person}{Zhenchang Xing}, \bibinfo{person}{Xiwei Xu},
  \bibinfo{person}{Liming Zhu}, {and} \bibinfo{person}{Qinghua Lu}.}
  \bibinfo{year}{2023}\natexlab{b}.
\newblock \showarticletitle{Prompt-Tuned Code Language Model as a Neural
  Knowledge Base for Type Inference in Statically-Typed Partial Code}. In
  \bibinfo{booktitle}{\emph{Proceedings of the 37th IEEE/ACM International
  Conference on Automated Software Engineering}} (Rochester, MI, USA)
  \emph{(\bibinfo{series}{ASE '22})}. \bibinfo{publisher}{Association for
  Computing Machinery}, \bibinfo{address}{New York, NY, USA}, Article
  \bibinfo{articleno}{79}, \bibinfo{numpages}{13}~pages.
\newblock
\showISBNx{9781450394758}
\href{https://doi.org/10.1145/3551349.3556912}{doi:\nolinkurl{10.1145/3551349.3556912}}


\bibitem[Inan et~al\mbox{.}(2021)]%
        {Inan:MS:21}
\bibfield{author}{\bibinfo{person}{Huseyin~Atahan Inan}, \bibinfo{person}{Osman
  Ramadan}, \bibinfo{person}{Lukas Wutschitz}, \bibinfo{person}{Daniel Jones},
  \bibinfo{person}{Victor Rühle}, \bibinfo{person}{James Withers}, {and}
  \bibinfo{person}{Robert Sim}.} \bibinfo{year}{2021}\natexlab{}.
\newblock \bibinfo{title}{Training Data Leakage Analysis in Language Models}.
  (\bibinfo{date}{February} \bibinfo{year}{2021}).
\newblock
\urldef\tempurl%
\url{https://www.microsoft.com/en-us/research/publication/training-data-leakage-analysis-in-language-models/}
\showURL{%
\tempurl}


\bibitem[Jiang et~al\mbox{.}(2024b)]%
        {Jiang:Arxiv:24}
\bibfield{author}{\bibinfo{person}{Bowen Jiang}, \bibinfo{person}{Yangxinyu
  Xie}, \bibinfo{person}{Zhuoqun Hao}, \bibinfo{person}{Xiaomeng Wang},
  \bibinfo{person}{Tanwi Mallick}, \bibinfo{person}{Weijie~J Su},
  \bibinfo{person}{Camillo~J Taylor}, {and} \bibinfo{person}{Dan Roth}.}
  \bibinfo{year}{2024}\natexlab{b}.
\newblock \showarticletitle{A Peek into Token Bias: Large Language Models Are
  Not Yet Genuine Reasoners}.
\newblock \bibinfo{journal}{\emph{arXiv preprint arXiv:2406.11050}}
  (\bibinfo{year}{2024}).
\newblock


\bibitem[Jiang et~al\mbox{.}(2024a)]%
        {Jiang:TOSEM:24}
\bibfield{author}{\bibinfo{person}{Xue Jiang}, \bibinfo{person}{Yihong Dong},
  \bibinfo{person}{Lecheng Wang}, \bibinfo{person}{Zheng Fang},
  \bibinfo{person}{Qiwei Shang}, \bibinfo{person}{Ge Li}, \bibinfo{person}{Zhi
  Jin}, {and} \bibinfo{person}{Wenpin Jiao}.} \bibinfo{year}{2024}\natexlab{a}.
\newblock \showarticletitle{Self-Planning Code Generation with Large Language
  Models}.
\newblock \bibinfo{journal}{\emph{ACM Trans. Softw. Eng. Methodol.}}
  \bibinfo{volume}{33}, \bibinfo{number}{7}, Article \bibinfo{articleno}{182}
  (\bibinfo{date}{Sept.} \bibinfo{year}{2024}), \bibinfo{numpages}{30}~pages.
\newblock
\showISSN{1049-331X}
\href{https://doi.org/10.1145/3672456}{doi:\nolinkurl{10.1145/3672456}}


\bibitem[Kabir et~al\mbox{.}(2025)]%
        {Kabir:TOSEM:24}
\bibfield{author}{\bibinfo{person}{Azmain Kabir}, \bibinfo{person}{Shaowei
  Wang}, \bibinfo{person}{Yuan Tian}, \bibinfo{person}{Tse-Hsun~(Peter) Chen},
  \bibinfo{person}{Muhammad Asaduzzaman}, {and} \bibinfo{person}{Wenbin
  Zhang}.} \bibinfo{year}{2025}\natexlab{}.
\newblock \showarticletitle{ZS4C: Zero-Shot Synthesis of Compilable Code for
  Incomplete Code Snippets Using LLMs}.
\newblock \bibinfo{journal}{\emph{ACM Trans. Softw. Eng. Methodol.}}
  \bibinfo{volume}{34}, \bibinfo{number}{4}, Article \bibinfo{articleno}{90}
  (\bibinfo{date}{April} \bibinfo{year}{2025}), \bibinfo{numpages}{30}~pages.
\newblock
\showISSN{1049-331X}
\href{https://doi.org/10.1145/3702979}{doi:\nolinkurl{10.1145/3702979}}


\bibitem[Kong et~al\mbox{.}(2025b)]%
        {Kong:TOSEM:25}
\bibfield{author}{\bibinfo{person}{Jiaolong Kong}, \bibinfo{person}{Xiaofei
  Xie}, \bibinfo{person}{Mingfei Cheng}, \bibinfo{person}{Shangqing Liu},
  \bibinfo{person}{Xiaoning Du}, {and} \bibinfo{person}{Qi Guo}.}
  \bibinfo{year}{2025}\natexlab{b}.
\newblock \showarticletitle{ContrastRepair: Enhancing Conversation-Based
  Automated Program Repair via Contrastive Test Case Pairs}.
\newblock \bibinfo{journal}{\emph{ACM Trans. Softw. Eng. Methodol.}}
  (\bibinfo{date}{March} \bibinfo{year}{2025}).
\newblock
\showISSN{1049-331X}
\href{https://doi.org/10.1145/3719345}{doi:\nolinkurl{10.1145/3719345}}
\newblock
\shownote{Just Accepted}.


\bibitem[Kong et~al\mbox{.}(2025a)]%
        {Kong:FSE:25}
\bibfield{author}{\bibinfo{person}{Jiaolong Kong}, \bibinfo{person}{Xiaofei
  Xie}, {and} \bibinfo{person}{Shangqing Liu}.}
  \bibinfo{year}{2025}\natexlab{a}.
\newblock \showarticletitle{Demystifying Memorization in LLM-Based Program
  Repair via a General Hypothesis Testing Framework}.
\newblock \bibinfo{journal}{\emph{Proc. ACM Softw. Eng.}} \bibinfo{volume}{2},
  \bibinfo{number}{FSE}, Article \bibinfo{articleno}{FSE120}
  (\bibinfo{date}{June} \bibinfo{year}{2025}), \bibinfo{numpages}{23}~pages.
\newblock
\href{https://doi.org/10.1145/3729390}{doi:\nolinkurl{10.1145/3729390}}


\bibitem[Li et~al\mbox{.}(2025)]%
        {Li:TOSEM:25}
\bibfield{author}{\bibinfo{person}{Jia Li}, \bibinfo{person}{Ge Li},
  \bibinfo{person}{Yongmin Li}, {and} \bibinfo{person}{Zhi Jin}.}
  \bibinfo{year}{2025}\natexlab{}.
\newblock \showarticletitle{Structured Chain-of-Thought Prompting for Code
  Generation}.
\newblock \bibinfo{journal}{\emph{ACM Trans. Softw. Eng. Methodol.}}
  \bibinfo{volume}{34}, \bibinfo{number}{2}, Article \bibinfo{articleno}{37}
  (\bibinfo{date}{Jan.} \bibinfo{year}{2025}), \bibinfo{numpages}{23}~pages.
\newblock
\showISSN{1049-331X}
\href{https://doi.org/10.1145/3690635}{doi:\nolinkurl{10.1145/3690635}}


\bibitem[Liu et~al\mbox{.}(2024)]%
        {Liu:NIPS:23}
\bibfield{author}{\bibinfo{person}{Yan Liu}, \bibinfo{person}{Xiaokang Chen},
  \bibinfo{person}{Yan Gao}, \bibinfo{person}{Zhe Su}, \bibinfo{person}{Fengji
  Zhang}, \bibinfo{person}{Daoguang Zan}, \bibinfo{person}{Jian-Guang Lou},
  \bibinfo{person}{Pin-Yu Chen}, {and} \bibinfo{person}{Tsung-Yi Ho}.}
  \bibinfo{year}{2024}\natexlab{}.
\newblock \showarticletitle{Uncovering and quantifying social biases in code
  generation}. In \bibinfo{booktitle}{\emph{Proceedings of the 37th
  International Conference on Neural Information Processing Systems}} (New
  Orleans, LA, USA) \emph{(\bibinfo{series}{NIPS '23})}.
  \bibinfo{publisher}{Curran Associates Inc.}, \bibinfo{address}{Red Hook, NY,
  USA}, Article \bibinfo{articleno}{110}, \bibinfo{numpages}{13}~pages.
\newblock


\bibitem[Lozhkov et~al\mbox{.}(2024)]%
        {starcoder2}
\bibfield{author}{\bibinfo{person}{Anton Lozhkov}, \bibinfo{person}{Raymond
  Li}, \bibinfo{person}{Loubna~Ben Allal}, \bibinfo{person}{Federico Cassano},
  \bibinfo{person}{Joel Lamy-Poirier}, \bibinfo{person}{Nouamane Tazi},
  \bibinfo{person}{Ao Tang}, \bibinfo{person}{Dmytro Pykhtar},
  \bibinfo{person}{Jiawei Liu}, \bibinfo{person}{Yuxiang Wei},
  \bibinfo{person}{Tianyang Liu}, \bibinfo{person}{Max Tian},
  \bibinfo{person}{Denis Kocetkov}, \bibinfo{person}{Arthur Zucker},
  \bibinfo{person}{Younes Belkada}, \bibinfo{person}{Zijian Wang},
  \bibinfo{person}{Qian Liu}, \bibinfo{person}{Dmitry Abulkhanov},
  \bibinfo{person}{Indraneil Paul}, \bibinfo{person}{Zhuang Li},
  \bibinfo{person}{Wen-Ding Li}, \bibinfo{person}{Megan Risdal},
  \bibinfo{person}{Jia Li}, \bibinfo{person}{Jian Zhu},
  \bibinfo{person}{Terry~Yue Zhuo}, \bibinfo{person}{Evgenii Zheltonozhskii},
  \bibinfo{person}{Nii Osae~Osae Dade}, \bibinfo{person}{Wenhao Yu},
  \bibinfo{person}{Lucas Krauß}, \bibinfo{person}{Naman Jain},
  \bibinfo{person}{Yixuan Su}, \bibinfo{person}{Xuanli He},
  \bibinfo{person}{Manan Dey}, \bibinfo{person}{Edoardo Abati},
  \bibinfo{person}{Yekun Chai}, \bibinfo{person}{Niklas Muennighoff},
  \bibinfo{person}{Xiangru Tang}, \bibinfo{person}{Muhtasham Oblokulov},
  \bibinfo{person}{Christopher Akiki}, \bibinfo{person}{Marc Marone},
  \bibinfo{person}{Chenghao Mou}, \bibinfo{person}{Mayank Mishra},
  \bibinfo{person}{Alex Gu}, \bibinfo{person}{Binyuan Hui},
  \bibinfo{person}{Tri Dao}, \bibinfo{person}{Armel Zebaze},
  \bibinfo{person}{Olivier Dehaene}, \bibinfo{person}{Nicolas Patry},
  \bibinfo{person}{Canwen Xu}, \bibinfo{person}{Julian McAuley},
  \bibinfo{person}{Han Hu}, \bibinfo{person}{Torsten Scholak},
  \bibinfo{person}{Sebastien Paquet}, \bibinfo{person}{Jennifer Robinson},
  \bibinfo{person}{Carolyn~Jane Anderson}, \bibinfo{person}{Nicolas Chapados},
  \bibinfo{person}{Mostofa Patwary}, \bibinfo{person}{Nima Tajbakhsh},
  \bibinfo{person}{Yacine Jernite}, \bibinfo{person}{Carlos~Muñoz Ferrandis},
  \bibinfo{person}{Lingming Zhang}, \bibinfo{person}{Sean Hughes},
  \bibinfo{person}{Thomas Wolf}, \bibinfo{person}{Arjun Guha},
  \bibinfo{person}{Leandro von Werra}, {and} \bibinfo{person}{Harm de Vries}.}
  \bibinfo{year}{2024}\natexlab{}.
\newblock \bibinfo{title}{StarCoder 2 and The Stack v2: The Next Generation}.
\newblock
\showeprint[arxiv]{2402.19173}~[cs.SE]
\urldef\tempurl%
\url{https://arxiv.org/abs/2402.19173}
\showURL{%
\tempurl}


\bibitem[Magar and Schwartz(2022)]%
        {Magar:ACL:22}
\bibfield{author}{\bibinfo{person}{Inbal Magar} {and} \bibinfo{person}{Roy
  Schwartz}.} \bibinfo{year}{2022}\natexlab{}.
\newblock \showarticletitle{Data Contamination: From Memorization to
  Exploitation}. In \bibinfo{booktitle}{\emph{Proceedings of the 60th Annual
  Meeting of the Association for Computational Linguistics (Volume 2: Short
  Papers)}}. \bibinfo{publisher}{Association for Computational Linguistics},
  \bibinfo{address}{Dublin, Ireland}, \bibinfo{pages}{157--165}.
\newblock
\href{https://doi.org/10.18653/v1/2022.acl-short.18}{doi:\nolinkurl{10.18653/v1/2022.acl-short.18}}


\bibitem[Marone and Van~Durme(2023)]%
        {Marone:NEURIPS:23}
\bibfield{author}{\bibinfo{person}{Marc Marone} {and} \bibinfo{person}{Benjamin
  Van~Durme}.} \bibinfo{year}{2023}\natexlab{}.
\newblock \showarticletitle{Data Portraits: Recording Foundation Model Training
  Data}. In \bibinfo{booktitle}{\emph{Advances in Neural Information Processing
  Systems}}, \bibfield{editor}{\bibinfo{person}{A.~Oh},
  \bibinfo{person}{T.~Naumann}, \bibinfo{person}{A.~Globerson},
  \bibinfo{person}{K.~Saenko}, \bibinfo{person}{M.~Hardt}, {and}
  \bibinfo{person}{S.~Levine}} (Eds.), Vol.~\bibinfo{volume}{36}.
  \bibinfo{publisher}{Curran Associates, Inc.}, \bibinfo{pages}{15121--15135}.
\newblock
\urldef\tempurl%
\url{https://proceedings.neurips.cc/paper_files/paper/2023/file/3112ee706d21d734c15532c1239773e1-Paper-Datasets_and_Benchmarks.pdf}
\showURL{%
\tempurl}


\bibitem[Meta(2024)]%
        {llama3}
\bibfield{author}{\bibinfo{person}{Meta}.} \bibinfo{year}{2024}\natexlab{}.
\newblock \bibinfo{title}{The Llama 3 Herd of Models}.
\newblock
\showeprint[arxiv]{2407.21783}~[cs.AI]
\urldef\tempurl%
\url{https://arxiv.org/abs/2407.21783}
\showURL{%
\tempurl}


\bibitem[Mirzadeh et~al\mbox{.}(2024)]%
        {Mirzadeh:Arxiv:24}
\bibfield{author}{\bibinfo{person}{Iman Mirzadeh}, \bibinfo{person}{Keivan
  Alizadeh}, \bibinfo{person}{Hooman Shahrokhi}, \bibinfo{person}{Oncel Tuzel},
  \bibinfo{person}{Samy Bengio}, {and} \bibinfo{person}{Mehrdad Farajtabar}.}
  \bibinfo{year}{2024}\natexlab{}.
\newblock \bibinfo{title}{GSM-Symbolic: Understanding the Limitations of
  Mathematical Reasoning in Large Language Models}.
\newblock
\showeprint[arxiv]{2410.05229}~[cs.LG]
\urldef\tempurl%
\url{https://arxiv.org/abs/2410.05229}
\showURL{%
\tempurl}


\bibitem[Mu et~al\mbox{.}(2024)]%
        {Mu:FSE:25}
\bibfield{author}{\bibinfo{person}{Fangwen Mu}, \bibinfo{person}{Lin Shi},
  \bibinfo{person}{Song Wang}, \bibinfo{person}{Zhuohao Yu},
  \bibinfo{person}{Binquan Zhang}, \bibinfo{person}{ChenXue Wang},
  \bibinfo{person}{Shichao Liu}, {and} \bibinfo{person}{Qing Wang}.}
  \bibinfo{year}{2024}\natexlab{}.
\newblock \showarticletitle{ClarifyGPT: A Framework for Enhancing LLM-Based
  Code Generation via Requirements Clarification}.
\newblock \bibinfo{journal}{\emph{Proc. ACM Softw. Eng.}} \bibinfo{volume}{1},
  \bibinfo{number}{FSE}, Article \bibinfo{articleno}{103} (\bibinfo{date}{July}
  \bibinfo{year}{2024}), \bibinfo{numpages}{23}~pages.
\newblock
\href{https://doi.org/10.1145/3660810}{doi:\nolinkurl{10.1145/3660810}}


\bibitem[Nasr et~al\mbox{.}(2023)]%
        {Nasr:Arxiv:23}
\bibfield{author}{\bibinfo{person}{Milad Nasr}, \bibinfo{person}{Nicholas
  Carlini}, \bibinfo{person}{Jonathan Hayase}, \bibinfo{person}{Matthew
  Jagielski}, \bibinfo{person}{A.~Feder Cooper}, \bibinfo{person}{Daphne
  Ippolito}, \bibinfo{person}{Christopher~A. Choquette-Choo},
  \bibinfo{person}{Eric Wallace}, \bibinfo{person}{Florian Tramèr}, {and}
  \bibinfo{person}{Katherine Lee}.} \bibinfo{year}{2023}\natexlab{}.
\newblock \bibinfo{title}{Scalable Extraction of Training Data from
  (Production) Language Models}.
\newblock
\showeprint[arxiv]{2311.17035}~[cs.LG]
\urldef\tempurl%
\url{https://arxiv.org/abs/2311.17035}
\showURL{%
\tempurl}


\bibitem[OpenAI(2024)]%
        {gpt4}
\bibfield{author}{\bibinfo{person}{OpenAI}.} \bibinfo{year}{2024}\natexlab{}.
\newblock \bibinfo{title}{GPT-4 Technical Report}.
\newblock
\showeprint[arxiv]{2303.08774}~[cs.CL]
\urldef\tempurl%
\url{https://arxiv.org/abs/2303.08774}
\showURL{%
\tempurl}


\bibitem[Ouyang et~al\mbox{.}(2024)]%
        {Ouyang:ISSTA:24}
\bibfield{author}{\bibinfo{person}{Yicheng Ouyang}, \bibinfo{person}{Jun Yang},
  {and} \bibinfo{person}{Lingming Zhang}.} \bibinfo{year}{2024}\natexlab{}.
\newblock \showarticletitle{Benchmarking Automated Program Repair: An Extensive
  Study on Both Real-World and Artificial Bugs}. In
  \bibinfo{booktitle}{\emph{Proceedings of the 33rd ACM SIGSOFT International
  Symposium on Software Testing and Analysis}} (Vienna, Austria)
  \emph{(\bibinfo{series}{ISSTA 2024})}. \bibinfo{publisher}{Association for
  Computing Machinery}, \bibinfo{address}{New York, NY, USA},
  \bibinfo{pages}{440–452}.
\newblock
\showISBNx{9798400706127}
\href{https://doi.org/10.1145/3650212.3652140}{doi:\nolinkurl{10.1145/3650212.3652140}}


\bibitem[Ozkaya(2023)]%
        {Ozkaya:Software:23}
\bibfield{author}{\bibinfo{person}{Ipek Ozkaya}.}
  \bibinfo{year}{2023}\natexlab{}.
\newblock \showarticletitle{Application of Large Language Models to Software
  Engineering Tasks: Opportunities, Risks, and Implications}.
\newblock \bibinfo{journal}{\emph{IEEE Software}} \bibinfo{volume}{40},
  \bibinfo{number}{3} (\bibinfo{year}{2023}), \bibinfo{pages}{4--8}.
\newblock
\href{https://doi.org/10.1109/MS.2023.3248401}{doi:\nolinkurl{10.1109/MS.2023.3248401}}


\bibitem[Peng et~al\mbox{.}(2022)]%
        {peng2022static}
\bibfield{author}{\bibinfo{person}{Yun Peng}, \bibinfo{person}{Cuiyun Gao},
  \bibinfo{person}{Zongjie Li}, \bibinfo{person}{Bowei Gao},
  \bibinfo{person}{David Lo}, \bibinfo{person}{Qirun Zhang}, {and}
  \bibinfo{person}{Michael Lyu}.} \bibinfo{year}{2022}\natexlab{}.
\newblock \showarticletitle{Static inference meets deep learning: a hybrid type
  inference approach for python}. In \bibinfo{booktitle}{\emph{Proceedings of
  the 44th International Conference on Software Engineering}}.
  \bibinfo{pages}{2019--2030}.
\newblock


\bibitem[Phan et~al\mbox{.}(2018)]%
        {Phan:ICSE:18}
\bibfield{author}{\bibinfo{person}{Hung Phan}, \bibinfo{person}{Hoan~Anh
  Nguyen}, \bibinfo{person}{Ngoc~M. Tran}, \bibinfo{person}{Linh~H. Truong},
  \bibinfo{person}{Anh~Tuan Nguyen}, {and} \bibinfo{person}{Tien~N. Nguyen}.}
  \bibinfo{year}{2018}\natexlab{}.
\newblock \showarticletitle{Statistical Learning of API Fully Qualified Names
  in Code Snippets of Online Forums}. In \bibinfo{booktitle}{\emph{Proceedings
  of the 40th International Conference on Software Engineering}} (Gothenburg,
  Sweden) \emph{(\bibinfo{series}{ICSE '18})}. \bibinfo{publisher}{Association
  for Computing Machinery}, \bibinfo{address}{New York, NY, USA},
  \bibinfo{pages}{632–642}.
\newblock
\showISBNx{9781450356381}
\href{https://doi.org/10.1145/3180155.3180230}{doi:\nolinkurl{10.1145/3180155.3180230}}


\bibitem[Pomian et~al\mbox{.}(2024)]%
        {Pomian:ICSME:24}
\bibfield{author}{\bibinfo{person}{Dorin Pomian}, \bibinfo{person}{Abhiram
  Bellur}, \bibinfo{person}{Malinda Dilhara}, \bibinfo{person}{Zarina
  Kurbatova}, \bibinfo{person}{Egor Bogomolov}, \bibinfo{person}{Timofey
  Bryksin}, {and} \bibinfo{person}{Danny Dig}.}
  \bibinfo{year}{2024}\natexlab{}.
\newblock \showarticletitle{Next-Generation Refactoring: Combining LLM Insights
  and IDE Capabilities for Extract Method}. In \bibinfo{booktitle}{\emph{2024
  IEEE International Conference on Software Maintenance and Evolution
  (ICSME)}}. \bibinfo{pages}{275--287}.
\newblock
\href{https://doi.org/10.1109/ICSME58944.2024.00034}{doi:\nolinkurl{10.1109/ICSME58944.2024.00034}}


\bibitem[Pradel et~al\mbox{.}(2020)]%
        {pradel2020typewriter}
\bibfield{author}{\bibinfo{person}{Michael Pradel}, \bibinfo{person}{Georgios
  Gousios}, \bibinfo{person}{Jason Liu}, {and} \bibinfo{person}{Satish
  Chandra}.} \bibinfo{year}{2020}\natexlab{}.
\newblock \showarticletitle{Typewriter: Neural type prediction with
  search-based validation}. In \bibinfo{booktitle}{\emph{Proceedings of the
  28th ACM Joint Meeting on European Software Engineering Conference and
  Symposium on the Foundations of Software Engineering}}.
  \bibinfo{pages}{209--220}.
\newblock


\bibitem[Saifullah et~al\mbox{.}(2020)]%
        {Saifullah:ASE:19}
\bibfield{author}{\bibinfo{person}{C~M~Khaled Saifullah},
  \bibinfo{person}{Muhammad Asaduzzaman}, {and} \bibinfo{person}{Chanchal~K.
  Roy}.} \bibinfo{year}{2020}\natexlab{}.
\newblock \showarticletitle{Learning from Examples to Find Fully Qualified
  Names of API Elements in Code Snippets}. In
  \bibinfo{booktitle}{\emph{Proceedings of the 34th IEEE/ACM International
  Conference on Automated Software Engineering}} (San Diego, California)
  \emph{(\bibinfo{series}{ASE '19})}. \bibinfo{publisher}{IEEE Press},
  \bibinfo{pages}{243–254}.
\newblock
\showISBNx{9781728125084}
\href{https://doi.org/10.1109/ASE.2019.00032}{doi:\nolinkurl{10.1109/ASE.2019.00032}}


\bibitem[Sainz et~al\mbox{.}(2023a)]%
        {Sainz:EMNLP:23}
\bibfield{author}{\bibinfo{person}{Oscar Sainz}, \bibinfo{person}{Jon Campos},
  \bibinfo{person}{Iker Garc{\'\i}a-Ferrero}, \bibinfo{person}{Julen Etxaniz},
  \bibinfo{person}{Oier~Lopez de Lacalle}, {and} \bibinfo{person}{Eneko
  Agirre}.} \bibinfo{year}{2023}\natexlab{a}.
\newblock \showarticletitle{{NLP} Evaluation in trouble: On the Need to Measure
  {LLM} Data Contamination for each Benchmark}. In
  \bibinfo{booktitle}{\emph{Findings of the Association for Computational
  Linguistics: EMNLP 2023}}, \bibfield{editor}{\bibinfo{person}{Houda Bouamor},
  \bibinfo{person}{Juan Pino}, {and} \bibinfo{person}{Kalika Bali}} (Eds.).
  \bibinfo{publisher}{Association for Computational Linguistics},
  \bibinfo{address}{Singapore}, \bibinfo{pages}{10776--10787}.
\newblock
\href{https://doi.org/10.18653/v1/2023.findings-emnlp.722}{doi:\nolinkurl{10.18653/v1/2023.findings-emnlp.722}}


\bibitem[Sainz et~al\mbox{.}(2023b)]%
        {Sainz:ChatGPT:cheet}
\bibfield{author}{\bibinfo{person}{Oscar Sainz}, \bibinfo{person}{Jon~Ander
  Campos}, \bibinfo{person}{Iker García-Ferrero}, \bibinfo{person}{Julen
  Etxaniz}, {and} \bibinfo{person}{Eneko Agirre}.}
  \bibinfo{year}{2023}\natexlab{b}.
\newblock \bibinfo{title}{Did ChatGPT cheat on your test?}
\newblock
\urldef\tempurl%
\url{https://hitz-zentroa.github.io/lm-contamination/blog}
\showURL{%
\tempurl}


\bibitem[Sallou et~al\mbox{.}(2024)]%
        {Sallou:ICSENIER:24}
\bibfield{author}{\bibinfo{person}{June Sallou}, \bibinfo{person}{Thomas
  Durieux}, {and} \bibinfo{person}{Annibale Panichella}.}
  \bibinfo{year}{2024}\natexlab{}.
\newblock \showarticletitle{Breaking the Silence: the Threats of Using LLMs in
  Software Engineering}. In \bibinfo{booktitle}{\emph{Proceedings of the 2024
  ACM/IEEE 44th International Conference on Software Engineering: New Ideas and
  Emerging Results}} (Lisbon, Portugal)
  \emph{(\bibinfo{series}{ICSE-NIER'24})}. \bibinfo{publisher}{Association for
  Computing Machinery}, \bibinfo{address}{New York, NY, USA},
  \bibinfo{pages}{102–106}.
\newblock
\showISBNx{9798400705007}
\href{https://doi.org/10.1145/3639476.3639764}{doi:\nolinkurl{10.1145/3639476.3639764}}


\bibitem[Semenkin et~al\mbox{.}(2025)]%
        {Semenkin:arxiv:25}
\bibfield{author}{\bibinfo{person}{Anton Semenkin}, \bibinfo{person}{Vitaliy
  Bibaev}, \bibinfo{person}{Yaroslav Sokolov}, \bibinfo{person}{Kirill Krylov},
  \bibinfo{person}{Alexey Kalina}, \bibinfo{person}{Anna Khannanova},
  \bibinfo{person}{Danila Savenkov}, \bibinfo{person}{Darya Rovdo},
  \bibinfo{person}{Igor Davidenko}, \bibinfo{person}{Kirill Karnaukhov},
  \bibinfo{person}{Maxim Vakhrushev}, \bibinfo{person}{Mikhail Kostyukov},
  \bibinfo{person}{Mikhail Podvitskii}, \bibinfo{person}{Petr Surkov},
  \bibinfo{person}{Yaroslav Golubev}, \bibinfo{person}{Nikita Povarov}, {and}
  \bibinfo{person}{Timofey Bryksin}.} \bibinfo{year}{2025}\natexlab{}.
\newblock \bibinfo{title}{Full Line Code Completion: Bringing AI to Desktop}.
\newblock
\showeprint[arxiv]{2405.08704}~[cs.SE]
\urldef\tempurl%
\url{https://arxiv.org/abs/2405.08704}
\showURL{%
\tempurl}


\bibitem[Shi et~al\mbox{.}(2023)]%
        {Shi:PMLR:23}
\bibfield{author}{\bibinfo{person}{Freda Shi}, \bibinfo{person}{Xinyun Chen},
  \bibinfo{person}{Kanishka Misra}, \bibinfo{person}{Nathan Scales},
  \bibinfo{person}{David Dohan}, \bibinfo{person}{Ed~H. Chi},
  \bibinfo{person}{Nathanael Sch\"{a}rli}, {and} \bibinfo{person}{Denny Zhou}.}
  \bibinfo{year}{2023}\natexlab{}.
\newblock \showarticletitle{Large Language Models Can Be Easily Distracted by
  Irrelevant Context}. In \bibinfo{booktitle}{\emph{Proceedings of the 40th
  International Conference on Machine Learning}}
  \emph{(\bibinfo{series}{Proceedings of Machine Learning Research},
  Vol.~\bibinfo{volume}{202})}, \bibfield{editor}{\bibinfo{person}{Andreas
  Krause}, \bibinfo{person}{Emma Brunskill}, \bibinfo{person}{Kyunghyun Cho},
  \bibinfo{person}{Barbara Engelhardt}, \bibinfo{person}{Sivan Sabato}, {and}
  \bibinfo{person}{Jonathan Scarlett}} (Eds.). \bibinfo{publisher}{PMLR},
  \bibinfo{pages}{31210--31227}.
\newblock
\urldef\tempurl%
\url{https://proceedings.mlr.press/v202/shi23a.html}
\showURL{%
\tempurl}


\bibitem[Shirafuji et~al\mbox{.}(2023)]%
        {Shirafuji:APSEC:23}
\bibfield{author}{\bibinfo{person}{Atsushi Shirafuji}, \bibinfo{person}{Yusuke
  Oda}, \bibinfo{person}{Jun Suzuki}, \bibinfo{person}{Makoto Morishita}, {and}
  \bibinfo{person}{Yutaka Watanobe}.} \bibinfo{year}{2023}\natexlab{}.
\newblock \showarticletitle{{ Refactoring Programs Using Large Language Models
  with Few-Shot Examples }}. In \bibinfo{booktitle}{\emph{2023 30th
  Asia-Pacific Software Engineering Conference (APSEC)}}.
  \bibinfo{publisher}{IEEE Computer Society}, \bibinfo{address}{Los Alamitos,
  CA, USA}, \bibinfo{pages}{151--160}.
\newblock
\href{https://doi.org/10.1109/APSEC60848.2023.00025}{doi:\nolinkurl{10.1109/APSEC60848.2023.00025}}


\bibitem[Sotiropoulos et~al\mbox{.}(2024)]%
        {Sotiropoulos:POPL:24}
\bibfield{author}{\bibinfo{person}{Thodoris Sotiropoulos},
  \bibinfo{person}{Stefanos Chaliasos}, {and} \bibinfo{person}{Zhendong Su}.}
  \bibinfo{year}{2024}\natexlab{}.
\newblock \showarticletitle{API-Driven Program Synthesis for Testing Static
  Typing Implementations}.
\newblock \bibinfo{journal}{\emph{Proc. ACM Program. Lang.}}
  \bibinfo{volume}{8}, \bibinfo{number}{POPL}, Article \bibinfo{articleno}{62}
  (\bibinfo{date}{Jan.} \bibinfo{year}{2024}), \bibinfo{numpages}{32}~pages.
\newblock
\href{https://doi.org/10.1145/3632904}{doi:\nolinkurl{10.1145/3632904}}


\bibitem[Subramanian et~al\mbox{.}(2014)]%
        {Subramanian:ICSE:14}
\bibfield{author}{\bibinfo{person}{Siddharth Subramanian},
  \bibinfo{person}{Laura Inozemtseva}, {and} \bibinfo{person}{Reid Holmes}.}
  \bibinfo{year}{2014}\natexlab{}.
\newblock \showarticletitle{Live API Documentation}. In
  \bibinfo{booktitle}{\emph{Proceedings of the 36th International Conference on
  Software Engineering}} (Hyderabad, India) \emph{(\bibinfo{series}{ICSE
  2014})}. \bibinfo{publisher}{Association for Computing Machinery},
  \bibinfo{address}{New York, NY, USA}, \bibinfo{pages}{643–652}.
\newblock
\showISBNx{9781450327565}
\href{https://doi.org/10.1145/2568225.2568313}{doi:\nolinkurl{10.1145/2568225.2568313}}


\bibitem[Terragni et~al\mbox{.}(2016)]%
        {Terragni:ISSTA:16}
\bibfield{author}{\bibinfo{person}{Valerio Terragni}, \bibinfo{person}{Yepang
  Liu}, {and} \bibinfo{person}{Shing-Chi Cheung}.}
  \bibinfo{year}{2016}\natexlab{}.
\newblock \showarticletitle{CSNIPPEX: Automated Synthesis of Compilable Code
  Snippets from Q\&A Sites}. In \bibinfo{booktitle}{\emph{Proceedings of the
  25th International Symposium on Software Testing and Analysis}}
  (Saarbr\"{u}cken, Germany) \emph{(\bibinfo{series}{ISSTA 2016})}.
  \bibinfo{publisher}{Association for Computing Machinery},
  \bibinfo{address}{New York, NY, USA}, \bibinfo{pages}{118–129}.
\newblock
\showISBNx{9781450343909}
\href{https://doi.org/10.1145/2931037.2931058}{doi:\nolinkurl{10.1145/2931037.2931058}}


\bibitem[Terragni and Salza(2022)]%
        {Terragni:ASE:21}
\bibfield{author}{\bibinfo{person}{Valerio Terragni} {and}
  \bibinfo{person}{Pasquale Salza}.} \bibinfo{year}{2022}\natexlab{}.
\newblock \showarticletitle{APIzation: generating reusable APIs from
  StackOverflow code snippets}. In \bibinfo{booktitle}{\emph{Proceedings of the
  36th IEEE/ACM International Conference on Automated Software Engineering}}
  (Melbourne, Australia) \emph{(\bibinfo{series}{ASE '21})}.
  \bibinfo{publisher}{IEEE Press}, \bibinfo{pages}{542–554}.
\newblock
\showISBNx{9781665403375}
\href{https://doi.org/10.1109/ASE51524.2021.9678576}{doi:\nolinkurl{10.1109/ASE51524.2021.9678576}}


\bibitem[Uddin et~al\mbox{.}(2025)]%
        {Uddin:ACL:25}
\bibfield{author}{\bibinfo{person}{Md~Nayem Uddin}, \bibinfo{person}{Amir
  Saeidi}, \bibinfo{person}{Divij Handa}, \bibinfo{person}{Agastya Seth},
  \bibinfo{person}{Tran~Cao Son}, \bibinfo{person}{Eduardo Blanco},
  \bibinfo{person}{Steven~R. Corman}, {and} \bibinfo{person}{Chitta Baral}.}
  \bibinfo{year}{2025}\natexlab{}.
\newblock \bibinfo{title}{UnSeenTimeQA: Time-Sensitive Question-Answering
  Beyond LLMs' Memorization}.
\newblock
\showeprint[arxiv]{2407.03525}~[cs.CL]
\urldef\tempurl%
\url{https://arxiv.org/abs/2407.03525}
\showURL{%
\tempurl}


\bibitem[Wilcoxon(1992)]%
        {Wilcoxon:92}
\bibfield{author}{\bibinfo{person}{Frank Wilcoxon}.}
  \bibinfo{year}{1992}\natexlab{}.
\newblock \bibinfo{booktitle}{\emph{Individual Comparisons by Ranking
  Methods}}.
\newblock \bibinfo{publisher}{Springer New York}, \bibinfo{address}{New York,
  NY}, \bibinfo{pages}{196--202}.
\newblock
\showISBNx{978-1-4612-4380-9}
\href{https://doi.org/10.1007/978-1-4612-4380-9_16}{doi:\nolinkurl{10.1007/978-1-4612-4380-9_16}}


\bibitem[Wu et~al\mbox{.}(2024)]%
        {Wu:ICML:24}
\bibfield{author}{\bibinfo{person}{Di Wu}, \bibinfo{person}{Wasi~Uddin Ahmad},
  \bibinfo{person}{Dejiao Zhang}, \bibinfo{person}{Murali~Krishna Ramanathan},
  {and} \bibinfo{person}{Xiaofei Ma}.} \bibinfo{year}{2024}\natexlab{}.
\newblock \showarticletitle{REPOFORMER: selective retrieval for
  repository-level code completion} \emph{(\bibinfo{series}{ICML'24})}.
  \bibinfo{publisher}{JMLR.org}, Article \bibinfo{articleno}{2183},
  \bibinfo{numpages}{21}~pages.
\newblock


\bibitem[Xia et~al\mbox{.}(2023)]%
        {Xia:ICSE:23}
\bibfield{author}{\bibinfo{person}{Chunqiu~Steven Xia},
  \bibinfo{person}{Yuxiang Wei}, {and} \bibinfo{person}{Lingming Zhang}.}
  \bibinfo{year}{2023}\natexlab{}.
\newblock \showarticletitle{Automated Program Repair in the Era of Large
  Pre-Trained Language Models}. In \bibinfo{booktitle}{\emph{Proceedings of the
  45th International Conference on Software Engineering}} (Melbourne, Victoria,
  Australia) \emph{(\bibinfo{series}{ICSE '23})}. \bibinfo{publisher}{IEEE
  Press}, \bibinfo{pages}{1482–1494}.
\newblock
\showISBNx{9781665457019}
\href{https://doi.org/10.1109/ICSE48619.2023.00129}{doi:\nolinkurl{10.1109/ICSE48619.2023.00129}}


\bibitem[Xia and Zhang(2024)]%
        {Xia:ISSTA:24}
\bibfield{author}{\bibinfo{person}{Chunqiu~Steven Xia} {and}
  \bibinfo{person}{Lingming Zhang}.} \bibinfo{year}{2024}\natexlab{}.
\newblock \showarticletitle{Automated Program Repair via Conversation: Fixing
  162 out of 337 Bugs for \$0.42 Each using ChatGPT}. In
  \bibinfo{booktitle}{\emph{Proceedings of the 33rd ACM SIGSOFT International
  Symposium on Software Testing and Analysis}} (Vienna, Austria)
  \emph{(\bibinfo{series}{ISSTA 2024})}. \bibinfo{publisher}{Association for
  Computing Machinery}, \bibinfo{address}{New York, NY, USA},
  \bibinfo{pages}{819–831}.
\newblock
\showISBNx{9798400706127}
\href{https://doi.org/10.1145/3650212.3680323}{doi:\nolinkurl{10.1145/3650212.3680323}}


\bibitem[Ye et~al\mbox{.}(2025)]%
        {Ye:ISSTA:25}
\bibfield{author}{\bibinfo{person}{He Ye}, \bibinfo{person}{Aidan~Z.H. Yang},
  \bibinfo{person}{Chang Hu}, \bibinfo{person}{Yanlin Wang},
  \bibinfo{person}{Tao Zhang}, {and} \bibinfo{person}{Claire Le~Goues}.}
  \bibinfo{year}{2025}\natexlab{}.
\newblock \showarticletitle{AdverIntent-Agent: Adversarial Reasoning for Repair
  Based on Inferred Program Intent}.
\newblock \bibinfo{journal}{\emph{Proc. ACM Softw. Eng.}} \bibinfo{volume}{2},
  \bibinfo{number}{ISSTA}, Article \bibinfo{articleno}{ISSTA062}
  (\bibinfo{date}{June} \bibinfo{year}{2025}), \bibinfo{numpages}{23}~pages.
\newblock
\href{https://doi.org/10.1145/3728939}{doi:\nolinkurl{10.1145/3728939}}


\bibitem[Yuan et~al\mbox{.}(2022)]%
        {Yuan:ISSTA:22}
\bibfield{author}{\bibinfo{person}{Wei Yuan}, \bibinfo{person}{Quanjun Zhang},
  \bibinfo{person}{Tieke He}, \bibinfo{person}{Chunrong Fang},
  \bibinfo{person}{Nguyen Quoc~Viet Hung}, \bibinfo{person}{Xiaodong Hao},
  {and} \bibinfo{person}{Hongzhi Yin}.} \bibinfo{year}{2022}\natexlab{}.
\newblock \showarticletitle{CIRCLE: continual repair across programming
  languages}. In \bibinfo{booktitle}{\emph{Proceedings of the 31st ACM SIGSOFT
  International Symposium on Software Testing and Analysis}} (Virtual, South
  Korea) \emph{(\bibinfo{series}{ISSTA 2022})}. \bibinfo{publisher}{Association
  for Computing Machinery}, \bibinfo{address}{New York, NY, USA},
  \bibinfo{pages}{678–690}.
\newblock
\showISBNx{9781450393799}
\href{https://doi.org/10.1145/3533767.3534219}{doi:\nolinkurl{10.1145/3533767.3534219}}


\bibitem[Zhang et~al\mbox{.}(2024)]%
        {Zhang:ASE:23}
\bibfield{author}{\bibinfo{person}{Quanjun Zhang}, \bibinfo{person}{Chunrong
  Fang}, \bibinfo{person}{Tongke Zhang}, \bibinfo{person}{Bowen Yu},
  \bibinfo{person}{Weisong Sun}, {and} \bibinfo{person}{Zhenyu Chen}.}
  \bibinfo{year}{2024}\natexlab{}.
\newblock \showarticletitle{Gamma: Revisiting Template-based Automated Program
  Repair via Mask Prediction}. In \bibinfo{booktitle}{\emph{Proceedings of the
  38th IEEE/ACM International Conference on Automated Software Engineering}}
  (Echternach, Luxembourg) \emph{(\bibinfo{series}{ASE '23})}.
  \bibinfo{publisher}{IEEE Press}, \bibinfo{pages}{535–547}.
\newblock
\showISBNx{9798350329964}
\href{https://doi.org/10.1109/ASE56229.2023.00063}{doi:\nolinkurl{10.1109/ASE56229.2023.00063}}


\end{thebibliography}

\end{document}